%% file: NANOG_PT.tex
\documentclass[aps,prl,reprint,superscriptaddress,floatfix,nofootinbib,showkeys,showpacs,preprintnumbers]{revtex4-2}
\usepackage{natbib}                          
\usepackage{slashed}
\usepackage{graphics}
\usepackage{multind}
\usepackage[latin1]{inputenc}
\usepackage{wrapfig}
\usepackage{amssymb}
\usepackage{latexsym}
\usepackage{mathtools}
\usepackage{dsfont}
\usepackage[colorlinks=true]{hyperref}
\usepackage{float}
\usepackage{amsfonts}
\usepackage{enumitem}
\usepackage[dvipsnames,table,xcdraw]{xcolor}
\usepackage[table,xcdraw]{xcolor}
\usepackage[framemethod=TikZ]{mdframed}
\usepackage{xstring} 
\usepackage{xspace} 
\usepackage[normalem]{ulem}
\usepackage{fontawesome}
\usepackage{aasmacros}

\newcommand{\yr}{\,{\rm yr}}

\newcommand{\MeV}{\,{\rm MeV}}
\newcommand{\GeV}{\,{\rm GeV}}

\newcommand{\GW}{{\rm GW}}

\newcommand{\eq}[1]{eq.~(\ref{#1})}
\newcommand{\Fig}[1]{Fig.~\ref{#1}}

\definecolor{lightblue}{rgb}{0.2,0.5,1}
\hypersetup{colorlinks=true,
linkcolor=lightblue,
citecolor=lightblue,
urlcolor=lightblue,
linktocpage=lightblue,
pdfproducer=lightblue}

\newcommand\Tstrut{\rule{0pt}{3.5ex}}         % = `top' strut
\newcommand\Bstrut{\rule[-2ex]{0pt}{0pt}}   % = `bottom' strut

\defcitealias{abb+15}{NG9}
\defcitealias{abb+16}{NG9gwb}
\defcitealias{abb+18a}{NG11}
\defcitealias{abb+18b}{NG11gwb}
\defcitealias{aab+20}{NG12}
\defcitealias{abb+20}{NG12gwb}

\begin{document}

\title{Searching For Gravitational Waves From Cosmological Phase Transitions \\With The NANOGrav 12.5-year dataset}
\input{authors_ng12p5_pt}

%\correspondingauthor{}
%\email{}
\collaboration{The NANOGrav Collaboration}
\noaffiliation

\begin{abstract}
We search for a first-order phase transition gravitational wave signal in $45$ pulsars from the NANOGrav 12.5 year dataset. We find that the data can be modeled in terms of a strong first order phase transition taking place at temperatures below the electroweak scale. However, we do not observe any strong preference for a phase-transition interpretation of the signal over the standard astrophysical interpretation in terms of supermassive black holes mergers; but we expect to gain additional discriminating power with future datasets, improving the signal to noise ratio and extending the sensitivity window to lower frequencies. An interesting open question is how well gravitational wave observatories could separate such signals. 
\end{abstract}

\maketitle

% %-------------------------------------------------------------------------------------------------
% 	         Introduction
% %-------------------------------------------------------------------------------------------------
\textit{Introduction ---}
The search for gravitational waves (GWs) spans many orders of magnitude and encapsulates a plethora of source phenomena. At very-low frequencies ($\sim1-100$~nHz), pulsar-timing arrays (PTAs; \cite{fb90,det79,saz78}) aim to detect GWs through the presence of correlated deviations to radio-pulse arrival times across an ensemble of precisely-timed Milky Way millisecond pulsars. There are three PTA collaborations that currently have decadal-length timing data from an ensemble of pulsars: The North American Nanohertz Observatory for Gravitational Waves (NANOGrav; \citep{ransom+19}), the European Pulsar Timing Array (EPTA; \citep{dcl+16}), and the Parkes Pulsar Timing Array (PPTA; \citep{krh+20}). These three, in addition to the Indian PTA (InPTA; \citep{InPTA}), are synthesized into the International Pulsar Timing Array (IPTA; \citealt{pdd+19}).  There are also emerging efforts in China (CPTA; \citep{CPTA}), as well as some telescope-centered timing programs (MeerKAT; \citep{MeerTime}; CHIME; \citep{CHIMEPulsar}).

The dominant GW signals at such low frequencies frequencies are expected to be from a cosmic population of tightly-bound inspiralling supermassive binary black holes (SMBHBs; \citep{shm+04, stc+19}), producing an aggregate incoherent signal that we search for as a stochastic GW background (GWB), and also individual binary signals that we attempt to resolve out of this stochastic confusion background. However, other more
speculative GW sources in the PTA frequency range include cosmic strings \citep{smc07, bos18}, a primordial GWB produced by quantum fluctuations of the gravitational field in the early universe, amplified by inflation \cite{g75, lms+16, Vagnozzi:2020gtf}, and cosmological phase transitions \citep{1973ApJ...182..919W,Hogan:1986qda,Deryagin:1986qq,cds10,klm+17}, the latter of which is the subject this study.

The most recent PTA results are from NANOGrav's analysis of  $12.5$ years of precision timing data from $47$ pulsars \citep[][hereafter \citetalias{aab+20}]{aab+20}, of which $45$ exceeded a timing baseline of $3$ years and were analysed in a search for a stochastic GWB \citep[][hereafter \citetalias{abb+20}]{abb+20}. NANOGrav reported strong evidence for a common-spectrum low-frequency stochastic process in its array of $45$ analyzed pulsars, where $\sim10$ of those pulsars are strongly supportive, most are ambivalent, and a few seem to disfavor the process (although not significantly). No evidence for the characteristic inter-pulsar correlation signature imparted by GWs was found. At low frequencies the shape of the characteristic strain spectrum was well matched to a power-law, with an amplitude and slope consistent with theoretical models of SMBHB populations. Under a model that assumes the origin of the GWB is a population of SMBHBs, the median characteristic strain amplitude at a frequency of $1/\mathrm{year}$ is $1.92\times10^{-15}$. Interpretations of this common-spectrum process as a GWB from SMBHBs have since appeared in the literature, showing that, if it is indeed so, robust evidence of the distinctive inter-pulsar correlations should accrue within the next several years, followed by characterization of the strain spectrum and astrophysical probes of the underlying population \citep{2021MNRAS.502L..99M,2020arXiv201011950P}. However, the Bayesian posterior probability distributions of the strain-spectrum amplitude and slope are broad enough to entertain a variety of different source interpretations, many of which have since appeared in the literature \citep[e.g.][]{ng12p5_pbh_1,ng12p5_pbh_2,ng12p5_cs_1,ng12p5_cs_2}.

In this Letter we consider gravitational waves produced by first-order cosmological phase transitions, both as an alternative origin of the common process measured in the NANOGrav 12.5 year Dataset \citep{2021arXiv210212428B,2021PhRvD.103L1302N,2021arXiv210108012L,2020arXiv201211969B,2020arXiv201006193A,2020arXiv200911875R,2020arXiv200910327A,2020arXiv200909754N}, and as a sub-dominant signal to that produced by SMBHBs.  The frequency range to which NANOGrav is sensitive corresponds to phase transitions at temperatures below the electroweak phase transitions of the Standard Model (\emph{i.e.} $T\lesssim 100\GeV$). This has led many to consider higher frequency GW observatories, such as LISA and LIGO, as the dominant instruments to search for phase transitions.  However, phase transitions may occur at much lower temperatures in particular in {\em hidden sectors} \cite{Strassler:2006im, Chacko:2004ky, Schwaller:2015tja}.  Hidden sectors/valleys feature rich dynamics, with multiple matter fields and forces, independent of the dynamics of the Standard Model.  They appear generically in top-down constructions like string theory, and in some solutions to the so-called hierarchy problem.  In many cases, they may be difficult to detect via their particle interactions with the Standard Model, but gravity is an irreducible messenger.   In this regard, PTAs provide a powerful complementary probe to the dynamics of hidden sectors already being explored through many terrestrial, astrophysical and cosmological probes (see Ref.~\cite{2017arXiv170704591B} for a recent summary). 

Previous studies on cosmological first order phase transition in the context of the NANOGrav results were carried out  in \cite{2020arXiv200911875R, Li:2021qer,Bian:2020bps,2021PhRvD.103L1302N}.  Our analysis presents two main novelties compared to these works: first, we properly include the relevant noise sources and discuss the impact of backgrounds (like the one generated by SMBHBs); second, we discuss how the results are affected by the theoretical uncertainties on the GW spectrum produced by first order phase transitions.

The outline of this Letter is as follows. In the next section we briefly summarize the signature of GWs from the dominant background of SMBH mergers. We then dive into the main subject of this Letter, GWs from a first-order phase transition, where we discuss the relevant parameters characterizing the signal. We then carry out an analysis with the NANOGrav 12.5 year dataset, finding that the data can be modeled in terms of a strong phase transition with a transition temperature around 10 MeV.  The dataset and data model for these analyses are exactly as described in \citetalias{aab+20} and \citetalias{abb+20}, respectively. All common processes (whether interpreted as being of SMBHB or phase-transition origin) are modeled within the five lowest sampling frequencies of the array time series, corresponding to $\sim2.5-12$~nHz. Finally, we discuss theoretical uncertainties, and compare the PT interpretation of the data to the standard interpretation in terms of SMBHB finding no strong preference for one over the other.

% %-----------------------------------------------
% 	     GW from SMBHBs mergers
% %-----------------------------------------------

\textit{GW from SMBHBs mergers ---}
Regardless of origin, the energy density of GWs as a fraction of closure density is related to the GW characteristic strain spectrum by \citep{2015CQGra..32a5014M}
\begin{equation}\label{eq:energy_density}
	\Omega_{\GW}(f) = \frac{2\pi^2}{3H_0^2} f^2 h_c^2(f),
\end{equation}
where $H_0$ is the Hubble constant (set here to be $67\, \mathrm{ km/s/Mpc}$ \citep{2020A&A...641A...6P}), and the GWB characteristic strain spectrum $h_c(f)$ is often described by a power-law function for astrophysical and cosmological sources:
\begin{equation}\label{eq:characteristic_strain}
	h_c(f) = A_{\mathrm{GWB}}\left(\frac{f}{\yr^{-1}}\right)^{\alpha}\, ,
\end{equation}
where $A_\mathrm{GWB}$ is the amplitude at a reference frequency of $1/\mathrm{year}$, and $\alpha$ is an exponent that depends on the origin of the GWB. For a population of inspiraling SMBHBs, this is $\alpha=-2/3$ \citep{p01}. The cross-power spectral density of GW-induced timing deviations between two pulsars $a$ and $b$ can be written as
\begin{align}\label{eq:cpsd}
	S_{ab}(f) = \Gamma_{ab} \frac{A_{\mathrm{GWB}}^2}{12\pi^2} \left(\frac{f}{\yr^{-1}}\right)^{-\gamma} \yr^3\,,
\end{align}
where $\gamma\equiv 3 - 2\alpha = 13/3$ for SMBHBs, and $\Gamma_{ab}$ is the Hellings-Downs \citep{hd83} correlation coefficient between pulsar $a$ and pulsar $b$. 

% %-----------------------------------
% 	     GW from PT
% %-----------------------------------
%-------------------------- Bubble spectrum parameters table ------------------%
\begin{table}[t]
  \centering
  \renewcommand{\arraystretch}{1}
  \setlength{\tabcolsep}{4pt}
  \setlength\extrarowheight{-10pt}
  \begin{tabular}{ccc}
	\hline\hline
		& {\bf Bubbles} \cite{1605.01403}  & {\bf Sound Waves} \cite{1704.05871}  \Tstrut\Bstrut       \\ \hline \\
	$\Delta(v_w)$ 	&	$\dfrac{0.48v_w^3}{1+5.3v_w^2+5v_w^4}$		&	$0.513\, v_w$		\Tstrut\Bstrut \\ 
	$\kappa$ 		& 	$\kappa_\phi$					& 	$ \kappa_{\rm sw}$		 \Tstrut\Bstrut \\ 
	$p$			&	$2$							&	$2$					 \Tstrut\Bstrut \\ 
	$q$			&	$2$							&	$1$					 \Tstrut\Bstrut \\
	$\mathcal{S}(x)$ &	$\dfrac{(a+b)^c}{\left[bx^{-a/c}+ax^{b/c}\right]^c}$	& $x^3\left(\dfrac{7}{4+3x^2}\right)^{7/2}$		 \\\\
	$f_*/\beta$			&	$\dfrac{0.35}{1+0.07v_w+0.69v_w^4}$	&	$\dfrac{0.536}{v_w}$	 \\\\ \hline\hline
\end{tabular}
\caption{\label{tab:params} Parameters for the gravitational wave spectrum of \eq{eq:spectrum}. The values of the parameters $(a,b,c)$ in the spectral shape of the bubble contribution are reported in Table \ref{tab:bubble_params}.}
\end{table} 
%-------------------------------------------------------------------------------------------%

%---------------- Spectral shape parameters -----------------%
\begin{table}
  \renewcommand{\arraystretch}{1}
  \setlength\extrarowheight{-15pt}
  \begin{tabular}{cccc}
	\hline\hline
				& {\bf Envelope}						& {\bf Semi-analytic}	\quad	&	{\bf Numerical}  \Tstrut\Bstrut       \\ \hline
	$a$			&	$3$								&	$1-2.2$ 				&	$1.6-0.7$	 \Tstrut\Bstrut\\ 
	$b$ 			&	$1$								&	$2.6-2.9$			&	$1.4-2.3$	\Tstrut\Bstrut \\ 
	$c$ 			& 	$1.5$							& 	$1.5-3.5$			&	$1$ \Tstrut\Bstrut \\ 
	$f_*/\beta$        &   $\dfrac{0.35}{1+0.07v_w+0.69v_w^2}$		&	$0.1$				&	$0.2$  \Tstrut\Bstrut \\ \\ \hline\hline
\end{tabular}
\caption{Comparison of the bubble spectral shape parameters derived using the envelope and thin wall approximation \cite{1605.01403} (left column), the semi-analytic approach of reference \cite{2012.07826} (middle column), and lattice simulations \cite{2005.13537} (right column). For numerical and semi-analytic results the values of the parameters depend on the choice of the scalar field potential, we report the range of values obtained for the different scalar field potentials considered in the above mentioned works. \label{tab:bubble_params} }
\end{table} 
%-----------------------------------------------------------------------%

\textit{GWs from first-order phase transition ---}  
A first-order phase transition (PT) occurs when the true minimum of a potential is separated from a false minimum by a barrier through which a field must locally tunnel.  This can occur in either weakly coupled (where a scalar field tunnels) or strongly coupled (where a vacuum condensate corresponds to the scalar field) theory. Such transitions are known to proceed through nucleation of bubbles of true vacuum which, if sufficiently large, expand in the background plasma (still in the false vacuum). Collisions of these bubbles, as well as interactions between the expanding bubble walls and the surrounding plasma, can be efficient sources of GWs.

We characterize the phase transition in terms of four parameters:
\begin{itemize}[leftmargin=*]
	\itemsep -4pt
	\item $T_*$ -- the Universe temperature at which the phase transition takes place.
	\item $\alpha_*$ -- the strength of the phase transition, defined as the ratio of the vacuum and relativistic energy density at the time of the phase transition.
	\item $\beta/H_*$ -- the bubble nucleation rate in units of the Hubble rate at the time of the phase transition, $H_*$.
	\item $v_w$ -- the velocity of the bubble walls.
\end{itemize}
The three main sources of GWs associated with a first-order phase transition are: (i) collisions of bubble walls, (ii) collisions of the sound waves generated in the background plasma by the bubbles expansion, and (iii) turbulence in the plasma generated by expansion and collisions of the sound-wave. However, in this analysis we will not include the turbulence contribution as it usually is subleading compared to the sound-wave one, and also affected by the largest theory uncertanties (see for example \cite{1512.06239, 1903.08585, 2102.12428} for recent developments).

The contribution to the total GW spectrum from bubbles and sound waves collisions can be parametrized as \cite{1201.0983, 1512.06239}
\begin{align}\label{eq:spectrum}
	h^2\Omega(f)=\mathcal{R}\;\Delta(v_w)\left(\frac{\kappa\,\alpha_*}{1+\alpha_*}\right)^p\left(\frac{H_*}{\beta}\right)^q\mathcal{S}\left(f/f_*^0\right)\,,
\end{align}
where the prefactor $\mathcal{R}\simeq 7.69 \times 10^{-5} g_*^{-1/3}$ accounts for the redshift of the GW energy density, $\mathcal{S}(\cdot)$ parametrizes the spectral shape, and  $\Delta(v_w)$ is a normalization factor which depends on the bubble wall velocity, $v_w$. The value of the peak frequency today, $f_*^0$, is related to the value of the peak frequency at emission, $f_*$, by:
\begin{equation}\label{eq:peak_f}
	f_*^0\simeq 1.13\times 10^{-10}{\rm \, Hz}\,\left(\frac{f_*}{\beta}\right)\left(\frac{\beta}{H_*}\right)\left(\frac{T_*}{\MeV}\right)\bigg(\frac{g_*}{10}\bigg)^{1/6}\,,
\end{equation}
where $g_*$ denotes the number of relativistic degrees of freedom at the time of the phase transition. 
The values of the peak frequency at emission, the spectral shape, the normalization factor, and the exponents $p$ and $q$ are reported in Table \ref{tab:params} for all the production mechanisms considered in this work. Due to the finite lifetime \cite{2007.08537, 2003.07360} of the sound waves, to derive $\Omega_{\rm sw}$ eq. \eqref{eq:spectrum} needs to be multiplied by a suppression factor $\Upsilon(\tau_{\rm sw})$ given by \cite{2007.08537}:
\begin{equation}
\Upsilon(\tau_{\rm sw}) = 1 - (1+2\tau_{\rm sw}H_*)^{-1/2}
\end{equation}
where the sound-wave lifetime is usually taken to be the timescale for the onset of turbulent behaviors in the plasma \cite{1705.01783}: $\tau_{\rm sw}\approx R_*/\bar U_f$, where the average bubble separation is given by $R_*=(8\pi)^{1/3}\beta^{-1}{\rm Max}(v_{\rm w},c_{s})$ \cite{1909.10040}, and $\bar U^2_f\approx3\kappa_{\rm sw}\alpha/[4(1+\alpha_*)]$ \cite{1705.01783}.

Generally both the production mechanism contribute to the GW spectrum. However, if the bubble walls interacts with the surrounding plasma most of the energy released in the PT is expected to be transferred to the plasma so that the sound waves (and possibly the turbulence) contribution dominates the GW spectrum. An exception to this scenario is provided by models in which the bubble walls do not interact with the plasma, or by models where the energy released in the PT is large enough that the friction exerted by the plasma is not enough to stop the walls from keep accelerating ({\it runaway scenario}). However, determining wether or not the runaway regime is realized is either model dependent or affected by large theoretical uncertainties. Therefore, we perform two separate of analyses. A {\it sound-wave-only} (SWO) {\it analysis}, where we assume that the runaway regime is not reached and that the sound wave and turbulence contributions dominate the GW spectrum; therefore we set $\kappa_{\phi}=0$, and use the results of reference \cite{1004.4187} to derive $\kappa_{\rm sw}$ as a function of $v_w$ and $\alpha_*$. A {\it bubble-only} (BO) {\it analysis}, where we assume that the runaway regime is reached and that bubble collisions dominate the GW spectrum; we then set $v_w=1$, $\kappa_{\rm sw}=0$ and $\kappa_{\phi}=1$. 

\begin{figure*}[t]
    \includegraphics[width=0.9\textwidth]{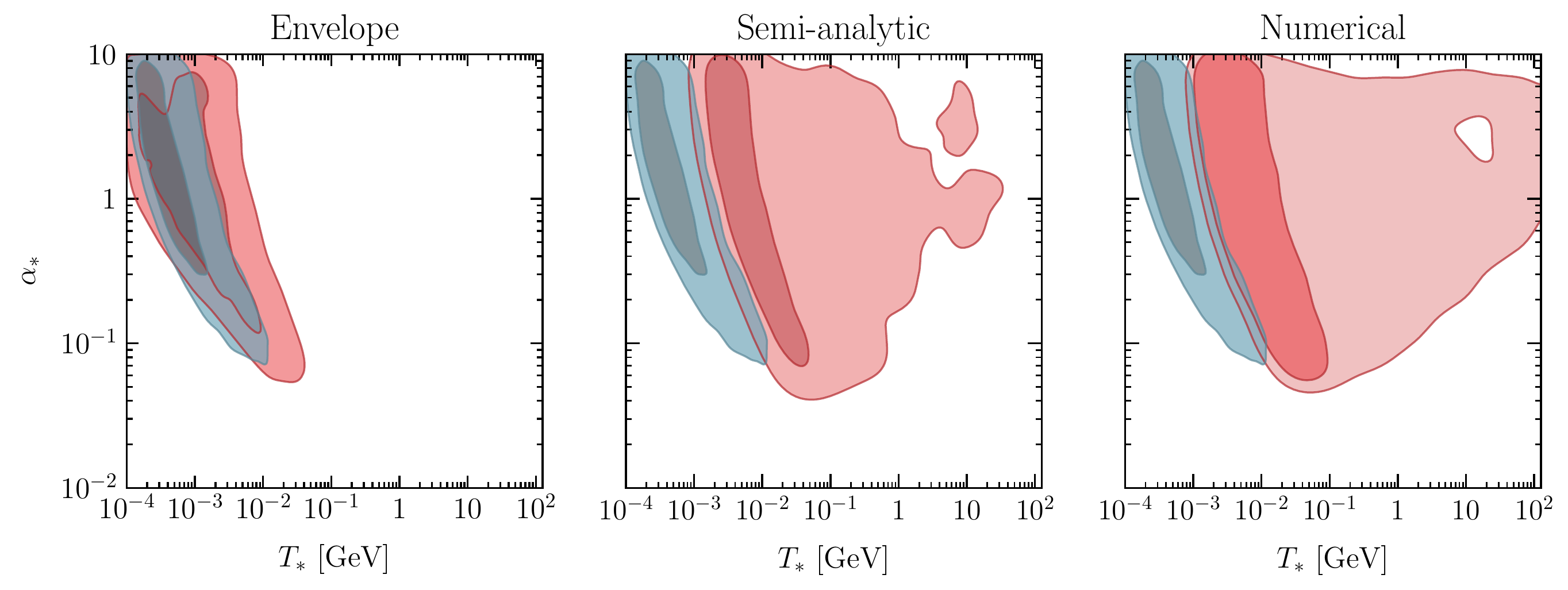}
    \caption{In red (blue) the 1-$\sigma$ ($68\%$ posterior credible level), and 2-$\sigma$ ($95\%$ posterior credible level) contours for the two-dimensional posterior distributions in the  $(T_*,\alpha_*)$ plane obtained in the BO (SWO). The BO analysis has been performed with the spectral shape computed by using the envelope approximation (left panel), semi-analytic results (central panel), and numerical results (right panel). Specifically, we use $(a,b,c)=(1, 2.61, 1.5)$ for the semi-analytic results, and  $(a,b,c)=(0.7, 2.3, 1)$ for the numerical results.}
    \label{fig:T_alpha_theo}
\end{figure*}

We conclude this section emphasizing that, despite recent progress, large theoretical uncertainties still affect the prediction of the GW signal produced in cosmological phase transitions. To get an idea of the impact that these uncertainties have on our results we will study how the BO analysis is impacted by them. Similar, if not larger, uncertainties affect the sound wave contribution and would impact the results of the SWO analysis. 

Assuming that the stress energy density of the expanding bubbles is localized in an infinitesimally thin shell near the bubble wall (thin shell approximation), and that it instantaneously decays to zero after two bubbles collide (envelope approximation), the bubble spectral shape can be derived analytically \cite{Kosowsky:1991ua,1605.01403}. The spectral shape parameters obtained in this way are reported in the left column of Table \ref{tab:bubble_params}. To go beyond these approximations, 3D lattice simulations are needed. These simulations are extremely expensive given the hierarchy between the large simulation volume needed to include multiple bubbles, and the small lattice spacing needed to resolve the thin walls. Because of the relativistic contraction of the wall width, this separation of scales becomes increasingly large for increasing wall velocities, making it impossible to simulate ultra-relativistic walls. However, the GW spectrum can be simulated at lower velocities and the results extrapolated to larger values. This is the approach taken in Refs.~\cite{1802.05712, 2005.13537}, where the authors show that at high frequencies the GW spectrum is much steeper than predicted by the envelope approximation ($b\sim1.4-2.3$ depending on the form of the scalar field potential). An alternative approach to the problem has been taken by the authors of Refs.~\citep{2007.04967, 2012.07826}.  In these works a parametric form for the evolution of the scalar field during bubble collisions is found by using two-bubble simulations. This parametric form is then used in many-bubble simulations to derive the GW spectrum. They also find a steeper high frequency slope ($b\sim2.6-2.9$) compared to the prediction of the envelope approximation. Similar discrepancies are found at low frequencies, where both the numerical and semi-analytic results find a shallower spectrum compared to the envelope approximation (see Tab. \ref{tab:bubble_params}). To probe the theoretical uncertainty associated with the bubble contribution, we will carry out three separate BO analyses utilizing each approach and compare the constraint on the phase transition temperature and strength.

% %-------------------------
% 	      Results                   
% %-------------------------
\begin{figure}[t]
    \includegraphics[width=0.48\textwidth]{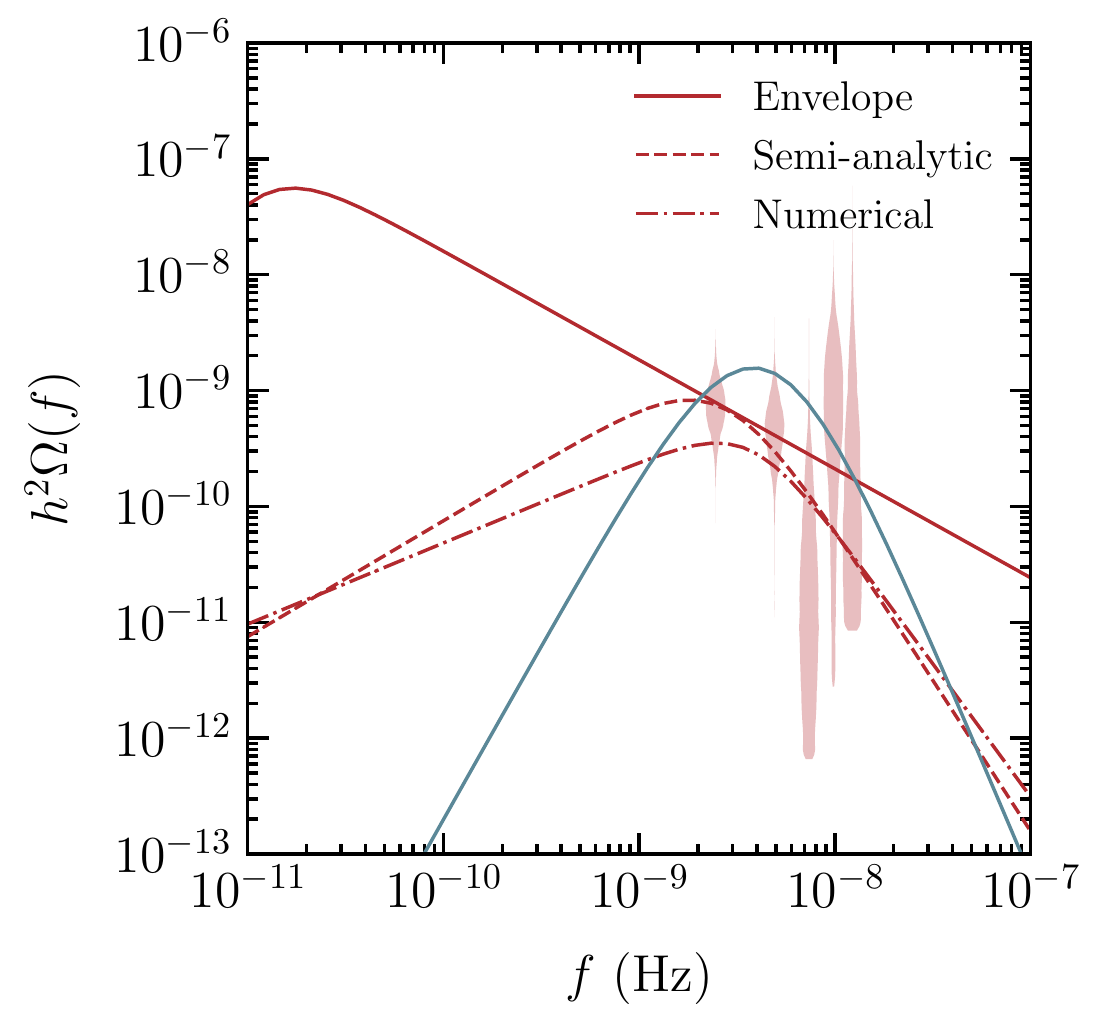}
    \caption{Maximum likelihood GWB fractional energy-density spectrum for the BO (red) and SWO (blue) analyses compared with the marginalized posterior for the free power spectrum (independent per-frequency characterization; red violin plot) derived in \citetalias{abb+20}. For the BO analysis we show the results derived by using the envelope (solid line), semi-analytic (dashed), and numerical (dot-dashed) spectral shapes. For the BO analyses the values of $(\alpha_*,T_*)$ for these maximum likelihood spectra are $(0.28,0.7\MeV)$ for the envelope results, $(1.2,3.4\MeV)$ for the semi-analytic results, and $(0.13,14.1\MeV)$ for the numerical results. While for the SO analysis we get $(6.0, 0.32\MeV)$.}
    \label{fig:best_fit}
\end{figure}

\textit{Results ---} We now report our results for the BO and SWO analyses. For either of them we consider both the case where the only GW signal is the one produced by the PT, and the one in which the PT signal is superimposed to an astrophysical background produced by SMBHB. This latter analyses will give an indication of how difficult it will be to disentangle a signal from a phase transition from the SMBHB background. The prior distributions for the model parameters of all these analyses, in addition to other noise characterization parameters, are listed in Table \ref{tab:priors}.

%-----------Prior Table-------------%
\begin{table*}[ht]
\begin{center}
\renewcommand{\arraystretch}{1.3}
\setlength{\tabcolsep}{6.5pt}
\small
\begin{tabular}{llll}
\hline\hline
\multicolumn{1}{c}{\bf{Parameter}}  & \multicolumn{1}{c}{\bf{Description}} & \multicolumn{1}{c}{\bf{Prior}} & \multicolumn{1}{c}{\bf{Comments}} \\
\hline

\multicolumn{4}{c}{\bf{White Noise}} \\[1pt]
$E_{k}$ & EFAC per backend/receiver system & Uniform $[0, 10]$ & single-pulsar analysis only \\
$Q_{k}$ [s] & EQUAD per backend/receiver system & log-Uniform $[-8.5, -5]$ & single-pulsar analysis only \\
$J_{k}$ [s] & ECORR per backend/receiver system & log-Uniform $[-8.5, -5]$ & single-pulsar analysis only \\
\hline

\multicolumn{4}{c}{\bf{Red Noise}} \\[1pt]
$A_{\rm red}$ & red-noise power-law amplitude & log-Uniform $[-20, -11]$ & one parameter per pulsar  \\
$\gamma_{\rm red}$ & red-noise power-law spectral index & Uniform $[0, 7]$ & one parameter per pulsar \\
\hline

\multicolumn{4}{c}{\bf{Phase Transition}} \\[1pt]
$T_*\,[\rm GeV]$ & phase transition temperature & log-Uniform $[-4, 3]$ & one parameter for PTA \\
$\alpha_*$ & phase transition strength & log-Uniform $[-1.3, 1]$ & one parameter for PTA \\
$H_*/\beta$ & bubble nucleation rate & log-Uniform $[-2, 0]$ & one parameter for PTA \\
$v_w$ & bubble wall velocity & log-Uniform $[-2, 1]$ & one parameter for PTA \\
\hline

\multicolumn{4}{c}{\bf{Supermassive Black Bole Binaries (SMBHB)}} \\[1pt]
$A_{\mathrm{GWB}}$ & common process strain amplitude & log-Uniform $[-18, -14]$ & one parameter for PTA \\
$\gamma_{\mathrm{GWB}}$ & common process power-law spectral index & delta function ($\gamma_\mathrm{GWB}=13/3$)& fixed \\
\hline\hline

\end{tabular}
\caption{Priors distributions for the parameters used in all the analyses in this work. The prior for the bubble wall velocity reported in this table is the one used for the SWO analysis, for the BO analyses we use $v_w=1$ as explained in the text. \label{tab:priors}}
\end{center}
\end{table*}
%------------------------------------------%

The two parameters that we can constrain the most are the transition temperature, $T_*$, and the phase transition strength, $\alpha_*$. Their 2D posterior distributions for the PT-only searches are shown in Fig.~\ref{fig:T_alpha_theo}. To assess the impact of theoretical uncertainties related to the bubble spectrum, for the BO analysis we report the results obtained by using the three different estimates of the bubble contribution to the GW spectrum described in the previous section (envelope, semi-analytic, and numerical). We can see that at the 1-$\sigma$ ($68\%$ posterior credible) level all the searches prefer a strong PT, $\alpha_*\gtrsim0.1$, with low transition temperature, $T_*\lesssim 10\MeV$. At 2-$\sigma$ ($95\%$ posterior credible) level the posteriors for the semi-analytical and numerical results have support at much higher temperatures, while the envelope results still prefer relatively low values. The preference for small values of $T_*$ at the 1-$\sigma$ level can be understood by noticing (see \Fig{fig:best_fit}) that the data prefer GW spectra that are peaked at frequencies below the NANOGrav sensitivity window (\emph{i.e.} $f_*^0\lesssim 10^{-9}\,{\rm Hz}$). And, by setting $\beta/H_*=1$ in \eqref{eq:peak_f}, we see that this requirement corresponds to $T_*\lesssim 10\MeV$. The low-frequency part of the numerical and semi-analytical GW spectra is shallow enough that, at the 2-$\sigma$ level, the data can be fitted also by spectra with peak frequencies above the NANOGrav sensitivity window. The same is not true for the envelope results, which have a much steeper low-frequency spectrum; this is the reason why the 2-$\sigma$ levels of the envelope results deviate substantially from the other two.

In \Fig{fig:best_fit} we show the GWB spectrum predicted for the maximum likelihood parameters of PT-only searches. To better illustrate our results, and how the different parameters and theoretical uncertainties affect the GWB spectrum, we release an interactive version of \Fig{fig:best_fit} at this \href{https://mybinder.org/v2/gh/AMitridate/bokeh/master?urlpath=%2Fproxy%2F5006%2Fbokeh-app}{link}. 

\begin{figure}[t]
\centering
    \includegraphics[width=0.48\textwidth]{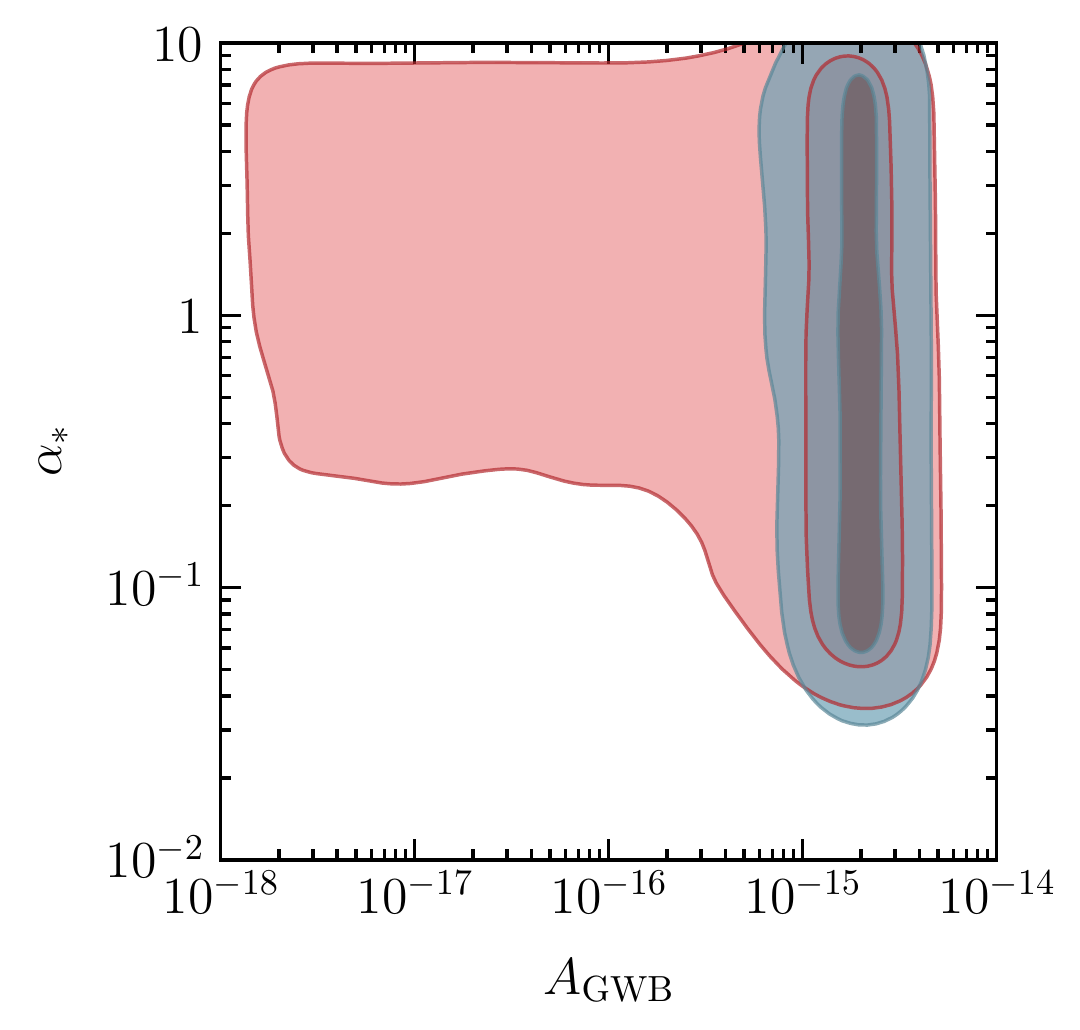}
    \caption{1-$\sigma$ ($68\%$ posterior credible level), and 2-$\sigma$ ($95\%$ posterior credible level) contours for the parameters $A_{\rm GWB}$ and $\alpha_*$ in the PT+SMBHB search. In red (blue) the results for the BO (SWO) analyses. In this figure we have used the semi-analytic results for the bubble spectrum. The posteriors do not extend to lower values of $\alpha_*$ because of our choice for the $\alpha_*$ prior: log-Uniform $[-1.3, 1]$.}
    \label{fig:SMBHB}
\end{figure}

To understand how the inclusion of the SMBHB background affects our results, in \Fig{fig:SMBHB} we show the posterior for the parameters $\alpha_*$ and $A_{\rm GWB}$ obtained in the PT+SMBHB search. As expected, with the inclusion of the SMBHB background, the posteriors for $\alpha_*$ stretch to lower values where most of the signal is provided by the SMBHB contribution.\footnote{The posterior stops around $\alpha_*\simeq0.05$ because of choice of the prior, otherwise it woiuld extend down to $\alpha_*\sim0$.} The Bayesian Information Criterion (BIC)~\cite{1978AnSta...6..461S}, defined to be $\mathrm{BIC}=k\ln n-2\ln \hat{\mathcal{L}}$ where $n=5$ is the number of data points in the frequency space, $k$ is the number of parameters in the model and $\hat{\mathcal{L}}$ is the maximum likelihood, is also computed. For the BO searches, the differences in BIC between the PT+SMBHB and SMBHB only searches are found to be $-0.92$, $3.04$ and $1.89$ for the envelope, semi-analytic and numerical results respectively; similarly the BIC differences between the PT-only and SMBHB-only searches are $-1.82$, $-3.18$, $-1.28$.  For the SWO analysis, the difference in BIC between the PT+SMBHB and SMBHB only searches is $-4.56$, while we get $-2.19$ for the difference between  the PT-only and SMBHB-only searches. We can then conclude that that the PT+SMBHB and PT-only models were neither strongly favored nor disfavored compared to the SMBHB only model~\cite{doi:10.1080/01621459.1995.10476572}.

\begin{figure*}[t]
    \centering
    \includegraphics[width=\textwidth]{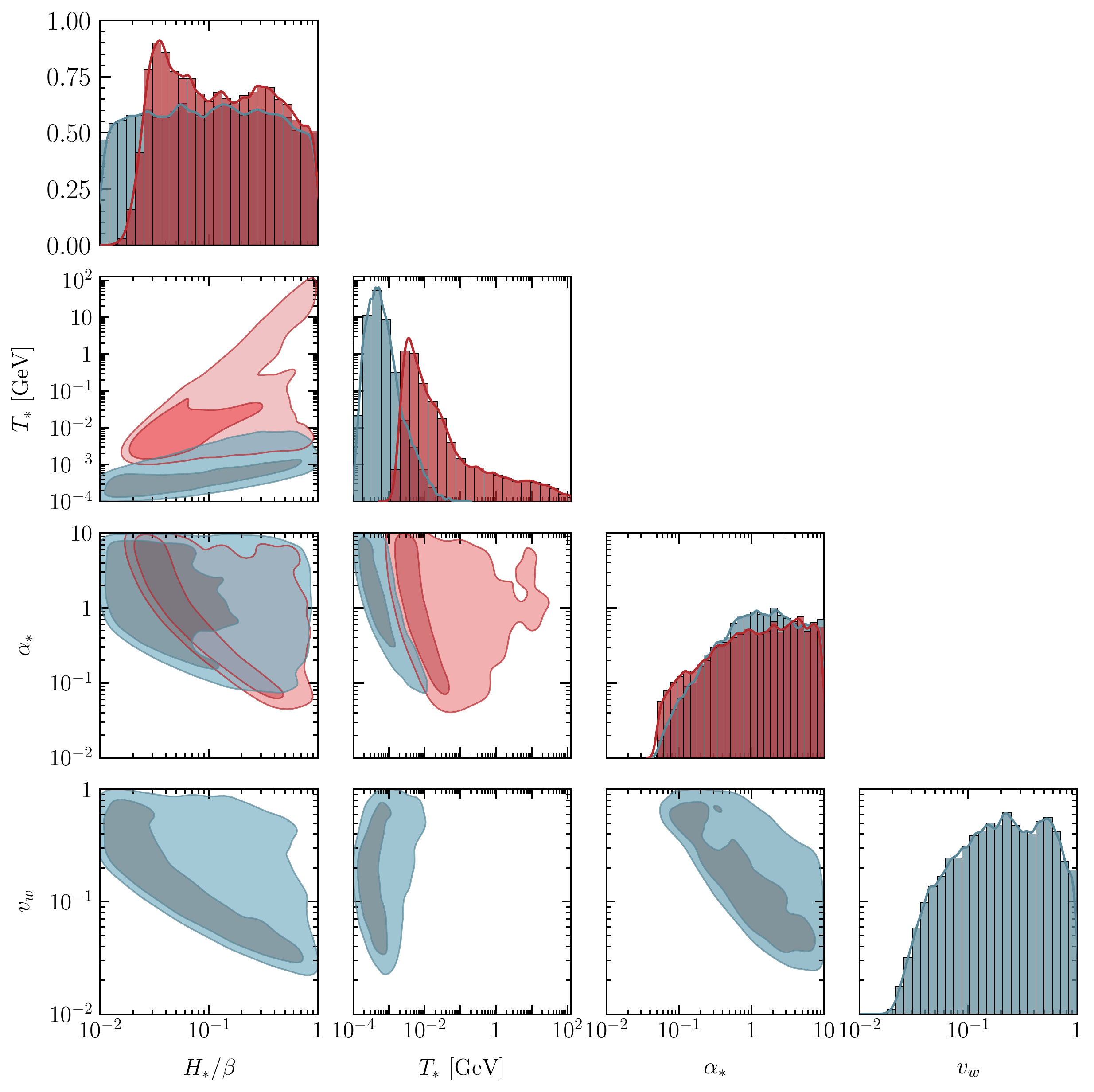}
    \caption{Corner plot showing the 1D and 2D posterior distributions for the parameters of the PT-only search. In red (blue) the results for the BO (SWO) analyses. In deriving these results we have used the semi-analytic bubble spectral shape with $(a, b, c) = (1, 2.61, 1.5)$.}
    \label{fig:corner}
\end{figure*}

A complete set of posteriors for the parameters of the PT-only searches (derived by using the semi-analytic spectrum for the BO analysis) are shown in Fig.~\ref{fig:corner}.  As noted previously, at 1-$\sigma$ level the data prefer a strongly first-order phase transition ($\alpha_*\gtrsim0.1$) taking place at temperature $T_*\lesssim 10\MeV$;  while no strong constraints on $v_w$ or $H_*/\beta$ is observed. We can also notice that the higher values of $T_*$ allowed in the 2-$\sigma$ region are accompanied by slower nucleation rates (large $H_*/\beta$). We should caution, however, that numerical simulation have been performed for phase transition strengths up to $\alpha_*\sim0.5$ \cite{1906.00480}, and that our results for $\alpha_*\gtrsim0.5$ are derived by extrapolating the results of these simulations. A similar remark should be made for $H_*/\beta$, numerical simulations with values of this parameter close to unity have not been performed yet. 

Given the low value of $T_*$, and the strong constraints on new physics at such low scales, we expect the phase transition to take place in a dark sector with only feeble interactions with the Standard Model (SM). In order to be consistent with the Hubble parameter constraints during the era of Big Bang Nucleosynthesis (BBN) \cite{2020PTEP.2020h3C01P}, the energy of this dark sector must be transferred to the SM before the onset of BBN at $T\sim 1\MeV$.  This leaves an allowed range of values for the transition temperature given by $T_*\sim1\MeV-100\GeV$. The next data release, which adds multiple years of observations and extends the the sensitivity window to lower frequency, should begin to resolve the peak of the spectrum or additionally shrink the range of allowed values for $T_*$.

\textit{Conclusions ---} We performed a search for a stochastic gravitational wave background from first-order phase transitions in the 12.5 year NANOGrav dataset. While previous NANOGrav analysis found no evidence yet for the inter-pulsar correlation signature of a GWB, the evidence for a common-spectrum process was significant. We found that the data can be modeled by a strong ($\alpha_*>0.1$) phase transition taking place at temperatures below the electroweak scale. However, the data do not show any strong preference between an SMBHB and a PT generated signal, but we expect to gain additional discriminating power with future datasets, improving the signal to noise ratio and extending the sensitivity window to lower frequencies. In particular, data from the International Pulsar Timing Array will allow the baseline of observations to be significantly extended, and the number of monitored pulsars to be greatly expanded. The present quality of the data is such that our results are not strongly affected by theoretical uncertainties on the GW spectral shape. However, methodological improvements on determining the origin of the GWB spectrum will be needed for future datasets in order to separate the signal from a first-order PT from the SMBHB background, as well as to constrain the microscopic origins of the PT. 

\textit{Author contributions ---} An alphabetical-order author list was used for this paper in recognition of the fact that a large, decade timescale project such as NANOGrav is necessarily the result of the work of many people. All authors contributed to the activities of the NANOGrav collaboration leading to the work presented here, and reviewed the manuscript, text, and figures prior to the paper's submission. Additional specific contributions to this paper are as follows. ZA, HB, PRB, HTC, MED, PBD, TD, JAE, RDF, ECF, EF, NG-D, PAG, DCG, MLJ, MTL, DRL, RSL, JL, MAM, CN, DJN, TTP, NSP, SMR, KS, IHS, RS, JKS, RS and SJV developed the 12.5-year dataset through a combination of observations, arrival time calculations, data checks and refinements, and timing model development and analysis; additional specific contributions to the dataset are summarized in NG12. KZ and SRT coordinated the writing of the paper. VL and AM performed all analyses presented in this paper. KZ, SRT, AM, and VL wrote the paper and collected the bibliography.

\textit{Acknowledgments ---} This work has been carried out by the NANOGrav collaboration, which is part of the International Pulsar Timing Array. %We thank the members of the IPTA Steering Committee whose comments helped improve and clarify the manuscript. %We also thank the anonymous reviewers for useful suggestions and comments, which improved the quality of the manuscript. 
The NANOGrav project receives support from National Science Foundation (NSF) Physics Frontiers Center award number 1430284. The Arecibo Observatory is a facility of the NSF operated under cooperative agreement (\#AST-1744119) by the University of Central Florida (UCF) in alliance with Universidad Ana G. M\'endez (UAGM) and Yang Enterprises (YEI), Inc. The Green Bank Observatory is a facility of the NSF operated under cooperative agreement by Associated Universities, Inc. The National Radio Astronomy Observatory is a facility of the NSF operated under cooperative agreement by Associated Universities, Inc. A majority of the computational work was performed on the Caltech High Performance Cluster, partially supported by a grant from the Gordon and Betty Moore Foundation.. This work made use of the Super Computing System (Spruce Knob) at WVU, which are funded in part by the National Science Foundation EPSCoR Research Infrastructure Improvement Cooperative Agreement \#1003907, the state of West Virginia (WVEPSCoR via the Higher Education Policy Commission) and WVU. Part of this research was carried out at the Jet Propulsion Laboratory, California Institute of Technology, under a contract with the National Aeronautics and Space Administration. Portions of this work performed at NRL were supported by Office of Naval Research 6.1 funding. The Flatiron Institute is supported by the Simons Foundation. Pulsar research at UBC is supported by an NSERC Discovery Grant and by the Canadian Institute for Advanced Research.  SRT
acknowledges support from NSF grant AST-\#2007993, and a Dean's Faculty Fellowship from Vanderbilt University's College of Arts \& Science. JS and MV acknowledge support from the JPL RTD program. SBS acknowledges support for this work from NSF grants \#1458952 and \#1815664. SBS is a CIFAR Azrieli Global Scholar in the Gravity and the Extreme Universe program. TTP acknowledges support from the MTA-ELTE Extragalactic Astrophysics Research Group, funded by the Hungarian Academy of Sciences (Magyar Tudom\'anyos Akad\'emia), that was used during the development of this research. TD and ML acknowledge NSF AAG award number 2009468. This work is supported in part by NASA under award number 80GSFC17M0002. VL, AM and KZ are supported by the U.S. Department of Energy, Office of Science, Office of High Energy Physics, under Award Number DE-SC0021431 and a Simons Investigator award

\textit{Facilities ---} Arecibo, GBT

\textit{Software ---} \texttt{ENTERPRISE} \citep{enterprise}, \texttt{enterprise\_extensions} \citep{enterprise_ext}, \texttt{HASASIA} \citep{hasasia}, \texttt{libstempo} \citep{libstempo}, \texttt{matplotlib} \citep{matplotlib}, \texttt{PTMCMC} \citep{ptmcmc}, \texttt{tempo} \citep{tempo}, \texttt{tempo2} \citep{tempo2}, \texttt{PINT} \citep{pint}

% %-------------------------------------------------------------------------------------------------
% 	         Bibliography
% %-------------------------------------------------------------------------------------------------
\bibliography{bib}

\end{document}

%% file: authors_ng12p5_pt.tex
% DO NOT EDIT THIS FILE. EDITS WILL BE OVERWRITTEN.
% AUTO-GENERATED WITH make-aastex62-author-list.py
% FROM authorship/author_list_12yr_data.txt, authorship/author_affil_and_orcid.txt, AND authorship/affil.txt
\author{Zaven Arzoumanian}
\affiliation{X-Ray Astrophysics Laboratory, NASA Goddard Space Flight Center, Code 662, Greenbelt, MD 20771, USA}
\author{Paul T. Baker}
\affiliation{Department of Physics and Astronomy, Widener University, One University Place, Chester, PA 19013, USA}
\author{Harsha Blumer}
\affiliation{Department of Physics and Astronomy, West Virginia University, P.O. Box 6315, Morgantown, WV 26506, USA}
\affiliation{Center for Gravitational Waves and Cosmology, West Virginia University, Chestnut Ridge Research Building, Morgantown, WV 26505, USA}
\author{Bence B\'{e}csy}
\affiliation{Department of Physics, Montana State University, Bozeman, MT 59717, USA}
\author{Adam Brazier}
\affiliation{Cornell Center for Astrophysics and Planetary Science and Department of Astronomy, Cornell University, Ithaca, NY 14853, USA}
\affiliation{Cornell Center for Advanced Computing, Cornell University, Ithaca, NY 14853, USA}
\author{Paul R. Brook}
\affiliation{Department of Physics and Astronomy, West Virginia University, P.O. Box 6315, Morgantown, WV 26506, USA}
\affiliation{Center for Gravitational Waves and Cosmology, West Virginia University, Chestnut Ridge Research Building, Morgantown, WV 26505, USA}
\author{Sarah Burke-Spolaor}
\affiliation{Department of Physics and Astronomy, West Virginia University, P.O. Box 6315, Morgantown, WV 26506, USA}
\affiliation{Center for Gravitational Waves and Cosmology, West Virginia University, Chestnut Ridge Research Building, Morgantown, WV 26505, USA}
\affiliation{CIFAR Azrieli Global Scholars program, CIFAR, Toronto, Canada}
\author{Maria Charisi}
\affiliation{Department of Physics and Astronomy, Vanderbilt University, 2301 Vanderbilt Place, Nashville, TN 37235, USA}
\author{Shami Chatterjee}
\affiliation{Cornell Center for Astrophysics and Planetary Science and Department of Astronomy, Cornell University, Ithaca, NY 14853, USA}
\author{Siyuan Chen}
\affiliation{Station de Radioastronomie de Nancay, Observatoire de Paris, Universite PSL, CNRS, Universite d'Orleans, 18330 Nancay, France}
\affiliation{FEMTO-ST Institut de recherche, Department of Time and Frequency, UBFC and CNRS, ENSMM, 25030 Besancon, France}
\affiliation{Laboratoire de Physique et Chimie de l'Environment et de l'Espace, LPC2E UMR7328, Universite d'Orleans, CNRS, 45071 Orleans, France}
\author{James M. Cordes}
\affiliation{Cornell Center for Astrophysics and Planetary Science and Department of Astronomy, Cornell University, Ithaca, NY 14853, USA}
\author{Neil J. Cornish}
\affiliation{Department of Physics, Montana State University, Bozeman, MT 59717, USA}
\author{Fronefield Crawford}
\affiliation{Department of Physics and Astronomy, Franklin \& Marshall College, P.O. Box 3003, Lancaster, PA 17604, USA}
\author{H. Thankful Cromartie}
\affiliation{Cornell Center for Astrophysics and Planetary Science and Department of Astronomy, Cornell University, Ithaca, NY 14853, USA}
\author{Megan E. DeCesar}
\altaffiliation{NANOGrav Physics Frontiers Center Postdoctoral Fellow}
\affiliation{Department of Physics, Lafayette College, Easton, PA 18042, USA}
\affiliation{George Mason University, Fairfax, VA 22030, resident at U.S. Naval Research Laboratory, Washington, D.C. 20375, USA}
\author{Paul B. Demorest}
\affiliation{National Radio Astronomy Observatory, 1003 Lopezville Rd., Socorro, NM 87801, USA}
\author{Timothy Dolch}
\affiliation{Department of Physics, Hillsdale College, 33 E. College Street, Hillsdale, MI 49242, USA}
\affiliation{Eureka Scientific, Inc. 2452 Delmer Street, Suite 100, Oakland, CA 94602-3017}
\author{Justin A. Ellis}
\affiliation{Infinia ML, 202 Rigsbee Avenue, Durham NC, 27701}
\author{Elizabeth C. Ferrara}
\affiliation{Department of Astronomy, University of Maryland, College Park, MD 20742}
\affiliation{Center for Research and Exploration in Space Science and Technology, NASA/GSFC, Greenbelt, MD 20771}
\affiliation{NASA Goddard Space Flight Center, Greenbelt, MD 20771, USA}
\author{William Fiore}
\affiliation{Department of Physics and Astronomy, West Virginia University, P.O. Box 6315, Morgantown, WV 26506, USA}
\affiliation{Center for Gravitational Waves and Cosmology, West Virginia University, Chestnut Ridge Research Building, Morgantown, WV 26505, USA}
\author{Emmanuel Fonseca}
\affiliation{Department of Physics, McGill University, 3600  University St., Montreal, QC H3A 2T8, Canada}
\author{Nathan Garver-Daniels}
\affiliation{Department of Physics and Astronomy, West Virginia University, P.O. Box 6315, Morgantown, WV 26506, USA}
\affiliation{Center for Gravitational Waves and Cosmology, West Virginia University, Chestnut Ridge Research Building, Morgantown, WV 26505, USA}
\author{Peter A. Gentile}
\affiliation{Department of Physics and Astronomy, West Virginia University, P.O. Box 6315, Morgantown, WV 26506, USA}
\affiliation{Center for Gravitational Waves and Cosmology, West Virginia University, Chestnut Ridge Research Building, Morgantown, WV 26505, USA}
\author{Deborah C. Good}
\affiliation{Department of Physics and Astronomy, University of British Columbia, 6224 Agricultural Road, Vancouver, BC V6T 1Z1, Canada}
\author{Jeffrey S. Hazboun}
\altaffiliation{NANOGrav Physics Frontiers Center Postdoctoral Fellow}
\affiliation{University of Washington Bothell, 18115 Campus Way NE, Bothell, WA 98011, USA}
\author{A. Miguel Holgado}
\affiliation{Department of Astronomy and National Center for Supercomputing Applications, University of Illinois at Urbana-Champaign, Urbana, IL 61801, USA}
\affiliation{McWilliams Center for Cosmology and Department of Physics, Carnegie Mellon University, Pittsburgh PA, 15213, USA}
\author{Kristina Islo}
\affiliation{Center for Gravitation, Cosmology and Astrophysics, Department of Physics, University of Wisconsin-Milwaukee,\\ P.O. Box 413, Milwaukee, WI 53201, USA}
\author{Ross J. Jennings}
\affiliation{Cornell Center for Astrophysics and Planetary Science and Department of Astronomy, Cornell University, Ithaca, NY 14853, USA}
\author{Megan L. Jones}
\affiliation{Center for Gravitation, Cosmology and Astrophysics, Department of Physics, University of Wisconsin-Milwaukee,\\ P.O. Box 413, Milwaukee, WI 53201, USA}
\author{Andrew R. Kaiser}
\affiliation{Department of Physics and Astronomy, West Virginia University, P.O. Box 6315, Morgantown, WV 26506, USA}
\affiliation{Center for Gravitational Waves and Cosmology, West Virginia University, Chestnut Ridge Research Building, Morgantown, WV 26505, USA}
\author{David L. Kaplan}
\affiliation{Center for Gravitation, Cosmology and Astrophysics, Department of Physics, University of Wisconsin-Milwaukee,\\ P.O. Box 413, Milwaukee, WI 53201, USA}
\author{Luke Zoltan Kelley}
\affiliation{Center for Interdisciplinary Exploration and Research in Astrophysics (CIERA), Northwestern University, Evanston, IL 60208}
\author{Joey Shapiro Key}
\affiliation{University of Washington Bothell, 18115 Campus Way NE, Bothell, WA 98011, USA}
\author{Nima Laal}
\affiliation{Department of Physics, Oregon State University, Corvallis, OR 97331, USA}
\author{Michael T. Lam}
\affiliation{School of Physics and Astronomy, Rochester Institute of Technology, Rochester, NY 14623, USA}
\affiliation{Laboratory for Multiwavelength Astrophysics, Rochester Institute of Technology, Rochester, NY 14623, USA}
\author{T. Joseph W. Lazio}
\affiliation{Jet Propulsion Laboratory, California Institute of Technology, 4800 Oak Grove Drive, Pasadena, CA 91109, USA}
\author{Vincent S. H. Lee}
\affiliation{Walter Burke Institute for Theoretical Physics, California Institute of Technology, Pasadena, CA}
\author{Duncan R. Lorimer}
\affiliation{Department of Physics and Astronomy, West Virginia University, P.O. Box 6315, Morgantown, WV 26506, USA}
\affiliation{Center for Gravitational Waves and Cosmology, West Virginia University, Chestnut Ridge Research Building, Morgantown, WV 26505, USA}
\author{Jing Luo}
\affiliation{Department of Astronomy \& Astrophysics, University of Toronto, 50 Saint George Street, Toronto, ON M5S 3H4, Canada}
\author{Ryan S. Lynch}
\affiliation{Green Bank Observatory, P.O. Box 2, Green Bank, WV 24944, USA}
\author{Dustin R. Madison}
\altaffiliation{NANOGrav Physics Frontiers Center Postdoctoral Fellow}
\affiliation{Department of Physics and Astronomy, West Virginia University, P.O. Box 6315, Morgantown, WV 26506, USA}
\affiliation{Center for Gravitational Waves and Cosmology, West Virginia University, Chestnut Ridge Research Building, Morgantown, WV 26505, USA}
\author{Maura A. McLaughlin}
\affiliation{Department of Physics and Astronomy, West Virginia University, P.O. Box 6315, Morgantown, WV 26506, USA}
\affiliation{Center for Gravitational Waves and Cosmology, West Virginia University, Chestnut Ridge Research Building, Morgantown, WV 26505, USA}
\author{Chiara M. F. Mingarelli}
\affiliation{Center for Computational Astrophysics, Flatiron Institute, 162 5th Avenue, New York, New York, 10010, USA}
\affiliation{Department of Physics, University of Connecticut, 196 Auditorium Road, U-3046, Storrs, CT 06269-3046, USA}
\author{Andrea Mitridate}
\affiliation{Walter Burke Institute for Theoretical Physics, California Institute of Technology, Pasadena, CA}
\author{Cherry Ng}
\affiliation{Dunlap Institute for Astronomy and Astrophysics, University of Toronto, 50 St. George St., Toronto, ON M5S 3H4, Canada}
\author{David J. Nice}
\affiliation{Department of Physics, Lafayette College, Easton, PA 18042, USA}
\author{Timothy T. Pennucci}
\altaffiliation{NANOGrav Physics Frontiers Center Postdoctoral Fellow}
\affiliation{National Radio Astronomy Observatory, 520 Edgemont Road, Charlottesville, VA 22903, USA}
\affiliation{Institute of Physics, E\"{o}tv\"{o}s Lor\'{a}nd University, P\'{a}zm\'{a}ny P. s. 1/A, 1117 Budapest, Hungary}
\author{Nihan S. Pol}
\affiliation{Department of Physics and Astronomy, West Virginia University, P.O. Box 6315, Morgantown, WV 26506, USA}
\affiliation{Center for Gravitational Waves and Cosmology, West Virginia University, Chestnut Ridge Research Building, Morgantown, WV 26505, USA}
\affiliation{Department of Physics and Astronomy, Vanderbilt University, 2301 Vanderbilt Place, Nashville, TN 37235, USA}
\author{Scott M. Ransom}
\affiliation{National Radio Astronomy Observatory, 520 Edgemont Road, Charlottesville, VA 22903, USA}
\author{Paul S. Ray}
\affiliation{Space Science Division, Naval Research Laboratory, Washington, DC 20375-5352, USA}
\author{Brent J. Shapiro-Albert}
\affiliation{Department of Physics and Astronomy, West Virginia University, P.O. Box 6315, Morgantown, WV 26506, USA}
\affiliation{Center for Gravitational Waves and Cosmology, West Virginia University, Chestnut Ridge Research Building, Morgantown, WV 26505, USA}
\author{Xavier Siemens}
\affiliation{Department of Physics, Oregon State University, Corvallis, OR 97331, USA}
\affiliation{Center for Gravitation, Cosmology and Astrophysics, Department of Physics, University of Wisconsin-Milwaukee,\\ P.O. Box 413, Milwaukee, WI 53201, USA}
\author{Joseph Simon}
\affiliation{Jet Propulsion Laboratory, California Institute of Technology, 4800 Oak Grove Drive, Pasadena, CA 91109, USA}
\affiliation{Department of Astrophysical and Planetary Sciences, University of Colorado, Boulder, CO 80309, USA}
\author{Ren\'{e}e Spiewak}
\affiliation{Centre for Astrophysics and Supercomputing, Swinburne University of Technology, P.O. Box 218, Hawthorn, Victoria 3122, Australia}
\author{Ingrid H. Stairs}
\affiliation{Department of Physics and Astronomy, University of British Columbia, 6224 Agricultural Road, Vancouver, BC V6T 1Z1, Canada}
\author{Daniel R. Stinebring}
\affiliation{Department of Physics and Astronomy, Oberlin College, Oberlin, OH 44074, USA}
\author{Kevin Stovall}
\affiliation{National Radio Astronomy Observatory, 1003 Lopezville Rd., Socorro, NM 87801, USA}
\author{Jerry P. Sun}
\affiliation{Department of Physics, Oregon State University, Corvallis, OR 97331, USA}
\author{Joseph K. Swiggum}
\altaffiliation{NANOGrav Physics Frontiers Center Postdoctoral Fellow}
\affiliation{Department of Physics, Lafayette College, Easton, PA 18042, USA}
\author{Stephen R. Taylor}
\affiliation{Department of Physics and Astronomy, Vanderbilt University, 2301 Vanderbilt Place, Nashville, TN 37235, USA}
\author{Jacob E. Turner}
\affiliation{Department of Physics and Astronomy, West Virginia University, P.O. Box 6315, Morgantown, WV 26506, USA}
\affiliation{Center for Gravitational Waves and Cosmology, West Virginia University, Chestnut Ridge Research Building, Morgantown, WV 26505, USA}
\author{Michele Vallisneri}
\affiliation{Jet Propulsion Laboratory, California Institute of Technology, 4800 Oak Grove Drive, Pasadena, CA 91109, USA}
\author{Sarah J. Vigeland}
\affiliation{Center for Gravitation, Cosmology and Astrophysics, Department of Physics, University of Wisconsin-Milwaukee,\\ P.O. Box 413, Milwaukee, WI 53201, USA}
\author{Caitlin A. Witt}
\affiliation{Department of Physics and Astronomy, West Virginia University, P.O. Box 6315, Morgantown, WV 26506, USA}
\affiliation{Center for Gravitational Waves and Cosmology, West Virginia University, Chestnut Ridge Research Building, Morgantown, WV 26505, USA}
\author{Kathryn M. Zurek}
\affiliation{Walter Burke Institute for Theoretical Physics, California Institute of Technology, Pasadena, CA}

%% file: NANOG_PT.bbl
%apsrev4-2.bst 2019-01-14 (MD) hand-edited version of apsrev4-1.bst
%Control: key (0)
%Control: author (8) initials jnrlst
%Control: editor formatted (1) identically to author
%Control: production of article title (0) allowed
%Control: page (0) single
%Control: year (1) truncated
%Control: production of eprint (0) enabled
\begin{thebibliography}{78}%
\makeatletter
\providecommand \@ifxundefined [1]{%
 \@ifx{#1\undefined}
}%
\providecommand \@ifnum [1]{%
 \ifnum #1\expandafter \@firstoftwo
 \else \expandafter \@secondoftwo
 \fi
}%
\providecommand \@ifx [1]{%
 \ifx #1\expandafter \@firstoftwo
 \else \expandafter \@secondoftwo
 \fi
}%
\providecommand \natexlab [1]{#1}%
\providecommand \enquote  [1]{``#1''}%
\providecommand \bibnamefont  [1]{#1}%
\providecommand \bibfnamefont [1]{#1}%
\providecommand \citenamefont [1]{#1}%
\providecommand \href@noop [0]{\@secondoftwo}%
\providecommand \href [0]{\begingroup \@sanitize@url \@href}%
\providecommand \@href[1]{\@@startlink{#1}\@@href}%
\providecommand \@@href[1]{\endgroup#1\@@endlink}%
\providecommand \@sanitize@url [0]{\catcode `\\12\catcode `\$12\catcode
  `\&12\catcode `\#12\catcode `\^12\catcode `\_12\catcode `\%12\relax}%
\providecommand \@@startlink[1]{}%
\providecommand \@@endlink[0]{}%
\providecommand \url  [0]{\begingroup\@sanitize@url \@url }%
\providecommand \@url [1]{\endgroup\@href {#1}{\urlprefix }}%
\providecommand \urlprefix  [0]{URL }%
\providecommand \Eprint [0]{\href }%
\providecommand \doibase [0]{https://doi.org/}%
\providecommand \selectlanguage [0]{\@gobble}%
\providecommand \bibinfo  [0]{\@secondoftwo}%
\providecommand \bibfield  [0]{\@secondoftwo}%
\providecommand \translation [1]{[#1]}%
\providecommand \BibitemOpen [0]{}%
\providecommand \bibitemStop [0]{}%
\providecommand \bibitemNoStop [0]{.\EOS\space}%
\providecommand \EOS [0]{\spacefactor3000\relax}%
\providecommand \BibitemShut  [1]{\csname bibitem#1\endcsname}%
\let\auto@bib@innerbib\@empty
%</preamble>
\bibitem [{\citenamefont {{Foster}}\ and\ \citenamefont
  {{Backer}}(1990)}]{fb90}%
  \BibitemOpen
  \bibfield  {author} {\bibinfo {author} {\bibfnamefont {R.~S.}\ \bibnamefont
  {{Foster}}}\ and\ \bibinfo {author} {\bibfnamefont {D.~C.}\ \bibnamefont
  {{Backer}}},\ }\bibfield  {title} {\bibinfo {title} {{Constructing a Pulsar
  Timing Array}},\ }\href {https://doi.org/10.1086/169195} {\bibfield
  {journal} {\bibinfo  {journal} {\apj}\ }\textbf {\bibinfo {volume} {361}},\
  \bibinfo {pages} {300} (\bibinfo {year} {1990})}\BibitemShut {NoStop}%
\bibitem [{\citenamefont {{Detweiler}}(1979)}]{det79}%
  \BibitemOpen
  \bibfield  {author} {\bibinfo {author} {\bibfnamefont {S.}~\bibnamefont
  {{Detweiler}}},\ }\bibfield  {title} {\bibinfo {title} {{Pulsar timing
  measurements and the search for gravitational waves}},\ }\href
  {https://doi.org/10.1086/157593} {\bibfield  {journal} {\bibinfo  {journal}
  {\apj}\ }\textbf {\bibinfo {volume} {234}},\ \bibinfo {pages} {1100}
  (\bibinfo {year} {1979})}\BibitemShut {NoStop}%
\bibitem [{\citenamefont {{Sazhin}}(1978)}]{saz78}%
  \BibitemOpen
  \bibfield  {author} {\bibinfo {author} {\bibfnamefont {M.~V.}\ \bibnamefont
  {{Sazhin}}},\ }\bibfield  {title} {\bibinfo {title} {{Opportunities for
  detecting ultralong gravitational waves}},\ }\href@noop {} {\bibfield
  {journal} {\bibinfo  {journal} {\sovast}\ }\textbf {\bibinfo {volume} {22}},\
  \bibinfo {pages} {36} (\bibinfo {year} {1978})}\BibitemShut {NoStop}%
\bibitem [{\citenamefont {{Ransom}}\ \emph {et~al.}(2019)\citenamefont
  {{Ransom}}, \citenamefont {{Brazier}}, \citenamefont {{Chatterjee}},
  \citenamefont {{Cohen}}, \citenamefont {{Cordes}}, \citenamefont {{DeCesar}},
  \citenamefont {{Demorest}}, \citenamefont {{Hazboun}}, \citenamefont {{Lam}},
  \citenamefont {{Lynch}}, \citenamefont {{McLaughlin}}, \citenamefont
  {{Ransom}}, \citenamefont {{Siemens}}, \citenamefont {{Taylor}},\ and\
  \citenamefont {{Vigeland}}}]{ransom+19}%
  \BibitemOpen
  \bibfield  {author} {\bibinfo {author} {\bibfnamefont {S.}~\bibnamefont
  {{Ransom}}}, \bibinfo {author} {\bibfnamefont {A.}~\bibnamefont {{Brazier}}},
  \bibinfo {author} {\bibfnamefont {S.}~\bibnamefont {{Chatterjee}}}, \bibinfo
  {author} {\bibfnamefont {T.}~\bibnamefont {{Cohen}}}, \bibinfo {author}
  {\bibfnamefont {J.~M.}\ \bibnamefont {{Cordes}}}, \bibinfo {author}
  {\bibfnamefont {M.~E.}\ \bibnamefont {{DeCesar}}}, \bibinfo {author}
  {\bibfnamefont {P.~B.}\ \bibnamefont {{Demorest}}}, \bibinfo {author}
  {\bibfnamefont {J.~S.}\ \bibnamefont {{Hazboun}}}, \bibinfo {author}
  {\bibfnamefont {M.~T.}\ \bibnamefont {{Lam}}}, \bibinfo {author}
  {\bibfnamefont {R.~S.}\ \bibnamefont {{Lynch}}}, \bibinfo {author}
  {\bibfnamefont {M.~A.}\ \bibnamefont {{McLaughlin}}}, \bibinfo {author}
  {\bibfnamefont {S.~M.}\ \bibnamefont {{Ransom}}}, \bibinfo {author}
  {\bibfnamefont {X.}~\bibnamefont {{Siemens}}}, \bibinfo {author}
  {\bibfnamefont {S.~R.}\ \bibnamefont {{Taylor}}},\ and\ \bibinfo {author}
  {\bibfnamefont {S.~J.}\ \bibnamefont {{Vigeland}}},\ }\bibfield  {title}
  {\bibinfo {title} {{The NANOGrav Program for Gravitational Waves and
  Fundamental Physics}},\ }in\ \href@noop {} {\emph {\bibinfo {booktitle}
  {\baas}}},\ Vol.~\bibinfo {volume} {51}\ (\bibinfo {year} {2019})\ p.\
  \bibinfo {pages} {195},\ \Eprint {https://arxiv.org/abs/1908.05356}
  {arXiv:1908.05356 [astro-ph.IM]} \BibitemShut {NoStop}%
\bibitem [{\citenamefont {{Desvignes}}\ \emph {et~al.}(2016)\citenamefont
  {{Desvignes}}, \citenamefont {{Caballero}}, \citenamefont {{Lentati}},
  \citenamefont {{Verbiest}}, \citenamefont {{Champion}}, \citenamefont
  {{Stappers}}, \citenamefont {{Janssen}}, \citenamefont {{Lazarus}},
  \citenamefont {{Os{\l}owski}}, \citenamefont {{Babak}}, \citenamefont
  {{Bassa}}, \citenamefont {{Brem}}, \citenamefont {{Burgay}}, \citenamefont
  {{Cognard}}, \citenamefont {{Gair}}, \citenamefont {{Graikou}}, \citenamefont
  {{Guillemot}}, \citenamefont {{Hessels}}, \citenamefont {{Jessner}},
  \citenamefont {{Jordan}}, \citenamefont {{Karuppusamy}}, \citenamefont
  {{Kramer}}, \citenamefont {{Lassus}}, \citenamefont {{Lazaridis}},
  \citenamefont {{Lee}}, \citenamefont {{Liu}}, \citenamefont {{Lyne}},
  \citenamefont {{McKee}}, \citenamefont {{Mingarelli}}, \citenamefont
  {{Perrodin}}, \citenamefont {{Petiteau}}, \citenamefont {{Possenti}},
  \citenamefont {{Purver}}, \citenamefont {{Rosado}}, \citenamefont
  {{Sanidas}}, \citenamefont {{Sesana}}, \citenamefont {{Shaifullah}},
  \citenamefont {{Smits}}, \citenamefont {{Taylor}}, \citenamefont
  {{Theureau}}, \citenamefont {{Tiburzi}}, \citenamefont {{van Haasteren}},\
  and\ \citenamefont {{Vecchio}}}]{dcl+16}%
  \BibitemOpen
  \bibfield  {author} {\bibinfo {author} {\bibfnamefont {G.}~\bibnamefont
  {{Desvignes}}}, \bibinfo {author} {\bibfnamefont {R.~N.}\ \bibnamefont
  {{Caballero}}}, \bibinfo {author} {\bibfnamefont {L.}~\bibnamefont
  {{Lentati}}}, \bibinfo {author} {\bibfnamefont {J.~P.~W.}\ \bibnamefont
  {{Verbiest}}}, \bibinfo {author} {\bibfnamefont {D.~J.}\ \bibnamefont
  {{Champion}}}, \bibinfo {author} {\bibfnamefont {B.~W.}\ \bibnamefont
  {{Stappers}}}, \bibinfo {author} {\bibfnamefont {G.~H.}\ \bibnamefont
  {{Janssen}}}, \bibinfo {author} {\bibfnamefont {P.}~\bibnamefont
  {{Lazarus}}}, \bibinfo {author} {\bibfnamefont {S.}~\bibnamefont
  {{Os{\l}owski}}}, \bibinfo {author} {\bibfnamefont {S.}~\bibnamefont
  {{Babak}}}, \bibinfo {author} {\bibfnamefont {C.~G.}\ \bibnamefont
  {{Bassa}}}, \bibinfo {author} {\bibfnamefont {P.}~\bibnamefont {{Brem}}},
  \bibinfo {author} {\bibfnamefont {M.}~\bibnamefont {{Burgay}}}, \bibinfo
  {author} {\bibfnamefont {I.}~\bibnamefont {{Cognard}}}, \bibinfo {author}
  {\bibfnamefont {J.~R.}\ \bibnamefont {{Gair}}}, \bibinfo {author}
  {\bibfnamefont {E.}~\bibnamefont {{Graikou}}}, \bibinfo {author}
  {\bibfnamefont {L.}~\bibnamefont {{Guillemot}}}, \bibinfo {author}
  {\bibfnamefont {J.~W.~T.}\ \bibnamefont {{Hessels}}}, \bibinfo {author}
  {\bibfnamefont {A.}~\bibnamefont {{Jessner}}}, \bibinfo {author}
  {\bibfnamefont {C.}~\bibnamefont {{Jordan}}}, \bibinfo {author}
  {\bibfnamefont {R.}~\bibnamefont {{Karuppusamy}}}, \bibinfo {author}
  {\bibfnamefont {M.}~\bibnamefont {{Kramer}}}, \bibinfo {author}
  {\bibfnamefont {A.}~\bibnamefont {{Lassus}}}, \bibinfo {author}
  {\bibfnamefont {K.}~\bibnamefont {{Lazaridis}}}, \bibinfo {author}
  {\bibfnamefont {K.~J.}\ \bibnamefont {{Lee}}}, \bibinfo {author}
  {\bibfnamefont {K.}~\bibnamefont {{Liu}}}, \bibinfo {author} {\bibfnamefont
  {A.~G.}\ \bibnamefont {{Lyne}}}, \bibinfo {author} {\bibfnamefont
  {J.}~\bibnamefont {{McKee}}}, \bibinfo {author} {\bibfnamefont {C.~M.~F.}\
  \bibnamefont {{Mingarelli}}}, \bibinfo {author} {\bibfnamefont
  {D.}~\bibnamefont {{Perrodin}}}, \bibinfo {author} {\bibfnamefont
  {A.}~\bibnamefont {{Petiteau}}}, \bibinfo {author} {\bibfnamefont
  {A.}~\bibnamefont {{Possenti}}}, \bibinfo {author} {\bibfnamefont {M.~B.}\
  \bibnamefont {{Purver}}}, \bibinfo {author} {\bibfnamefont {P.~A.}\
  \bibnamefont {{Rosado}}}, \bibinfo {author} {\bibfnamefont {S.}~\bibnamefont
  {{Sanidas}}}, \bibinfo {author} {\bibfnamefont {A.}~\bibnamefont {{Sesana}}},
  \bibinfo {author} {\bibfnamefont {G.}~\bibnamefont {{Shaifullah}}}, \bibinfo
  {author} {\bibfnamefont {R.}~\bibnamefont {{Smits}}}, \bibinfo {author}
  {\bibfnamefont {S.~R.}\ \bibnamefont {{Taylor}}}, \bibinfo {author}
  {\bibfnamefont {G.}~\bibnamefont {{Theureau}}}, \bibinfo {author}
  {\bibfnamefont {C.}~\bibnamefont {{Tiburzi}}}, \bibinfo {author}
  {\bibfnamefont {R.}~\bibnamefont {{van Haasteren}}},\ and\ \bibinfo {author}
  {\bibfnamefont {A.}~\bibnamefont {{Vecchio}}},\ }\bibfield  {title} {\bibinfo
  {title} {{High-precision timing of 42 millisecond pulsars with the European
  Pulsar Timing Array}},\ }\href {https://doi.org/10.1093/mnras/stw483}
  {\bibfield  {journal} {\bibinfo  {journal} {\mnras}\ }\textbf {\bibinfo
  {volume} {458}},\ \bibinfo {pages} {3341} (\bibinfo {year} {2016})},\ \Eprint
  {https://arxiv.org/abs/1602.08511} {arXiv:1602.08511 [astro-ph.HE]}
  \BibitemShut {NoStop}%
\bibitem [{\citenamefont {{Kerr}}\ \emph {et~al.}(2020)\citenamefont {{Kerr}},
  \citenamefont {{Reardon}}, \citenamefont {{Hobbs}}, \citenamefont
  {{Shannon}}, \citenamefont {{Manchester}}, \citenamefont {{Dai}},
  \citenamefont {{Russell}}, \citenamefont {{Zhang}}, \citenamefont {{van
  Straten}}, \citenamefont {{Os{\l}owski}}, \citenamefont {{Parthasarathy}},
  \citenamefont {{Spiewak}}, \citenamefont {{Bailes}}, \citenamefont {{Bhat}},
  \citenamefont {{Cameron}}, \citenamefont {{Coles}}, \citenamefont
  {{Dempsey}}, \citenamefont {{Deng}}, \citenamefont {{Goncharov}},
  \citenamefont {{Kaczmarek}}, \citenamefont {{Keith}}, \citenamefont
  {{Lasky}}, \citenamefont {{Lower}}, \citenamefont {{Preisig}}, \citenamefont
  {{Sarkissian}}, \citenamefont {{Toomey}}, \citenamefont {{Wang}},
  \citenamefont {{Wang}}, \citenamefont {{Zhang}},\ and\ \citenamefont
  {{Zhu}}}]{krh+20}%
  \BibitemOpen
  \bibfield  {author} {\bibinfo {author} {\bibfnamefont {M.}~\bibnamefont
  {{Kerr}}}, \bibinfo {author} {\bibfnamefont {D.~J.}\ \bibnamefont
  {{Reardon}}}, \bibinfo {author} {\bibfnamefont {G.}~\bibnamefont {{Hobbs}}},
  \bibinfo {author} {\bibfnamefont {R.~M.}\ \bibnamefont {{Shannon}}}, \bibinfo
  {author} {\bibfnamefont {R.~N.}\ \bibnamefont {{Manchester}}}, \bibinfo
  {author} {\bibfnamefont {S.}~\bibnamefont {{Dai}}}, \bibinfo {author}
  {\bibfnamefont {C.~J.}\ \bibnamefont {{Russell}}}, \bibinfo {author}
  {\bibfnamefont {S.~B.}\ \bibnamefont {{Zhang}}}, \bibinfo {author}
  {\bibfnamefont {W.}~\bibnamefont {{van Straten}}}, \bibinfo {author}
  {\bibfnamefont {S.}~\bibnamefont {{Os{\l}owski}}}, \bibinfo {author}
  {\bibfnamefont {A.}~\bibnamefont {{Parthasarathy}}}, \bibinfo {author}
  {\bibfnamefont {R.}~\bibnamefont {{Spiewak}}}, \bibinfo {author}
  {\bibfnamefont {M.}~\bibnamefont {{Bailes}}}, \bibinfo {author}
  {\bibfnamefont {N.~D.~R.}\ \bibnamefont {{Bhat}}}, \bibinfo {author}
  {\bibfnamefont {A.~D.}\ \bibnamefont {{Cameron}}}, \bibinfo {author}
  {\bibfnamefont {W.~A.}\ \bibnamefont {{Coles}}}, \bibinfo {author}
  {\bibfnamefont {J.}~\bibnamefont {{Dempsey}}}, \bibinfo {author}
  {\bibfnamefont {X.}~\bibnamefont {{Deng}}}, \bibinfo {author} {\bibfnamefont
  {B.}~\bibnamefont {{Goncharov}}}, \bibinfo {author} {\bibfnamefont {J.~F.}\
  \bibnamefont {{Kaczmarek}}}, \bibinfo {author} {\bibfnamefont {M.~J.}\
  \bibnamefont {{Keith}}}, \bibinfo {author} {\bibfnamefont {P.~D.}\
  \bibnamefont {{Lasky}}}, \bibinfo {author} {\bibfnamefont {M.~E.}\
  \bibnamefont {{Lower}}}, \bibinfo {author} {\bibfnamefont {B.}~\bibnamefont
  {{Preisig}}}, \bibinfo {author} {\bibfnamefont {J.~M.}\ \bibnamefont
  {{Sarkissian}}}, \bibinfo {author} {\bibfnamefont {L.}~\bibnamefont
  {{Toomey}}}, \bibinfo {author} {\bibfnamefont {H.}~\bibnamefont {{Wang}}},
  \bibinfo {author} {\bibfnamefont {J.}~\bibnamefont {{Wang}}}, \bibinfo
  {author} {\bibfnamefont {L.}~\bibnamefont {{Zhang}}},\ and\ \bibinfo {author}
  {\bibfnamefont {X.}~\bibnamefont {{Zhu}}},\ }\bibfield  {title} {\bibinfo
  {title} {{The Parkes Pulsar Timing Array Project: Second data release}},\
  }\href@noop {} {\bibfield  {journal} {\bibinfo  {journal} {arXiv e-prints}\
  ,\ \bibinfo {eid} {arXiv:2003.09780}} (\bibinfo {year} {2020})},\ \Eprint
  {https://arxiv.org/abs/2003.09780} {arXiv:2003.09780 [astro-ph.IM]}
  \BibitemShut {NoStop}%
\bibitem [{\citenamefont {{Joshi}}\ \emph {et~al.}(2018)\citenamefont
  {{Joshi}}, \citenamefont {{Arumugasamy}}, \citenamefont {{Bagchi}},
  \citenamefont {{Bandyopadhyay}}, \citenamefont {{Basu}}, \citenamefont
  {{Dhand a Batra}}, \citenamefont {{Bethapudi}}, \citenamefont {{Choudhary}},
  \citenamefont {{De}}, \citenamefont {{Dey}}, \citenamefont {{Gopakumar}},
  \citenamefont {{Gupta}}, \citenamefont {{Krishnakumar}}, \citenamefont
  {{Maan}}, \citenamefont {{Manoharan}}, \citenamefont {{Naidu}}, \citenamefont
  {{Nandi}}, \citenamefont {{Pathak}}, \citenamefont {{Surnis}},\ and\
  \citenamefont {{Susobhanan}}}]{InPTA}%
  \BibitemOpen
  \bibfield  {author} {\bibinfo {author} {\bibfnamefont {B.~C.}\ \bibnamefont
  {{Joshi}}}, \bibinfo {author} {\bibfnamefont {P.}~\bibnamefont
  {{Arumugasamy}}}, \bibinfo {author} {\bibfnamefont {M.}~\bibnamefont
  {{Bagchi}}}, \bibinfo {author} {\bibfnamefont {D.}~\bibnamefont
  {{Bandyopadhyay}}}, \bibinfo {author} {\bibfnamefont {A.}~\bibnamefont
  {{Basu}}}, \bibinfo {author} {\bibfnamefont {N.}~\bibnamefont {{Dhand a
  Batra}}}, \bibinfo {author} {\bibfnamefont {S.}~\bibnamefont {{Bethapudi}}},
  \bibinfo {author} {\bibfnamefont {A.}~\bibnamefont {{Choudhary}}}, \bibinfo
  {author} {\bibfnamefont {K.}~\bibnamefont {{De}}}, \bibinfo {author}
  {\bibfnamefont {L.}~\bibnamefont {{Dey}}}, \bibinfo {author} {\bibfnamefont
  {A.}~\bibnamefont {{Gopakumar}}}, \bibinfo {author} {\bibfnamefont
  {Y.}~\bibnamefont {{Gupta}}}, \bibinfo {author} {\bibfnamefont {M.~A.}\
  \bibnamefont {{Krishnakumar}}}, \bibinfo {author} {\bibfnamefont
  {Y.}~\bibnamefont {{Maan}}}, \bibinfo {author} {\bibfnamefont {P.~K.}\
  \bibnamefont {{Manoharan}}}, \bibinfo {author} {\bibfnamefont
  {A.}~\bibnamefont {{Naidu}}}, \bibinfo {author} {\bibfnamefont
  {R.}~\bibnamefont {{Nandi}}}, \bibinfo {author} {\bibfnamefont
  {D.}~\bibnamefont {{Pathak}}}, \bibinfo {author} {\bibfnamefont
  {M.}~\bibnamefont {{Surnis}}},\ and\ \bibinfo {author} {\bibfnamefont
  {A.}~\bibnamefont {{Susobhanan}}},\ }\bibfield  {title} {\bibinfo {title}
  {{Precision pulsar timing with the ORT and the GMRT and its applications in
  pulsar astrophysics}},\ }\href {https://doi.org/10.1007/s12036-018-9549-y}
  {\bibfield  {journal} {\bibinfo  {journal} {Journal of Astrophysics and
  Astronomy}\ }\textbf {\bibinfo {volume} {39}},\ \bibinfo {eid} {51} (\bibinfo
  {year} {2018})}\BibitemShut {NoStop}%
\bibitem [{\citenamefont {{Perera}}\ \emph {et~al.}(2019)\citenamefont
  {{Perera}}, \citenamefont {{DeCesar}}, \citenamefont {{Demorest}},
  \citenamefont {{Kerr}}, \citenamefont {{Lentati}}, \citenamefont {{Nice}},
  \citenamefont {{Os{\l}owski}}, \citenamefont {{Ransom}}, \citenamefont
  {{Keith}}, \citenamefont {{Arzoumanian}}, \citenamefont {{Bailes}},
  \citenamefont {{Baker}}, \citenamefont {{Bassa}}, \citenamefont {{Bhat}},
  \citenamefont {{Brazier}}, \citenamefont {{Burgay}}, \citenamefont
  {{Burke-Spolaor}}, \citenamefont {{Caballero}}, \citenamefont {{Champion}},
  \citenamefont {{Chatterjee}}, \citenamefont {{Chen}}, \citenamefont
  {{Cognard}}, \citenamefont {{Cordes}}, \citenamefont {{Crowter}},
  \citenamefont {{Dai}}, \citenamefont {{Desvignes}}, \citenamefont {{Dolch}},
  \citenamefont {{Ferdman}}, \citenamefont {{Ferrara}}, \citenamefont
  {{Fonseca}}, \citenamefont {{Goldstein}}, \citenamefont {{Graikou}},
  \citenamefont {{Guillemot}}, \citenamefont {{Hazboun}}, \citenamefont
  {{Hobbs}}, \citenamefont {{Hu}}, \citenamefont {{Islo}}, \citenamefont
  {{Janssen}}, \citenamefont {{Karuppusamy}}, \citenamefont {{Kramer}},
  \citenamefont {{Lam}}, \citenamefont {{Lee}}, \citenamefont {{Liu}},
  \citenamefont {{Luo}}, \citenamefont {{Lyne}}, \citenamefont {{Manchester}},
  \citenamefont {{McKee}}, \citenamefont {{McLaughlin}}, \citenamefont
  {{Mingarelli}}, \citenamefont {{Parthasarathy}}, \citenamefont {{Pennucci}},
  \citenamefont {{Perrodin}}, \citenamefont {{Possenti}}, \citenamefont
  {{Reardon}}, \citenamefont {{Russell}}, \citenamefont {{Sanidas}},
  \citenamefont {{Sesana}}, \citenamefont {{Shaifullah}}, \citenamefont
  {{Shannon}}, \citenamefont {{Siemens}}, \citenamefont {{Simon}},
  \citenamefont {{Spiewak}}, \citenamefont {{Stairs}}, \citenamefont
  {{Stappers}}, \citenamefont {{Swiggum}}, \citenamefont {{Taylor}},
  \citenamefont {{Theureau}}, \citenamefont {{Tiburzi}}, \citenamefont
  {{Vallisneri}}, \citenamefont {{Vecchio}}, \citenamefont {{Wang}},
  \citenamefont {{Zhang}}, \citenamefont {{Zhang}}, \citenamefont {{Zhu}},\
  and\ \citenamefont {{Zhu}}}]{pdd+19}%
  \BibitemOpen
  \bibfield  {author} {\bibinfo {author} {\bibfnamefont {B.~B.~P.}\
  \bibnamefont {{Perera}}}, \bibinfo {author} {\bibfnamefont {M.~E.}\
  \bibnamefont {{DeCesar}}}, \bibinfo {author} {\bibfnamefont {P.~B.}\
  \bibnamefont {{Demorest}}}, \bibinfo {author} {\bibfnamefont
  {M.}~\bibnamefont {{Kerr}}}, \bibinfo {author} {\bibfnamefont
  {L.}~\bibnamefont {{Lentati}}}, \bibinfo {author} {\bibfnamefont {D.~J.}\
  \bibnamefont {{Nice}}}, \bibinfo {author} {\bibfnamefont {S.}~\bibnamefont
  {{Os{\l}owski}}}, \bibinfo {author} {\bibfnamefont {S.~M.}\ \bibnamefont
  {{Ransom}}}, \bibinfo {author} {\bibfnamefont {M.~J.}\ \bibnamefont
  {{Keith}}}, \bibinfo {author} {\bibfnamefont {Z.}~\bibnamefont
  {{Arzoumanian}}}, \bibinfo {author} {\bibfnamefont {M.}~\bibnamefont
  {{Bailes}}}, \bibinfo {author} {\bibfnamefont {P.~T.}\ \bibnamefont
  {{Baker}}}, \bibinfo {author} {\bibfnamefont {C.~G.}\ \bibnamefont
  {{Bassa}}}, \bibinfo {author} {\bibfnamefont {N.~D.~R.}\ \bibnamefont
  {{Bhat}}}, \bibinfo {author} {\bibfnamefont {A.}~\bibnamefont {{Brazier}}},
  \bibinfo {author} {\bibfnamefont {M.}~\bibnamefont {{Burgay}}}, \bibinfo
  {author} {\bibfnamefont {S.}~\bibnamefont {{Burke-Spolaor}}}, \bibinfo
  {author} {\bibfnamefont {R.~N.}\ \bibnamefont {{Caballero}}}, \bibinfo
  {author} {\bibfnamefont {D.~J.}\ \bibnamefont {{Champion}}}, \bibinfo
  {author} {\bibfnamefont {S.}~\bibnamefont {{Chatterjee}}}, \bibinfo {author}
  {\bibfnamefont {S.}~\bibnamefont {{Chen}}}, \bibinfo {author} {\bibfnamefont
  {I.}~\bibnamefont {{Cognard}}}, \bibinfo {author} {\bibfnamefont {J.~M.}\
  \bibnamefont {{Cordes}}}, \bibinfo {author} {\bibfnamefont {K.}~\bibnamefont
  {{Crowter}}}, \bibinfo {author} {\bibfnamefont {S.}~\bibnamefont {{Dai}}},
  \bibinfo {author} {\bibfnamefont {G.}~\bibnamefont {{Desvignes}}}, \bibinfo
  {author} {\bibfnamefont {T.}~\bibnamefont {{Dolch}}}, \bibinfo {author}
  {\bibfnamefont {R.~D.}\ \bibnamefont {{Ferdman}}}, \bibinfo {author}
  {\bibfnamefont {E.~C.}\ \bibnamefont {{Ferrara}}}, \bibinfo {author}
  {\bibfnamefont {E.}~\bibnamefont {{Fonseca}}}, \bibinfo {author}
  {\bibfnamefont {J.~M.}\ \bibnamefont {{Goldstein}}}, \bibinfo {author}
  {\bibfnamefont {E.}~\bibnamefont {{Graikou}}}, \bibinfo {author}
  {\bibfnamefont {L.}~\bibnamefont {{Guillemot}}}, \bibinfo {author}
  {\bibfnamefont {J.~S.}\ \bibnamefont {{Hazboun}}}, \bibinfo {author}
  {\bibfnamefont {G.}~\bibnamefont {{Hobbs}}}, \bibinfo {author} {\bibfnamefont
  {H.}~\bibnamefont {{Hu}}}, \bibinfo {author} {\bibfnamefont {K.}~\bibnamefont
  {{Islo}}}, \bibinfo {author} {\bibfnamefont {G.~H.}\ \bibnamefont
  {{Janssen}}}, \bibinfo {author} {\bibfnamefont {R.}~\bibnamefont
  {{Karuppusamy}}}, \bibinfo {author} {\bibfnamefont {M.}~\bibnamefont
  {{Kramer}}}, \bibinfo {author} {\bibfnamefont {M.~T.}\ \bibnamefont {{Lam}}},
  \bibinfo {author} {\bibfnamefont {K.~J.}\ \bibnamefont {{Lee}}}, \bibinfo
  {author} {\bibfnamefont {K.}~\bibnamefont {{Liu}}}, \bibinfo {author}
  {\bibfnamefont {J.}~\bibnamefont {{Luo}}}, \bibinfo {author} {\bibfnamefont
  {A.~G.}\ \bibnamefont {{Lyne}}}, \bibinfo {author} {\bibfnamefont {R.~N.}\
  \bibnamefont {{Manchester}}}, \bibinfo {author} {\bibfnamefont {J.~W.}\
  \bibnamefont {{McKee}}}, \bibinfo {author} {\bibfnamefont {M.~A.}\
  \bibnamefont {{McLaughlin}}}, \bibinfo {author} {\bibfnamefont {C.~M.~F.}\
  \bibnamefont {{Mingarelli}}}, \bibinfo {author} {\bibfnamefont {A.~P.}\
  \bibnamefont {{Parthasarathy}}}, \bibinfo {author} {\bibfnamefont {T.~T.}\
  \bibnamefont {{Pennucci}}}, \bibinfo {author} {\bibfnamefont
  {D.}~\bibnamefont {{Perrodin}}}, \bibinfo {author} {\bibfnamefont
  {A.}~\bibnamefont {{Possenti}}}, \bibinfo {author} {\bibfnamefont {D.~J.}\
  \bibnamefont {{Reardon}}}, \bibinfo {author} {\bibfnamefont {C.~J.}\
  \bibnamefont {{Russell}}}, \bibinfo {author} {\bibfnamefont {S.~A.}\
  \bibnamefont {{Sanidas}}}, \bibinfo {author} {\bibfnamefont {A.}~\bibnamefont
  {{Sesana}}}, \bibinfo {author} {\bibfnamefont {G.}~\bibnamefont
  {{Shaifullah}}}, \bibinfo {author} {\bibfnamefont {R.~M.}\ \bibnamefont
  {{Shannon}}}, \bibinfo {author} {\bibfnamefont {X.}~\bibnamefont
  {{Siemens}}}, \bibinfo {author} {\bibfnamefont {J.}~\bibnamefont {{Simon}}},
  \bibinfo {author} {\bibfnamefont {R.}~\bibnamefont {{Spiewak}}}, \bibinfo
  {author} {\bibfnamefont {I.~H.}\ \bibnamefont {{Stairs}}}, \bibinfo {author}
  {\bibfnamefont {B.~W.}\ \bibnamefont {{Stappers}}}, \bibinfo {author}
  {\bibfnamefont {J.~K.}\ \bibnamefont {{Swiggum}}}, \bibinfo {author}
  {\bibfnamefont {S.~R.}\ \bibnamefont {{Taylor}}}, \bibinfo {author}
  {\bibfnamefont {G.}~\bibnamefont {{Theureau}}}, \bibinfo {author}
  {\bibfnamefont {C.}~\bibnamefont {{Tiburzi}}}, \bibinfo {author}
  {\bibfnamefont {M.}~\bibnamefont {{Vallisneri}}}, \bibinfo {author}
  {\bibfnamefont {A.}~\bibnamefont {{Vecchio}}}, \bibinfo {author}
  {\bibfnamefont {J.~B.}\ \bibnamefont {{Wang}}}, \bibinfo {author}
  {\bibfnamefont {S.~B.}\ \bibnamefont {{Zhang}}}, \bibinfo {author}
  {\bibfnamefont {L.}~\bibnamefont {{Zhang}}}, \bibinfo {author} {\bibfnamefont
  {W.~W.}\ \bibnamefont {{Zhu}}},\ and\ \bibinfo {author} {\bibfnamefont
  {X.~J.}\ \bibnamefont {{Zhu}}},\ }\bibfield  {title} {\bibinfo {title} {{The
  International Pulsar Timing Array: second data release}},\ }\href
  {https://doi.org/10.1093/mnras/stz2857} {\bibfield  {journal} {\bibinfo
  {journal} {\mnras}\ }\textbf {\bibinfo {volume} {490}},\ \bibinfo {pages}
  {4666} (\bibinfo {year} {2019})},\ \Eprint {https://arxiv.org/abs/1909.04534}
  {arXiv:1909.04534 [astro-ph.HE]} \BibitemShut {NoStop}%
\bibitem [{\citenamefont {{Lee}}(2016)}]{CPTA}%
  \BibitemOpen
  \bibfield  {author} {\bibinfo {author} {\bibfnamefont {K.~J.}\ \bibnamefont
  {{Lee}}},\ }\bibfield  {title} {\bibinfo {title} {{Prospects of Gravitational
  Wave Detection Using Pulsar Timing Array for Chinese Future Telescopes}},\
  }in\ \href@noop {} {\emph {\bibinfo {booktitle} {Frontiers in Radio Astronomy
  and FAST Early Sciences Symposium 2015}}},\ \bibinfo {series} {Astronomical
  Society of the Pacific Conference Series}, Vol.\ \bibinfo {volume} {502},\
  \bibinfo {editor} {edited by\ \bibinfo {editor} {\bibfnamefont
  {L.}~\bibnamefont {{Qain}}}\ and\ \bibinfo {editor} {\bibfnamefont
  {D.}~\bibnamefont {{Li}}}}\ (\bibinfo {year} {2016})\ p.~\bibinfo {pages}
  {19}\BibitemShut {NoStop}%
\bibitem [{\citenamefont {{Bailes}}\ \emph {et~al.}(2018)\citenamefont
  {{Bailes}}, \citenamefont {{Barr}}, \citenamefont {{Bhat}}, \citenamefont
  {{Brink}}, \citenamefont {{Buchner}}, \citenamefont {{Burgay}}, \citenamefont
  {{Camilo}}, \citenamefont {{Champion}}, \citenamefont {{Hessels}},
  \citenamefont {{Janssen}}, \citenamefont {{Jameson}}, \citenamefont
  {{Johnston}}, \citenamefont {{Karastergiou}}, \citenamefont {{Karuppusamy}},
  \citenamefont {{Kaspi}}, \citenamefont {{Keith}}, \citenamefont {{Kramer}},
  \citenamefont {{McLaughlin}}, \citenamefont {{Moodley}}, \citenamefont
  {{Oslowski}}, \citenamefont {{Possenti}}, \citenamefont {{Ransom}},
  \citenamefont {{Rasio}}, \citenamefont {{Sievers}}, \citenamefont
  {{Serylak}}, \citenamefont {{Stappers}}, \citenamefont {{Stairs}},
  \citenamefont {{Theureau}}, \citenamefont {{van Straten}}, \citenamefont
  {{Weltevrede}},\ and\ \citenamefont {{Wex}}}]{MeerTime}%
  \BibitemOpen
  \bibfield  {author} {\bibinfo {author} {\bibfnamefont {M.}~\bibnamefont
  {{Bailes}}}, \bibinfo {author} {\bibfnamefont {E.}~\bibnamefont {{Barr}}},
  \bibinfo {author} {\bibfnamefont {N.~D.~R.}\ \bibnamefont {{Bhat}}}, \bibinfo
  {author} {\bibfnamefont {J.}~\bibnamefont {{Brink}}}, \bibinfo {author}
  {\bibfnamefont {S.}~\bibnamefont {{Buchner}}}, \bibinfo {author}
  {\bibfnamefont {M.}~\bibnamefont {{Burgay}}}, \bibinfo {author}
  {\bibfnamefont {F.}~\bibnamefont {{Camilo}}}, \bibinfo {author}
  {\bibfnamefont {D.~J.}\ \bibnamefont {{Champion}}}, \bibinfo {author}
  {\bibfnamefont {J.}~\bibnamefont {{Hessels}}}, \bibinfo {author}
  {\bibfnamefont {G.~H.}\ \bibnamefont {{Janssen}}}, \bibinfo {author}
  {\bibfnamefont {A.}~\bibnamefont {{Jameson}}}, \bibinfo {author}
  {\bibfnamefont {S.}~\bibnamefont {{Johnston}}}, \bibinfo {author}
  {\bibfnamefont {A.}~\bibnamefont {{Karastergiou}}}, \bibinfo {author}
  {\bibfnamefont {R.}~\bibnamefont {{Karuppusamy}}}, \bibinfo {author}
  {\bibfnamefont {V.}~\bibnamefont {{Kaspi}}}, \bibinfo {author} {\bibfnamefont
  {M.~J.}\ \bibnamefont {{Keith}}}, \bibinfo {author} {\bibfnamefont
  {M.}~\bibnamefont {{Kramer}}}, \bibinfo {author} {\bibfnamefont {M.~A.}\
  \bibnamefont {{McLaughlin}}}, \bibinfo {author} {\bibfnamefont
  {K.}~\bibnamefont {{Moodley}}}, \bibinfo {author} {\bibfnamefont
  {S.}~\bibnamefont {{Oslowski}}}, \bibinfo {author} {\bibfnamefont
  {A.}~\bibnamefont {{Possenti}}}, \bibinfo {author} {\bibfnamefont {S.~M.}\
  \bibnamefont {{Ransom}}}, \bibinfo {author} {\bibfnamefont {F.~A.}\
  \bibnamefont {{Rasio}}}, \bibinfo {author} {\bibfnamefont {J.}~\bibnamefont
  {{Sievers}}}, \bibinfo {author} {\bibfnamefont {M.}~\bibnamefont
  {{Serylak}}}, \bibinfo {author} {\bibfnamefont {B.~W.}\ \bibnamefont
  {{Stappers}}}, \bibinfo {author} {\bibfnamefont {I.~H.}\ \bibnamefont
  {{Stairs}}}, \bibinfo {author} {\bibfnamefont {G.}~\bibnamefont
  {{Theureau}}}, \bibinfo {author} {\bibfnamefont {W.}~\bibnamefont {{van
  Straten}}}, \bibinfo {author} {\bibfnamefont {P.}~\bibnamefont
  {{Weltevrede}}},\ and\ \bibinfo {author} {\bibfnamefont {N.}~\bibnamefont
  {{Wex}}},\ }\bibfield  {title} {\bibinfo {title} {{MeerTime - the MeerKAT Key
  Science Program on Pulsar Timing}},\ }\href@noop {} {\bibfield  {journal}
  {\bibinfo  {journal} {arXiv e-prints}\ ,\ \bibinfo {eid} {arXiv:1803.07424}}
  (\bibinfo {year} {2018})},\ \Eprint {https://arxiv.org/abs/1803.07424}
  {arXiv:1803.07424 [astro-ph.IM]} \BibitemShut {NoStop}%
\bibitem [{\citenamefont {{Ng}}(2018)}]{CHIMEPulsar}%
  \BibitemOpen
  \bibfield  {author} {\bibinfo {author} {\bibfnamefont {C.}~\bibnamefont
  {{Ng}}},\ }\bibfield  {title} {\bibinfo {title} {{Pulsar science with the
  CHIME telescope}},\ }in\ \href {https://doi.org/10.1017/S1743921317010638}
  {\emph {\bibinfo {booktitle} {Pulsar Astrophysics the Next Fifty Years}}},\
  \bibinfo {series} {IAU Symposium}, Vol.\ \bibinfo {volume} {337},\ \bibinfo
  {editor} {edited by\ \bibinfo {editor} {\bibfnamefont {P.}~\bibnamefont
  {{Weltevrede}}}, \bibinfo {editor} {\bibfnamefont {B.~B.~P.}\ \bibnamefont
  {{Perera}}}, \bibinfo {editor} {\bibfnamefont {L.~L.}\ \bibnamefont
  {{Preston}}},\ and\ \bibinfo {editor} {\bibfnamefont {S.}~\bibnamefont
  {{Sanidas}}}}\ (\bibinfo {year} {2018})\ pp.\ \bibinfo {pages} {179--182},\
  \Eprint {https://arxiv.org/abs/1711.02104} {arXiv:1711.02104 [astro-ph.IM]}
  \BibitemShut {NoStop}%
\bibitem [{\citenamefont {{Sesana}}\ \emph {et~al.}(2004)\citenamefont
  {{Sesana}}, \citenamefont {{Haardt}}, \citenamefont {{Madau}},\ and\
  \citenamefont {{Volonteri}}}]{shm+04}%
  \BibitemOpen
  \bibfield  {author} {\bibinfo {author} {\bibfnamefont {A.}~\bibnamefont
  {{Sesana}}}, \bibinfo {author} {\bibfnamefont {F.}~\bibnamefont {{Haardt}}},
  \bibinfo {author} {\bibfnamefont {P.}~\bibnamefont {{Madau}}},\ and\ \bibinfo
  {author} {\bibfnamefont {M.}~\bibnamefont {{Volonteri}}},\ }\bibfield
  {title} {\bibinfo {title} {{Low-Frequency Gravitational Radiation from
  Coalescing Massive Black Hole Binaries in Hierarchical Cosmologies}},\ }\href
  {https://doi.org/10.1086/422185} {\bibfield  {journal} {\bibinfo  {journal}
  {\apj}\ }\textbf {\bibinfo {volume} {611}},\ \bibinfo {pages} {623} (\bibinfo
  {year} {2004})},\ \Eprint {https://arxiv.org/abs/astro-ph/0401543}
  {astro-ph/0401543} \BibitemShut {NoStop}%
\bibitem [{\citenamefont {{Burke-Spolaor}}\ \emph {et~al.}(2019)\citenamefont
  {{Burke-Spolaor}}, \citenamefont {{Taylor}}, \citenamefont {{Charisi}},
  \citenamefont {{Dolch}}, \citenamefont {{Hazboun}}, \citenamefont
  {{Holgado}}, \citenamefont {{Kelley}}, \citenamefont {{Lazio}}, \citenamefont
  {{Madison}}, \citenamefont {{McMann}}, \citenamefont {{Mingarelli}},
  \citenamefont {{Rasskazov}}, \citenamefont {{Siemens}}, \citenamefont
  {{Simon}},\ and\ \citenamefont {{Smith}}}]{stc+19}%
  \BibitemOpen
  \bibfield  {author} {\bibinfo {author} {\bibfnamefont {S.}~\bibnamefont
  {{Burke-Spolaor}}}, \bibinfo {author} {\bibfnamefont {S.~R.}\ \bibnamefont
  {{Taylor}}}, \bibinfo {author} {\bibfnamefont {M.}~\bibnamefont {{Charisi}}},
  \bibinfo {author} {\bibfnamefont {T.}~\bibnamefont {{Dolch}}}, \bibinfo
  {author} {\bibfnamefont {J.~S.}\ \bibnamefont {{Hazboun}}}, \bibinfo {author}
  {\bibfnamefont {A.~M.}\ \bibnamefont {{Holgado}}}, \bibinfo {author}
  {\bibfnamefont {L.~Z.}\ \bibnamefont {{Kelley}}}, \bibinfo {author}
  {\bibfnamefont {T.~J.~W.}\ \bibnamefont {{Lazio}}}, \bibinfo {author}
  {\bibfnamefont {D.~R.}\ \bibnamefont {{Madison}}}, \bibinfo {author}
  {\bibfnamefont {N.}~\bibnamefont {{McMann}}}, \bibinfo {author}
  {\bibfnamefont {C.~M.~F.}\ \bibnamefont {{Mingarelli}}}, \bibinfo {author}
  {\bibfnamefont {A.}~\bibnamefont {{Rasskazov}}}, \bibinfo {author}
  {\bibfnamefont {X.}~\bibnamefont {{Siemens}}}, \bibinfo {author}
  {\bibfnamefont {J.~J.}\ \bibnamefont {{Simon}}},\ and\ \bibinfo {author}
  {\bibfnamefont {T.~L.}\ \bibnamefont {{Smith}}},\ }\bibfield  {title}
  {\bibinfo {title} {{The astrophysics of nanohertz gravitational waves}},\
  }\href {https://doi.org/10.1007/s00159-019-0115-7} {\bibfield  {journal}
  {\bibinfo  {journal} {\aapr}\ }\textbf {\bibinfo {volume} {27}},\ \bibinfo
  {eid} {5} (\bibinfo {year} {2019})},\ \Eprint
  {https://arxiv.org/abs/1811.08826} {arXiv:1811.08826 [astro-ph.HE]}
  \BibitemShut {NoStop}%
\bibitem [{\citenamefont {{Siemens}}\ \emph {et~al.}(2007)\citenamefont
  {{Siemens}}, \citenamefont {{Mandic}},\ and\ \citenamefont
  {{Creighton}}}]{smc07}%
  \BibitemOpen
  \bibfield  {author} {\bibinfo {author} {\bibfnamefont {X.}~\bibnamefont
  {{Siemens}}}, \bibinfo {author} {\bibfnamefont {V.}~\bibnamefont
  {{Mandic}}},\ and\ \bibinfo {author} {\bibfnamefont {J.}~\bibnamefont
  {{Creighton}}},\ }\bibfield  {title} {\bibinfo {title} {{Gravitational-Wave
  Stochastic Background from Cosmic Strings}},\ }\href
  {https://doi.org/10.1103/PhysRevLett.98.111101} {\bibfield  {journal}
  {\bibinfo  {journal} {\prl}\ }\textbf {\bibinfo {volume} {98}},\ \bibinfo
  {eid} {111101} (\bibinfo {year} {2007})},\ \Eprint
  {https://arxiv.org/abs/astro-ph/0610920} {arXiv:astro-ph/0610920 [astro-ph]}
  \BibitemShut {NoStop}%
\bibitem [{\citenamefont {{Blanco-Pillado}}\ \emph {et~al.}(2018)\citenamefont
  {{Blanco-Pillado}}, \citenamefont {{Olum}},\ and\ \citenamefont
  {{Siemens}}}]{bos18}%
  \BibitemOpen
  \bibfield  {author} {\bibinfo {author} {\bibfnamefont {J.~J.}\ \bibnamefont
  {{Blanco-Pillado}}}, \bibinfo {author} {\bibfnamefont {K.~D.}\ \bibnamefont
  {{Olum}}},\ and\ \bibinfo {author} {\bibfnamefont {X.}~\bibnamefont
  {{Siemens}}},\ }\bibfield  {title} {\bibinfo {title} {{New limits on cosmic
  strings from gravitational wave observation}},\ }\href
  {https://doi.org/10.1016/j.physletb.2018.01.050} {\bibfield  {journal}
  {\bibinfo  {journal} {Physics Letters B}\ }\textbf {\bibinfo {volume}
  {778}},\ \bibinfo {pages} {392} (\bibinfo {year} {2018})},\ \Eprint
  {https://arxiv.org/abs/1709.02434} {arXiv:1709.02434 [astro-ph.CO]}
  \BibitemShut {NoStop}%
\bibitem [{\citenamefont {{Grishchuk}}(1975)}]{g75}%
  \BibitemOpen
  \bibfield  {author} {\bibinfo {author} {\bibfnamefont {L.~P.}\ \bibnamefont
  {{Grishchuk}}},\ }\bibfield  {title} {\bibinfo {title} {{Amplification of
  gravitational waves in an isotropic universe}},\ }\href@noop {} {\bibfield
  {journal} {\bibinfo  {journal} {Soviet Journal of Experimental and
  Theoretical Physics}\ }\textbf {\bibinfo {volume} {40}},\ \bibinfo {pages}
  {409} (\bibinfo {year} {1975})}\BibitemShut {NoStop}%
\bibitem [{\citenamefont {{Lasky}}\ \emph {et~al.}(2016)\citenamefont
  {{Lasky}}, \citenamefont {{Mingarelli}}, \citenamefont {{Smith}},
  \citenamefont {{Giblin}}, \citenamefont {{Thrane}}, \citenamefont
  {{Reardon}}, \citenamefont {{Caldwell}}, \citenamefont {{Bailes}},
  \citenamefont {{Bhat}}, \citenamefont {{Burke-Spolaor}}, \citenamefont
  {{Dai}}, \citenamefont {{Dempsey}}, \citenamefont {{Hobbs}}, \citenamefont
  {{Kerr}}, \citenamefont {{Levin}}, \citenamefont {{Manchester}},
  \citenamefont {{Os{\l}owski}}, \citenamefont {{Ravi}}, \citenamefont
  {{Rosado}}, \citenamefont {{Shannon}}, \citenamefont {{Spiewak}},
  \citenamefont {{van Straten}}, \citenamefont {{Toomey}}, \citenamefont
  {{Wang}}, \citenamefont {{Wen}}, \citenamefont {{You}},\ and\ \citenamefont
  {{Zhu}}}]{lms+16}%
  \BibitemOpen
  \bibfield  {author} {\bibinfo {author} {\bibfnamefont {P.~D.}\ \bibnamefont
  {{Lasky}}}, \bibinfo {author} {\bibfnamefont {C.~M.~F.}\ \bibnamefont
  {{Mingarelli}}}, \bibinfo {author} {\bibfnamefont {T.~L.}\ \bibnamefont
  {{Smith}}}, \bibinfo {author} {\bibfnamefont {J.~T.}\ \bibnamefont
  {{Giblin}}}, \bibinfo {author} {\bibfnamefont {E.}~\bibnamefont {{Thrane}}},
  \bibinfo {author} {\bibfnamefont {D.~J.}\ \bibnamefont {{Reardon}}}, \bibinfo
  {author} {\bibfnamefont {R.}~\bibnamefont {{Caldwell}}}, \bibinfo {author}
  {\bibfnamefont {M.}~\bibnamefont {{Bailes}}}, \bibinfo {author}
  {\bibfnamefont {N.~D.~R.}\ \bibnamefont {{Bhat}}}, \bibinfo {author}
  {\bibfnamefont {S.}~\bibnamefont {{Burke-Spolaor}}}, \bibinfo {author}
  {\bibfnamefont {S.}~\bibnamefont {{Dai}}}, \bibinfo {author} {\bibfnamefont
  {J.}~\bibnamefont {{Dempsey}}}, \bibinfo {author} {\bibfnamefont
  {G.}~\bibnamefont {{Hobbs}}}, \bibinfo {author} {\bibfnamefont
  {M.}~\bibnamefont {{Kerr}}}, \bibinfo {author} {\bibfnamefont
  {Y.}~\bibnamefont {{Levin}}}, \bibinfo {author} {\bibfnamefont {R.~N.}\
  \bibnamefont {{Manchester}}}, \bibinfo {author} {\bibfnamefont
  {S.}~\bibnamefont {{Os{\l}owski}}}, \bibinfo {author} {\bibfnamefont
  {V.}~\bibnamefont {{Ravi}}}, \bibinfo {author} {\bibfnamefont {P.~A.}\
  \bibnamefont {{Rosado}}}, \bibinfo {author} {\bibfnamefont {R.~M.}\
  \bibnamefont {{Shannon}}}, \bibinfo {author} {\bibfnamefont {R.}~\bibnamefont
  {{Spiewak}}}, \bibinfo {author} {\bibfnamefont {W.}~\bibnamefont {{van
  Straten}}}, \bibinfo {author} {\bibfnamefont {L.}~\bibnamefont {{Toomey}}},
  \bibinfo {author} {\bibfnamefont {J.}~\bibnamefont {{Wang}}}, \bibinfo
  {author} {\bibfnamefont {L.}~\bibnamefont {{Wen}}}, \bibinfo {author}
  {\bibfnamefont {X.}~\bibnamefont {{You}}},\ and\ \bibinfo {author}
  {\bibfnamefont {X.}~\bibnamefont {{Zhu}}},\ }\bibfield  {title} {\bibinfo
  {title} {{Gravitational-Wave Cosmology across 29 Decades in Frequency}},\
  }\href {https://doi.org/10.1103/PhysRevX.6.011035} {\bibfield  {journal}
  {\bibinfo  {journal} {Physical Review X}\ }\textbf {\bibinfo {volume} {6}},\
  \bibinfo {eid} {011035} (\bibinfo {year} {2016})},\ \Eprint
  {https://arxiv.org/abs/1511.05994} {arXiv:1511.05994 [astro-ph.CO]}
  \BibitemShut {NoStop}%
\bibitem [{\citenamefont {Vagnozzi}(2021)}]{Vagnozzi:2020gtf}%
  \BibitemOpen
  \bibfield  {author} {\bibinfo {author} {\bibfnamefont {S.}~\bibnamefont
  {Vagnozzi}},\ }\bibfield  {title} {\bibinfo {title} {{Implications of the
  NANOGrav results for inflation}},\ }\href
  {https://doi.org/10.1093/mnrasl/slaa203} {\bibfield  {journal} {\bibinfo
  {journal} {Mon. Not. Roy. Astron. Soc.}\ }\textbf {\bibinfo {volume} {502}},\
  \bibinfo {pages} {L11} (\bibinfo {year} {2021})},\ \Eprint
  {https://arxiv.org/abs/2009.13432} {arXiv:2009.13432 [astro-ph.CO]}
  \BibitemShut {NoStop}%
\bibitem [{\citenamefont {{Winicour}}(1973)}]{1973ApJ...182..919W}%
  \BibitemOpen
  \bibfield  {author} {\bibinfo {author} {\bibfnamefont {J.}~\bibnamefont
  {{Winicour}}},\ }\bibfield  {title} {\bibinfo {title} {{Gravitational
  Radiation from Relativistic Phase Transitions}},\ }\href
  {https://doi.org/10.1086/152193} {\bibfield  {journal} {\bibinfo  {journal}
  {\apj}\ }\textbf {\bibinfo {volume} {182}},\ \bibinfo {pages} {919} (\bibinfo
  {year} {1973})}\BibitemShut {NoStop}%
\bibitem [{\citenamefont {Hogan}(1986)}]{Hogan:1986qda}%
  \BibitemOpen
  \bibfield  {author} {\bibinfo {author} {\bibfnamefont {C.~J.}\ \bibnamefont
  {Hogan}},\ }\bibfield  {title} {\bibinfo {title} {{Gravitational radiation
  from cosmological phase transitions}},\ }\href@noop {} {\bibfield  {journal}
  {\bibinfo  {journal} {Mon. Not. Roy. Astron. Soc.}\ }\textbf {\bibinfo
  {volume} {218}},\ \bibinfo {pages} {629} (\bibinfo {year}
  {1986})}\BibitemShut {NoStop}%
\bibitem [{\citenamefont {Deryagin}\ \emph {et~al.}(1986)\citenamefont
  {Deryagin}, \citenamefont {Grigoriev}, \citenamefont {Rubakov},\ and\
  \citenamefont {Sazhin}}]{Deryagin:1986qq}%
  \BibitemOpen
  \bibfield  {author} {\bibinfo {author} {\bibfnamefont {D.~V.}\ \bibnamefont
  {Deryagin}}, \bibinfo {author} {\bibfnamefont {D.~Y.}\ \bibnamefont
  {Grigoriev}}, \bibinfo {author} {\bibfnamefont {V.~A.}\ \bibnamefont
  {Rubakov}},\ and\ \bibinfo {author} {\bibfnamefont {M.~V.}\ \bibnamefont
  {Sazhin}},\ }\bibfield  {title} {\bibinfo {title} {{Possible Anisotropic
  Phases in the Early Universe and Gravitational Wave Background}},\ }\href
  {https://doi.org/10.1142/S0217732386000750} {\bibfield  {journal} {\bibinfo
  {journal} {Mod. Phys. Lett. A}\ }\textbf {\bibinfo {volume} {1}},\ \bibinfo
  {pages} {593} (\bibinfo {year} {1986})}\BibitemShut {NoStop}%
\bibitem [{\citenamefont {{Caprini}}\ \emph {et~al.}(2010)\citenamefont
  {{Caprini}}, \citenamefont {{Durrer}},\ and\ \citenamefont
  {{Siemens}}}]{cds10}%
  \BibitemOpen
  \bibfield  {author} {\bibinfo {author} {\bibfnamefont {C.}~\bibnamefont
  {{Caprini}}}, \bibinfo {author} {\bibfnamefont {R.}~\bibnamefont
  {{Durrer}}},\ and\ \bibinfo {author} {\bibfnamefont {X.}~\bibnamefont
  {{Siemens}}},\ }\bibfield  {title} {\bibinfo {title} {{Detection of
  gravitational waves from the QCD phase transition with pulsar timing
  arrays}},\ }\href {https://doi.org/10.1103/PhysRevD.82.063511} {\bibfield
  {journal} {\bibinfo  {journal} {\prd}\ }\textbf {\bibinfo {volume} {82}},\
  \bibinfo {eid} {063511} (\bibinfo {year} {2010})},\ \Eprint
  {https://arxiv.org/abs/1007.1218} {arXiv:1007.1218 [astro-ph.CO]}
  \BibitemShut {NoStop}%
\bibitem [{\citenamefont {{Kobakhidze}}\ \emph {et~al.}(2017)\citenamefont
  {{Kobakhidze}}, \citenamefont {{Lagger}}, \citenamefont {{Manning}},\ and\
  \citenamefont {{Yue}}}]{klm+17}%
  \BibitemOpen
  \bibfield  {author} {\bibinfo {author} {\bibfnamefont {A.}~\bibnamefont
  {{Kobakhidze}}}, \bibinfo {author} {\bibfnamefont {C.}~\bibnamefont
  {{Lagger}}}, \bibinfo {author} {\bibfnamefont {A.}~\bibnamefont
  {{Manning}}},\ and\ \bibinfo {author} {\bibfnamefont {J.}~\bibnamefont
  {{Yue}}},\ }\bibfield  {title} {\bibinfo {title} {Gravitational waves from a
  supercooled electroweak phase transition and their detection with pulsar
  timing arrays},\ }\href {https://doi.org/10.1140/epjc/s10052-017-5132-y}
  {\bibfield  {journal} {\bibinfo  {journal} {The European Physical Journal C}\
  }\textbf {\bibinfo {volume} {77}} (\bibinfo {year} {2017})}\BibitemShut
  {NoStop}%
\bibitem [{\citenamefont {{Alam}}\ \emph {et~al.}(2020)\citenamefont {{Alam}},
  \citenamefont {{Arzoumanian}}, \citenamefont {{Baker}}, \citenamefont
  {{Blumer}}, \citenamefont {{Bohler}}, \citenamefont {{Brazier}},
  \citenamefont {{Brook}}, \citenamefont {{Burke-Spolaor}}, \citenamefont
  {{Caballero}}, \citenamefont {{Camuccio}}, \citenamefont {{Chamberlain}},
  \citenamefont {{Chatterjee}}, \citenamefont {{Cordes}}, \citenamefont
  {{Cornish}}, \citenamefont {{Crawford}}, \citenamefont {{Cromartie}},
  \citenamefont {{DeCesar}}, \citenamefont {{Demorest}}, \citenamefont
  {{Dolch}}, \citenamefont {{Ellis}}, \citenamefont {{Ferdman}}, \citenamefont
  {{Ferrara}}, \citenamefont {{Fiore}}, \citenamefont {{Fonseca}},
  \citenamefont {{Garcia}}, \citenamefont {{Garver-Daniels}}, \citenamefont
  {{Gentile}}, \citenamefont {{Good}}, \citenamefont {{Gusdorff}},
  \citenamefont {{Halmrast}}, \citenamefont {{Hazboun}}, \citenamefont
  {{Islo}}, \citenamefont {{Jennings}}, \citenamefont {{Jessup}}, \citenamefont
  {{Jones}}, \citenamefont {{Kaiser}}, \citenamefont {{Kaplan}}, \citenamefont
  {{Kelley}}, \citenamefont {{Shapiro Key}}, \citenamefont {{Lam}},
  \citenamefont {{Lazio}}, \citenamefont {{Lorimer}}, \citenamefont {{Luo}},
  \citenamefont {{Lynch}}, \citenamefont {{Madison}}, \citenamefont
  {{Maraccini}}, \citenamefont {{McLaughlin}}, \citenamefont {{Mingarelli}},
  \citenamefont {{Ng}}, \citenamefont {{Nguyen}}, \citenamefont {{Nice}},
  \citenamefont {{Pennucci}}, \citenamefont {{Pol}}, \citenamefont {{Ramette}},
  \citenamefont {{Ransom}}, \citenamefont {{Ray}}, \citenamefont
  {{Shapiro-Albert}}, \citenamefont {{Siemens}}, \citenamefont {{Simon}},
  \citenamefont {{Spiewak}}, \citenamefont {{Stairs}}, \citenamefont
  {{Stinebring}}, \citenamefont {{Stovall}}, \citenamefont {{Swiggum}},
  \citenamefont {{Taylor}}, \citenamefont {{Tripepi}}, \citenamefont
  {{Vallisneri}}, \citenamefont {{Vigeland}}, \citenamefont {{Witt}},\ and\
  \citenamefont {{Zhu}}}]{aab+20}%
  \BibitemOpen
  \bibfield  {author} {\bibinfo {author} {\bibfnamefont {M.~F.}\ \bibnamefont
  {{Alam}}}, \bibinfo {author} {\bibfnamefont {Z.}~\bibnamefont
  {{Arzoumanian}}}, \bibinfo {author} {\bibfnamefont {P.~T.}\ \bibnamefont
  {{Baker}}}, \bibinfo {author} {\bibfnamefont {H.}~\bibnamefont {{Blumer}}},
  \bibinfo {author} {\bibfnamefont {K.~E.}\ \bibnamefont {{Bohler}}}, \bibinfo
  {author} {\bibfnamefont {A.}~\bibnamefont {{Brazier}}}, \bibinfo {author}
  {\bibfnamefont {P.~R.}\ \bibnamefont {{Brook}}}, \bibinfo {author}
  {\bibfnamefont {S.}~\bibnamefont {{Burke-Spolaor}}}, \bibinfo {author}
  {\bibfnamefont {K.}~\bibnamefont {{Caballero}}}, \bibinfo {author}
  {\bibfnamefont {R.~S.}\ \bibnamefont {{Camuccio}}}, \bibinfo {author}
  {\bibfnamefont {R.~L.}\ \bibnamefont {{Chamberlain}}}, \bibinfo {author}
  {\bibfnamefont {S.}~\bibnamefont {{Chatterjee}}}, \bibinfo {author}
  {\bibfnamefont {J.~M.}\ \bibnamefont {{Cordes}}}, \bibinfo {author}
  {\bibfnamefont {N.~J.}\ \bibnamefont {{Cornish}}}, \bibinfo {author}
  {\bibfnamefont {F.}~\bibnamefont {{Crawford}}}, \bibinfo {author}
  {\bibfnamefont {H.~T.}\ \bibnamefont {{Cromartie}}}, \bibinfo {author}
  {\bibfnamefont {M.~E.}\ \bibnamefont {{DeCesar}}}, \bibinfo {author}
  {\bibfnamefont {P.~B.}\ \bibnamefont {{Demorest}}}, \bibinfo {author}
  {\bibfnamefont {T.}~\bibnamefont {{Dolch}}}, \bibinfo {author} {\bibfnamefont
  {J.~A.}\ \bibnamefont {{Ellis}}}, \bibinfo {author} {\bibfnamefont {R.~D.}\
  \bibnamefont {{Ferdman}}}, \bibinfo {author} {\bibfnamefont {E.~C.}\
  \bibnamefont {{Ferrara}}}, \bibinfo {author} {\bibfnamefont {W.}~\bibnamefont
  {{Fiore}}}, \bibinfo {author} {\bibfnamefont {E.}~\bibnamefont {{Fonseca}}},
  \bibinfo {author} {\bibfnamefont {Y.}~\bibnamefont {{Garcia}}}, \bibinfo
  {author} {\bibfnamefont {N.}~\bibnamefont {{Garver-Daniels}}}, \bibinfo
  {author} {\bibfnamefont {P.~A.}\ \bibnamefont {{Gentile}}}, \bibinfo {author}
  {\bibfnamefont {D.~C.}\ \bibnamefont {{Good}}}, \bibinfo {author}
  {\bibfnamefont {J.~A.}\ \bibnamefont {{Gusdorff}}}, \bibinfo {author}
  {\bibfnamefont {D.}~\bibnamefont {{Halmrast}}}, \bibinfo {author}
  {\bibfnamefont {J.}~\bibnamefont {{Hazboun}}}, \bibinfo {author}
  {\bibfnamefont {K.}~\bibnamefont {{Islo}}}, \bibinfo {author} {\bibfnamefont
  {R.~J.}\ \bibnamefont {{Jennings}}}, \bibinfo {author} {\bibfnamefont
  {C.}~\bibnamefont {{Jessup}}}, \bibinfo {author} {\bibfnamefont {M.~L.}\
  \bibnamefont {{Jones}}}, \bibinfo {author} {\bibfnamefont {A.~R.}\
  \bibnamefont {{Kaiser}}}, \bibinfo {author} {\bibfnamefont {D.~L.}\
  \bibnamefont {{Kaplan}}}, \bibinfo {author} {\bibfnamefont {L.~Z.}\
  \bibnamefont {{Kelley}}}, \bibinfo {author} {\bibfnamefont {J.}~\bibnamefont
  {{Shapiro Key}}}, \bibinfo {author} {\bibfnamefont {M.~T.}\ \bibnamefont
  {{Lam}}}, \bibinfo {author} {\bibfnamefont {T.~J.~W.}\ \bibnamefont
  {{Lazio}}}, \bibinfo {author} {\bibfnamefont {D.~R.}\ \bibnamefont
  {{Lorimer}}}, \bibinfo {author} {\bibfnamefont {J.}~\bibnamefont {{Luo}}},
  \bibinfo {author} {\bibfnamefont {R.~S.}\ \bibnamefont {{Lynch}}}, \bibinfo
  {author} {\bibfnamefont {D.}~\bibnamefont {{Madison}}}, \bibinfo {author}
  {\bibfnamefont {K.}~\bibnamefont {{Maraccini}}}, \bibinfo {author}
  {\bibfnamefont {M.~A.}\ \bibnamefont {{McLaughlin}}}, \bibinfo {author}
  {\bibfnamefont {C.~M.~F.}\ \bibnamefont {{Mingarelli}}}, \bibinfo {author}
  {\bibfnamefont {C.}~\bibnamefont {{Ng}}}, \bibinfo {author} {\bibfnamefont
  {B.~M.~X.}\ \bibnamefont {{Nguyen}}}, \bibinfo {author} {\bibfnamefont
  {D.~J.}\ \bibnamefont {{Nice}}}, \bibinfo {author} {\bibfnamefont {T.~T.}\
  \bibnamefont {{Pennucci}}}, \bibinfo {author} {\bibfnamefont {N.~S.}\
  \bibnamefont {{Pol}}}, \bibinfo {author} {\bibfnamefont {J.}~\bibnamefont
  {{Ramette}}}, \bibinfo {author} {\bibfnamefont {S.~M.}\ \bibnamefont
  {{Ransom}}}, \bibinfo {author} {\bibfnamefont {P.~S.}\ \bibnamefont {{Ray}}},
  \bibinfo {author} {\bibfnamefont {B.~J.}\ \bibnamefont {{Shapiro-Albert}}},
  \bibinfo {author} {\bibfnamefont {X.}~\bibnamefont {{Siemens}}}, \bibinfo
  {author} {\bibfnamefont {J.}~\bibnamefont {{Simon}}}, \bibinfo {author}
  {\bibfnamefont {R.}~\bibnamefont {{Spiewak}}}, \bibinfo {author}
  {\bibfnamefont {I.~H.}\ \bibnamefont {{Stairs}}}, \bibinfo {author}
  {\bibfnamefont {D.~R.}\ \bibnamefont {{Stinebring}}}, \bibinfo {author}
  {\bibfnamefont {K.}~\bibnamefont {{Stovall}}}, \bibinfo {author}
  {\bibfnamefont {J.~K.}\ \bibnamefont {{Swiggum}}}, \bibinfo {author}
  {\bibfnamefont {S.~R.}\ \bibnamefont {{Taylor}}}, \bibinfo {author}
  {\bibfnamefont {M.}~\bibnamefont {{Tripepi}}}, \bibinfo {author}
  {\bibfnamefont {M.}~\bibnamefont {{Vallisneri}}}, \bibinfo {author}
  {\bibfnamefont {S.~J.}\ \bibnamefont {{Vigeland}}}, \bibinfo {author}
  {\bibfnamefont {C.~A.}\ \bibnamefont {{Witt}}},\ and\ \bibinfo {author}
  {\bibfnamefont {W.}~\bibnamefont {{Zhu}}},\ }\bibfield  {title} {\bibinfo
  {title} {{The NANOGrav 12.5-year Data Set: Observations and Narrowband Timing
  of 47 Millisecond Pulsars}},\ }\href@noop {} {\bibfield  {journal} {\bibinfo
  {journal} {arXiv e-prints}\ ,\ \bibinfo {eid} {arXiv:2005.06490}} (\bibinfo
  {year} {2020})},\ \Eprint {https://arxiv.org/abs/2005.06490}
  {arXiv:2005.06490 [astro-ph.HE]} \BibitemShut {NoStop}%
\bibitem [{\citenamefont {{Arzoumanian}}\ \emph {et~al.}(2020)\citenamefont
  {{Arzoumanian}}, \citenamefont {{Baker}}, \citenamefont {{Blumer}},
  \citenamefont {{B{\'e}csy}}, \citenamefont {{Brazier}}, \citenamefont
  {{Brook}}, \citenamefont {{Burke-Spolaor}}, \citenamefont {{Chatterjee}},
  \citenamefont {{Chen}}, \citenamefont {{Cordes}}, \citenamefont {{Cornish}},
  \citenamefont {{Crawford}}, \citenamefont {{Cromartie}}, \citenamefont
  {{Decesar}}, \citenamefont {{Demorest}}, \citenamefont {{Dolch}},
  \citenamefont {{Ellis}}, \citenamefont {{Ferrara}}, \citenamefont {{Fiore}},
  \citenamefont {{Fonseca}}, \citenamefont {{Garver-Daniels}}, \citenamefont
  {{Gentile}}, \citenamefont {{Good}}, \citenamefont {{Hazboun}}, \citenamefont
  {{Holgado}}, \citenamefont {{Islo}}, \citenamefont {{Jennings}},
  \citenamefont {{Jones}}, \citenamefont {{Kaiser}}, \citenamefont {{Kaplan}},
  \citenamefont {{Kelley}}, \citenamefont {{Key}}, \citenamefont {{Laal}},
  \citenamefont {{Lam}}, \citenamefont {{Lazio}}, \citenamefont {{Lorimer}},
  \citenamefont {{Luo}}, \citenamefont {{Lynch}}, \citenamefont {{Madison}},
  \citenamefont {{McLaughlin}}, \citenamefont {{Mingarelli}}, \citenamefont
  {{Ng}}, \citenamefont {{Nice}}, \citenamefont {{Pennucci}}, \citenamefont
  {{Pol}}, \citenamefont {{Ransom}}, \citenamefont {{Ray}}, \citenamefont
  {{Shapiro-Albert}}, \citenamefont {{Siemens}}, \citenamefont {{Simon}},
  \citenamefont {{Spiewak}}, \citenamefont {{Stairs}}, \citenamefont
  {{Stinebring}}, \citenamefont {{Stovall}}, \citenamefont {{Sun}},
  \citenamefont {{Swiggum}}, \citenamefont {{Taylor}}, \citenamefont
  {{Turner}}, \citenamefont {{Vallisneri}}, \citenamefont {{Vigeland}},
  \citenamefont {{Witt}},\ and\ \citenamefont {{Nanograv
  Collaboration}}}]{abb+20}%
  \BibitemOpen
  \bibfield  {author} {\bibinfo {author} {\bibfnamefont {Z.}~\bibnamefont
  {{Arzoumanian}}}, \bibinfo {author} {\bibfnamefont {P.~T.}\ \bibnamefont
  {{Baker}}}, \bibinfo {author} {\bibfnamefont {H.}~\bibnamefont {{Blumer}}},
  \bibinfo {author} {\bibfnamefont {B.}~\bibnamefont {{B{\'e}csy}}}, \bibinfo
  {author} {\bibfnamefont {A.}~\bibnamefont {{Brazier}}}, \bibinfo {author}
  {\bibfnamefont {P.~R.}\ \bibnamefont {{Brook}}}, \bibinfo {author}
  {\bibfnamefont {S.}~\bibnamefont {{Burke-Spolaor}}}, \bibinfo {author}
  {\bibfnamefont {S.}~\bibnamefont {{Chatterjee}}}, \bibinfo {author}
  {\bibfnamefont {S.}~\bibnamefont {{Chen}}}, \bibinfo {author} {\bibfnamefont
  {J.~M.}\ \bibnamefont {{Cordes}}}, \bibinfo {author} {\bibfnamefont {N.~J.}\
  \bibnamefont {{Cornish}}}, \bibinfo {author} {\bibfnamefont {F.}~\bibnamefont
  {{Crawford}}}, \bibinfo {author} {\bibfnamefont {H.~T.}\ \bibnamefont
  {{Cromartie}}}, \bibinfo {author} {\bibfnamefont {M.~E.}\ \bibnamefont
  {{Decesar}}}, \bibinfo {author} {\bibfnamefont {P.~B.}\ \bibnamefont
  {{Demorest}}}, \bibinfo {author} {\bibfnamefont {T.}~\bibnamefont {{Dolch}}},
  \bibinfo {author} {\bibfnamefont {J.~A.}\ \bibnamefont {{Ellis}}}, \bibinfo
  {author} {\bibfnamefont {E.~C.}\ \bibnamefont {{Ferrara}}}, \bibinfo {author}
  {\bibfnamefont {W.}~\bibnamefont {{Fiore}}}, \bibinfo {author} {\bibfnamefont
  {E.}~\bibnamefont {{Fonseca}}}, \bibinfo {author} {\bibfnamefont
  {N.}~\bibnamefont {{Garver-Daniels}}}, \bibinfo {author} {\bibfnamefont
  {P.~A.}\ \bibnamefont {{Gentile}}}, \bibinfo {author} {\bibfnamefont {D.~C.}\
  \bibnamefont {{Good}}}, \bibinfo {author} {\bibfnamefont {J.~S.}\
  \bibnamefont {{Hazboun}}}, \bibinfo {author} {\bibfnamefont {A.~M.}\
  \bibnamefont {{Holgado}}}, \bibinfo {author} {\bibfnamefont {K.}~\bibnamefont
  {{Islo}}}, \bibinfo {author} {\bibfnamefont {R.~J.}\ \bibnamefont
  {{Jennings}}}, \bibinfo {author} {\bibfnamefont {M.~L.}\ \bibnamefont
  {{Jones}}}, \bibinfo {author} {\bibfnamefont {A.~R.}\ \bibnamefont
  {{Kaiser}}}, \bibinfo {author} {\bibfnamefont {D.~L.}\ \bibnamefont
  {{Kaplan}}}, \bibinfo {author} {\bibfnamefont {L.~Z.}\ \bibnamefont
  {{Kelley}}}, \bibinfo {author} {\bibfnamefont {J.~S.}\ \bibnamefont {{Key}}},
  \bibinfo {author} {\bibfnamefont {N.}~\bibnamefont {{Laal}}}, \bibinfo
  {author} {\bibfnamefont {M.~T.}\ \bibnamefont {{Lam}}}, \bibinfo {author}
  {\bibfnamefont {T.~J.~W.}\ \bibnamefont {{Lazio}}}, \bibinfo {author}
  {\bibfnamefont {D.~R.}\ \bibnamefont {{Lorimer}}}, \bibinfo {author}
  {\bibfnamefont {J.}~\bibnamefont {{Luo}}}, \bibinfo {author} {\bibfnamefont
  {R.~S.}\ \bibnamefont {{Lynch}}}, \bibinfo {author} {\bibfnamefont {D.~R.}\
  \bibnamefont {{Madison}}}, \bibinfo {author} {\bibfnamefont {M.~A.}\
  \bibnamefont {{McLaughlin}}}, \bibinfo {author} {\bibfnamefont {C.~M.~F.}\
  \bibnamefont {{Mingarelli}}}, \bibinfo {author} {\bibfnamefont
  {C.}~\bibnamefont {{Ng}}}, \bibinfo {author} {\bibfnamefont {D.~J.}\
  \bibnamefont {{Nice}}}, \bibinfo {author} {\bibfnamefont {T.~T.}\
  \bibnamefont {{Pennucci}}}, \bibinfo {author} {\bibfnamefont {N.~S.}\
  \bibnamefont {{Pol}}}, \bibinfo {author} {\bibfnamefont {S.~M.}\ \bibnamefont
  {{Ransom}}}, \bibinfo {author} {\bibfnamefont {P.~S.}\ \bibnamefont {{Ray}}},
  \bibinfo {author} {\bibfnamefont {B.~J.}\ \bibnamefont {{Shapiro-Albert}}},
  \bibinfo {author} {\bibfnamefont {X.}~\bibnamefont {{Siemens}}}, \bibinfo
  {author} {\bibfnamefont {J.}~\bibnamefont {{Simon}}}, \bibinfo {author}
  {\bibfnamefont {R.}~\bibnamefont {{Spiewak}}}, \bibinfo {author}
  {\bibfnamefont {I.~H.}\ \bibnamefont {{Stairs}}}, \bibinfo {author}
  {\bibfnamefont {D.~R.}\ \bibnamefont {{Stinebring}}}, \bibinfo {author}
  {\bibfnamefont {K.}~\bibnamefont {{Stovall}}}, \bibinfo {author}
  {\bibfnamefont {J.~P.}\ \bibnamefont {{Sun}}}, \bibinfo {author}
  {\bibfnamefont {J.~K.}\ \bibnamefont {{Swiggum}}}, \bibinfo {author}
  {\bibfnamefont {S.~R.}\ \bibnamefont {{Taylor}}}, \bibinfo {author}
  {\bibfnamefont {J.~E.}\ \bibnamefont {{Turner}}}, \bibinfo {author}
  {\bibfnamefont {M.}~\bibnamefont {{Vallisneri}}}, \bibinfo {author}
  {\bibfnamefont {S.~J.}\ \bibnamefont {{Vigeland}}}, \bibinfo {author}
  {\bibfnamefont {C.~A.}\ \bibnamefont {{Witt}}},\ and\ \bibinfo {author}
  {\bibnamefont {{Nanograv Collaboration}}},\ }\bibfield  {title} {\bibinfo
  {title} {{The NANOGrav 12.5 yr Data Set: Search for an Isotropic Stochastic
  Gravitational-wave Background}},\ }\href
  {https://doi.org/10.3847/2041-8213/abd401} {\bibfield  {journal} {\bibinfo
  {journal} {\apjl}\ }\textbf {\bibinfo {volume} {905}},\ \bibinfo {eid} {L34}
  (\bibinfo {year} {2020})},\ \Eprint {https://arxiv.org/abs/2009.04496}
  {arXiv:2009.04496 [astro-ph.HE]} \BibitemShut {NoStop}%
\bibitem [{\citenamefont {{Middleton}}\ \emph {et~al.}(2021)\citenamefont
  {{Middleton}}, \citenamefont {{Sesana}}, \citenamefont {{Chen}},
  \citenamefont {{Vecchio}}, \citenamefont {{Del Pozzo}},\ and\ \citenamefont
  {{Rosado}}}]{2021MNRAS.502L..99M}%
  \BibitemOpen
  \bibfield  {author} {\bibinfo {author} {\bibfnamefont {H.}~\bibnamefont
  {{Middleton}}}, \bibinfo {author} {\bibfnamefont {A.}~\bibnamefont
  {{Sesana}}}, \bibinfo {author} {\bibfnamefont {S.}~\bibnamefont {{Chen}}},
  \bibinfo {author} {\bibfnamefont {A.}~\bibnamefont {{Vecchio}}}, \bibinfo
  {author} {\bibfnamefont {W.}~\bibnamefont {{Del Pozzo}}},\ and\ \bibinfo
  {author} {\bibfnamefont {P.~A.}\ \bibnamefont {{Rosado}}},\ }\bibfield
  {title} {\bibinfo {title} {{Massive black hole binary systems and the
  NANOGrav 12.5 yr results}},\ }\href {https://doi.org/10.1093/mnrasl/slab008}
  {\bibfield  {journal} {\bibinfo  {journal} {\mnras}\ }\textbf {\bibinfo
  {volume} {502}},\ \bibinfo {pages} {L99} (\bibinfo {year} {2021})},\ \Eprint
  {https://arxiv.org/abs/2011.01246} {arXiv:2011.01246 [astro-ph.HE]}
  \BibitemShut {NoStop}%
\bibitem [{\citenamefont {{Pol}}\ \emph {et~al.}(2020)\citenamefont {{Pol}},
  \citenamefont {{Taylor}}, \citenamefont {{Kelley}}, \citenamefont
  {{Vigeland}}, \citenamefont {{Simon}}, \citenamefont {{Chen}}, \citenamefont
  {{Arzoumanian}}, \citenamefont {{Baker}}, \citenamefont {{B{\'e}csy}},
  \citenamefont {{Brazier}}, \citenamefont {{Brook}}, \citenamefont
  {{Burke-Spolaor}}, \citenamefont {{Chatterjee}}, \citenamefont {{Cordes}},
  \citenamefont {{Cornish}}, \citenamefont {{Crawford}}, \citenamefont
  {{Cromartie}}, \citenamefont {{DeCesar}}, \citenamefont {{Demorest}},
  \citenamefont {{Dolch}}, \citenamefont {{Ferrara}}, \citenamefont {{Fiore}},
  \citenamefont {{Fonseca}}, \citenamefont {{Garver-Daniels}}, \citenamefont
  {{Good}}, \citenamefont {{Hazboun}}, \citenamefont {{Jennings}},
  \citenamefont {{Jones}}, \citenamefont {{Kaiser}}, \citenamefont {{Kaplan}},
  \citenamefont {{Shapiro Key}}, \citenamefont {{Lam}}, \citenamefont
  {{Lazio}}, \citenamefont {{Luo}}, \citenamefont {{Lynch}}, \citenamefont
  {{Madison}}, \citenamefont {{McEwen}}, \citenamefont {{McLaughlin}},
  \citenamefont {{Mingarelli}}, \citenamefont {{Ng}}, \citenamefont {{Nice}},
  \citenamefont {{Pennucci}}, \citenamefont {{Ransom}}, \citenamefont {{Ray}},
  \citenamefont {{Shapiro-Albert}}, \citenamefont {{Siemens}}, \citenamefont
  {{Stairs}}, \citenamefont {{Stinebring}}, \citenamefont {{Swiggum}},
  \citenamefont {{Vallisneri}}, \citenamefont {{Wahl}},\ and\ \citenamefont
  {{Witt}}}]{2020arXiv201011950P}%
  \BibitemOpen
  \bibfield  {author} {\bibinfo {author} {\bibfnamefont {N.~S.}\ \bibnamefont
  {{Pol}}}, \bibinfo {author} {\bibfnamefont {S.~R.}\ \bibnamefont {{Taylor}}},
  \bibinfo {author} {\bibfnamefont {L.~Z.}\ \bibnamefont {{Kelley}}}, \bibinfo
  {author} {\bibfnamefont {S.~J.}\ \bibnamefont {{Vigeland}}}, \bibinfo
  {author} {\bibfnamefont {J.}~\bibnamefont {{Simon}}}, \bibinfo {author}
  {\bibfnamefont {S.}~\bibnamefont {{Chen}}}, \bibinfo {author} {\bibfnamefont
  {Z.}~\bibnamefont {{Arzoumanian}}}, \bibinfo {author} {\bibfnamefont {P.~T.}\
  \bibnamefont {{Baker}}}, \bibinfo {author} {\bibfnamefont {B.}~\bibnamefont
  {{B{\'e}csy}}}, \bibinfo {author} {\bibfnamefont {A.}~\bibnamefont
  {{Brazier}}}, \bibinfo {author} {\bibfnamefont {P.~R.}\ \bibnamefont
  {{Brook}}}, \bibinfo {author} {\bibfnamefont {S.}~\bibnamefont
  {{Burke-Spolaor}}}, \bibinfo {author} {\bibfnamefont {S.}~\bibnamefont
  {{Chatterjee}}}, \bibinfo {author} {\bibfnamefont {J.~M.}\ \bibnamefont
  {{Cordes}}}, \bibinfo {author} {\bibfnamefont {N.~J.}\ \bibnamefont
  {{Cornish}}}, \bibinfo {author} {\bibfnamefont {F.}~\bibnamefont
  {{Crawford}}}, \bibinfo {author} {\bibfnamefont {H.~T.}\ \bibnamefont
  {{Cromartie}}}, \bibinfo {author} {\bibfnamefont {M.~E.}\ \bibnamefont
  {{DeCesar}}}, \bibinfo {author} {\bibfnamefont {P.~B.}\ \bibnamefont
  {{Demorest}}}, \bibinfo {author} {\bibfnamefont {T.}~\bibnamefont {{Dolch}}},
  \bibinfo {author} {\bibfnamefont {E.~C.}\ \bibnamefont {{Ferrara}}}, \bibinfo
  {author} {\bibfnamefont {W.}~\bibnamefont {{Fiore}}}, \bibinfo {author}
  {\bibfnamefont {E.}~\bibnamefont {{Fonseca}}}, \bibinfo {author}
  {\bibfnamefont {N.}~\bibnamefont {{Garver-Daniels}}}, \bibinfo {author}
  {\bibfnamefont {D.~C.}\ \bibnamefont {{Good}}}, \bibinfo {author}
  {\bibfnamefont {J.~S.}\ \bibnamefont {{Hazboun}}}, \bibinfo {author}
  {\bibfnamefont {R.~J.}\ \bibnamefont {{Jennings}}}, \bibinfo {author}
  {\bibfnamefont {M.~L.}\ \bibnamefont {{Jones}}}, \bibinfo {author}
  {\bibfnamefont {A.~R.}\ \bibnamefont {{Kaiser}}}, \bibinfo {author}
  {\bibfnamefont {D.~L.}\ \bibnamefont {{Kaplan}}}, \bibinfo {author}
  {\bibfnamefont {J.}~\bibnamefont {{Shapiro Key}}}, \bibinfo {author}
  {\bibfnamefont {M.~T.}\ \bibnamefont {{Lam}}}, \bibinfo {author}
  {\bibfnamefont {T.~J.~W.}\ \bibnamefont {{Lazio}}}, \bibinfo {author}
  {\bibfnamefont {J.}~\bibnamefont {{Luo}}}, \bibinfo {author} {\bibfnamefont
  {R.~S.}\ \bibnamefont {{Lynch}}}, \bibinfo {author} {\bibfnamefont {D.~R.}\
  \bibnamefont {{Madison}}}, \bibinfo {author} {\bibfnamefont {A.}~\bibnamefont
  {{McEwen}}}, \bibinfo {author} {\bibfnamefont {M.~A.}\ \bibnamefont
  {{McLaughlin}}}, \bibinfo {author} {\bibfnamefont {C.~M.~F.}\ \bibnamefont
  {{Mingarelli}}}, \bibinfo {author} {\bibfnamefont {C.}~\bibnamefont {{Ng}}},
  \bibinfo {author} {\bibfnamefont {D.~J.}\ \bibnamefont {{Nice}}}, \bibinfo
  {author} {\bibfnamefont {T.~T.}\ \bibnamefont {{Pennucci}}}, \bibinfo
  {author} {\bibfnamefont {S.~M.}\ \bibnamefont {{Ransom}}}, \bibinfo {author}
  {\bibfnamefont {P.~S.}\ \bibnamefont {{Ray}}}, \bibinfo {author}
  {\bibfnamefont {B.~J.}\ \bibnamefont {{Shapiro-Albert}}}, \bibinfo {author}
  {\bibfnamefont {X.}~\bibnamefont {{Siemens}}}, \bibinfo {author}
  {\bibfnamefont {I.~H.}\ \bibnamefont {{Stairs}}}, \bibinfo {author}
  {\bibfnamefont {D.~R.}\ \bibnamefont {{Stinebring}}}, \bibinfo {author}
  {\bibfnamefont {J.~K.}\ \bibnamefont {{Swiggum}}}, \bibinfo {author}
  {\bibfnamefont {M.}~\bibnamefont {{Vallisneri}}}, \bibinfo {author}
  {\bibfnamefont {H.}~\bibnamefont {{Wahl}}},\ and\ \bibinfo {author}
  {\bibfnamefont {C.~A.}\ \bibnamefont {{Witt}}},\ }\bibfield  {title}
  {\bibinfo {title} {{Astrophysics Milestones For Pulsar Timing Array
  Gravitational Wave Detection}},\ }\href@noop {} {\bibfield  {journal}
  {\bibinfo  {journal} {arXiv e-prints}\ ,\ \bibinfo {eid} {arXiv:2010.11950}}
  (\bibinfo {year} {2020})},\ \Eprint {https://arxiv.org/abs/2010.11950}
  {arXiv:2010.11950 [astro-ph.HE]} \BibitemShut {NoStop}%
\bibitem [{\citenamefont {Vaskonen}\ and\ \citenamefont
  {Veerm\"ae}(2021)}]{ng12p5_pbh_1}%
  \BibitemOpen
  \bibfield  {author} {\bibinfo {author} {\bibfnamefont {V.}~\bibnamefont
  {Vaskonen}}\ and\ \bibinfo {author} {\bibfnamefont {H.}~\bibnamefont
  {Veerm\"ae}},\ }\bibfield  {title} {\bibinfo {title} {Did nanograv see a
  signal from primordial black hole formation?},\ }\href
  {https://doi.org/10.1103/PhysRevLett.126.051303} {\bibfield  {journal}
  {\bibinfo  {journal} {Phys. Rev. Lett.}\ }\textbf {\bibinfo {volume} {126}},\
  \bibinfo {pages} {051303} (\bibinfo {year} {2021})}\BibitemShut {NoStop}%
\bibitem [{\citenamefont {De~Luca}\ \emph {et~al.}(2021)\citenamefont
  {De~Luca}, \citenamefont {Franciolini},\ and\ \citenamefont
  {Riotto}}]{ng12p5_pbh_2}%
  \BibitemOpen
  \bibfield  {author} {\bibinfo {author} {\bibfnamefont {V.}~\bibnamefont
  {De~Luca}}, \bibinfo {author} {\bibfnamefont {G.}~\bibnamefont
  {Franciolini}},\ and\ \bibinfo {author} {\bibfnamefont {A.}~\bibnamefont
  {Riotto}},\ }\bibfield  {title} {\bibinfo {title} {Nanograv data hints at
  primordial black holes as dark matter},\ }\href
  {https://doi.org/10.1103/PhysRevLett.126.041303} {\bibfield  {journal}
  {\bibinfo  {journal} {Phys. Rev. Lett.}\ }\textbf {\bibinfo {volume} {126}},\
  \bibinfo {pages} {041303} (\bibinfo {year} {2021})}\BibitemShut {NoStop}%
\bibitem [{\citenamefont {Ellis}\ and\ \citenamefont
  {Lewicki}(2021)}]{ng12p5_cs_1}%
  \BibitemOpen
  \bibfield  {author} {\bibinfo {author} {\bibfnamefont {J.}~\bibnamefont
  {Ellis}}\ and\ \bibinfo {author} {\bibfnamefont {M.}~\bibnamefont
  {Lewicki}},\ }\bibfield  {title} {\bibinfo {title} {Cosmic string
  interpretation of nanograv pulsar timing data},\ }\href
  {https://doi.org/10.1103/PhysRevLett.126.041304} {\bibfield  {journal}
  {\bibinfo  {journal} {Phys. Rev. Lett.}\ }\textbf {\bibinfo {volume} {126}},\
  \bibinfo {pages} {041304} (\bibinfo {year} {2021})}\BibitemShut {NoStop}%
\bibitem [{\citenamefont {Blasi}\ \emph {et~al.}(2021)\citenamefont {Blasi},
  \citenamefont {Brdar},\ and\ \citenamefont {Schmitz}}]{ng12p5_cs_2}%
  \BibitemOpen
  \bibfield  {author} {\bibinfo {author} {\bibfnamefont {S.}~\bibnamefont
  {Blasi}}, \bibinfo {author} {\bibfnamefont {V.}~\bibnamefont {Brdar}},\ and\
  \bibinfo {author} {\bibfnamefont {K.}~\bibnamefont {Schmitz}},\ }\bibfield
  {title} {\bibinfo {title} {Has nanograv found first evidence for cosmic
  strings?},\ }\href {https://doi.org/10.1103/PhysRevLett.126.041305}
  {\bibfield  {journal} {\bibinfo  {journal} {Phys. Rev. Lett.}\ }\textbf
  {\bibinfo {volume} {126}},\ \bibinfo {pages} {041305} (\bibinfo {year}
  {2021})}\BibitemShut {NoStop}%
\bibitem [{\citenamefont {{Brandenburg}}\ \emph {et~al.}(2021)\citenamefont
  {{Brandenburg}}, \citenamefont {{Clarke}}, \citenamefont {{He}},\ and\
  \citenamefont {{Kahniashvili}}}]{2021arXiv210212428B}%
  \BibitemOpen
  \bibfield  {author} {\bibinfo {author} {\bibfnamefont {A.}~\bibnamefont
  {{Brandenburg}}}, \bibinfo {author} {\bibfnamefont {E.}~\bibnamefont
  {{Clarke}}}, \bibinfo {author} {\bibfnamefont {Y.}~\bibnamefont {{He}}},\
  and\ \bibinfo {author} {\bibfnamefont {T.}~\bibnamefont {{Kahniashvili}}},\
  }\bibfield  {title} {\bibinfo {title} {{Can we observe the QCD phase
  transition-generated gravitational waves through pulsar timing arrays?}},\
  }\href@noop {} {\bibfield  {journal} {\bibinfo  {journal} {arXiv e-prints}\
  ,\ \bibinfo {eid} {arXiv:2102.12428}} (\bibinfo {year} {2021})},\ \Eprint
  {https://arxiv.org/abs/2102.12428} {arXiv:2102.12428 [astro-ph.CO]}
  \BibitemShut {NoStop}%
\bibitem [{\citenamefont {{Neronov}}\ \emph {et~al.}(2021)\citenamefont
  {{Neronov}}, \citenamefont {{Pol}}, \citenamefont {{Caprini}},\ and\
  \citenamefont {{Semikoz}}}]{2021PhRvD.103L1302N}%
  \BibitemOpen
  \bibfield  {author} {\bibinfo {author} {\bibfnamefont {A.}~\bibnamefont
  {{Neronov}}}, \bibinfo {author} {\bibfnamefont {A.~R.}\ \bibnamefont
  {{Pol}}}, \bibinfo {author} {\bibfnamefont {C.}~\bibnamefont {{Caprini}}},\
  and\ \bibinfo {author} {\bibfnamefont {D.}~\bibnamefont {{Semikoz}}},\
  }\bibfield  {title} {\bibinfo {title} {{NANOGrav signal from
  magnetohydrodynamic turbulence at the QCD phase transition in the early
  Universe}},\ }\href {https://doi.org/10.1103/PhysRevD.103.L041302} {\bibfield
   {journal} {\bibinfo  {journal} {\prd}\ }\textbf {\bibinfo {volume} {103}},\
  \bibinfo {eid} {L041302} (\bibinfo {year} {2021})},\ \Eprint
  {https://arxiv.org/abs/2009.14174} {arXiv:2009.14174 [astro-ph.CO]}
  \BibitemShut {NoStop}%
\bibitem [{\citenamefont {{Li}}\ \emph {et~al.}(2021)\citenamefont {{Li}},
  \citenamefont {{Shao}}, \citenamefont {{Wu}},\ and\ \citenamefont
  {{Yu}}}]{2021arXiv210108012L}%
  \BibitemOpen
  \bibfield  {author} {\bibinfo {author} {\bibfnamefont {S.-L.}\ \bibnamefont
  {{Li}}}, \bibinfo {author} {\bibfnamefont {L.}~\bibnamefont {{Shao}}},
  \bibinfo {author} {\bibfnamefont {P.}~\bibnamefont {{Wu}}},\ and\ \bibinfo
  {author} {\bibfnamefont {H.}~\bibnamefont {{Yu}}},\ }\bibfield  {title}
  {\bibinfo {title} {{NANOGrav Signal from First-Order
  Confinement/Deconfinement Phase Transition in Different QCD Matters}},\
  }\href@noop {} {\bibfield  {journal} {\bibinfo  {journal} {arXiv e-prints}\
  ,\ \bibinfo {eid} {arXiv:2101.08012}} (\bibinfo {year} {2021})},\ \Eprint
  {https://arxiv.org/abs/2101.08012} {arXiv:2101.08012 [astro-ph.CO]}
  \BibitemShut {NoStop}%
\bibitem [{\citenamefont {{Barman}}\ \emph {et~al.}(2020)\citenamefont
  {{Barman}}, \citenamefont {{Dutta Banik}},\ and\ \citenamefont
  {{Paul}}}]{2020arXiv201211969B}%
  \BibitemOpen
  \bibfield  {author} {\bibinfo {author} {\bibfnamefont {B.}~\bibnamefont
  {{Barman}}}, \bibinfo {author} {\bibfnamefont {A.}~\bibnamefont {{Dutta
  Banik}}},\ and\ \bibinfo {author} {\bibfnamefont {A.}~\bibnamefont
  {{Paul}}},\ }\bibfield  {title} {\bibinfo {title} {{Implications of NANOGrav
  results and UV freeze-in in a fast-expanding Universe}},\ }\href@noop {}
  {\bibfield  {journal} {\bibinfo  {journal} {arXiv e-prints}\ ,\ \bibinfo
  {eid} {arXiv:2012.11969}} (\bibinfo {year} {2020})},\ \Eprint
  {https://arxiv.org/abs/2012.11969} {arXiv:2012.11969 [astro-ph.CO]}
  \BibitemShut {NoStop}%
\bibitem [{\citenamefont {{Abe}}\ \emph {et~al.}(2020)\citenamefont {{Abe}},
  \citenamefont {{Tada}},\ and\ \citenamefont {{Ueda}}}]{2020arXiv201006193A}%
  \BibitemOpen
  \bibfield  {author} {\bibinfo {author} {\bibfnamefont {K.~T.}\ \bibnamefont
  {{Abe}}}, \bibinfo {author} {\bibfnamefont {Y.}~\bibnamefont {{Tada}}},\ and\
  \bibinfo {author} {\bibfnamefont {I.}~\bibnamefont {{Ueda}}},\ }\bibfield
  {title} {\bibinfo {title} {{Induced gravitational waves as a cosmological
  probe of the sound speed during the QCD phase transition}},\ }\href@noop {}
  {\bibfield  {journal} {\bibinfo  {journal} {arXiv e-prints}\ ,\ \bibinfo
  {eid} {arXiv:2010.06193}} (\bibinfo {year} {2020})},\ \Eprint
  {https://arxiv.org/abs/2010.06193} {arXiv:2010.06193 [astro-ph.CO]}
  \BibitemShut {NoStop}%
\bibitem [{\citenamefont {{Ratzinger}}\ and\ \citenamefont
  {{Schwaller}}(2020)}]{2020arXiv200911875R}%
  \BibitemOpen
  \bibfield  {author} {\bibinfo {author} {\bibfnamefont {W.}~\bibnamefont
  {{Ratzinger}}}\ and\ \bibinfo {author} {\bibfnamefont {P.}~\bibnamefont
  {{Schwaller}}},\ }\bibfield  {title} {\bibinfo {title} {{Whispers from the
  dark side: Confronting light new physics with NANOGrav data}},\ }\href@noop
  {} {\bibfield  {journal} {\bibinfo  {journal} {arXiv e-prints}\ ,\ \bibinfo
  {eid} {arXiv:2009.11875}} (\bibinfo {year} {2020})},\ \Eprint
  {https://arxiv.org/abs/2009.11875} {arXiv:2009.11875 [astro-ph.CO]}
  \BibitemShut {NoStop}%
\bibitem [{\citenamefont {{Addazi}}\ \emph {et~al.}(2020)\citenamefont
  {{Addazi}}, \citenamefont {{Cai}}, \citenamefont {{Gan}}, \citenamefont
  {{Marciano}},\ and\ \citenamefont {{Zeng}}}]{2020arXiv200910327A}%
  \BibitemOpen
  \bibfield  {author} {\bibinfo {author} {\bibfnamefont {A.}~\bibnamefont
  {{Addazi}}}, \bibinfo {author} {\bibfnamefont {Y.-F.}\ \bibnamefont {{Cai}}},
  \bibinfo {author} {\bibfnamefont {Q.}~\bibnamefont {{Gan}}}, \bibinfo
  {author} {\bibfnamefont {A.}~\bibnamefont {{Marciano}}},\ and\ \bibinfo
  {author} {\bibfnamefont {K.}~\bibnamefont {{Zeng}}},\ }\bibfield  {title}
  {\bibinfo {title} {{NANOGrav results and Dark First Order Phase
  Transitions}},\ }\href@noop {} {\bibfield  {journal} {\bibinfo  {journal}
  {arXiv e-prints}\ ,\ \bibinfo {eid} {arXiv:2009.10327}} (\bibinfo {year}
  {2020})},\ \Eprint {https://arxiv.org/abs/2009.10327} {arXiv:2009.10327
  [hep-ph]} \BibitemShut {NoStop}%
\bibitem [{\citenamefont {{Nakai}}\ \emph {et~al.}(2020)\citenamefont
  {{Nakai}}, \citenamefont {{Suzuki}}, \citenamefont {{Takahashi}},\ and\
  \citenamefont {{Yamada}}}]{2020arXiv200909754N}%
  \BibitemOpen
  \bibfield  {author} {\bibinfo {author} {\bibfnamefont {Y.}~\bibnamefont
  {{Nakai}}}, \bibinfo {author} {\bibfnamefont {M.}~\bibnamefont {{Suzuki}}},
  \bibinfo {author} {\bibfnamefont {F.}~\bibnamefont {{Takahashi}}},\ and\
  \bibinfo {author} {\bibfnamefont {M.}~\bibnamefont {{Yamada}}},\ }\bibfield
  {title} {\bibinfo {title} {{Gravitational Waves and Dark Radiation from Dark
  Phase Transition: Connecting NANOGrav Pulsar Timing Data and Hubble
  Tension}},\ }\href@noop {} {\bibfield  {journal} {\bibinfo  {journal} {arXiv
  e-prints}\ ,\ \bibinfo {eid} {arXiv:2009.09754}} (\bibinfo {year} {2020})},\
  \Eprint {https://arxiv.org/abs/2009.09754} {arXiv:2009.09754 [astro-ph.CO]}
  \BibitemShut {NoStop}%
\bibitem [{\citenamefont {Strassler}\ and\ \citenamefont
  {Zurek}(2007)}]{Strassler:2006im}%
  \BibitemOpen
  \bibfield  {author} {\bibinfo {author} {\bibfnamefont {M.~J.}\ \bibnamefont
  {Strassler}}\ and\ \bibinfo {author} {\bibfnamefont {K.~M.}\ \bibnamefont
  {Zurek}},\ }\bibfield  {title} {\bibinfo {title} {{Echoes of a hidden valley
  at hadron colliders}},\ }\href
  {https://doi.org/10.1016/j.physletb.2007.06.055} {\bibfield  {journal}
  {\bibinfo  {journal} {Phys. Lett. B}\ }\textbf {\bibinfo {volume} {651}},\
  \bibinfo {pages} {374} (\bibinfo {year} {2007})},\ \Eprint
  {https://arxiv.org/abs/hep-ph/0604261} {arXiv:hep-ph/0604261} \BibitemShut
  {NoStop}%
\bibitem [{\citenamefont {Chacko}\ \emph {et~al.}(2004)\citenamefont {Chacko},
  \citenamefont {Hall},\ and\ \citenamefont {Nomura}}]{Chacko:2004ky}%
  \BibitemOpen
  \bibfield  {author} {\bibinfo {author} {\bibfnamefont {Z.}~\bibnamefont
  {Chacko}}, \bibinfo {author} {\bibfnamefont {L.~J.}\ \bibnamefont {Hall}},\
  and\ \bibinfo {author} {\bibfnamefont {Y.}~\bibnamefont {Nomura}},\
  }\bibfield  {title} {\bibinfo {title} {{Acceleressence: dark energy from a
  phase transition at the seesaw scale}},\ }\href
  {https://doi.org/10.1088/1475-7516/2004/10/011} {\bibfield  {journal}
  {\bibinfo  {journal} {JCAP}\ }\textbf {\bibinfo {volume} {10}},\ \bibinfo
  {pages} {011}},\ \Eprint {https://arxiv.org/abs/astro-ph/0405596}
  {arXiv:astro-ph/0405596} \BibitemShut {NoStop}%
\bibitem [{\citenamefont {Schwaller}(2015)}]{Schwaller:2015tja}%
  \BibitemOpen
  \bibfield  {author} {\bibinfo {author} {\bibfnamefont {P.}~\bibnamefont
  {Schwaller}},\ }\bibfield  {title} {\bibinfo {title} {{Gravitational Waves
  from a Dark Phase Transition}},\ }\href
  {https://doi.org/10.1103/PhysRevLett.115.181101} {\bibfield  {journal}
  {\bibinfo  {journal} {Phys. Rev. Lett.}\ }\textbf {\bibinfo {volume} {115}},\
  \bibinfo {pages} {181101} (\bibinfo {year} {2015})},\ \Eprint
  {https://arxiv.org/abs/1504.07263} {arXiv:1504.07263 [hep-ph]} \BibitemShut
  {NoStop}%
\bibitem [{\citenamefont {{Battaglieri}}\ \emph {et~al.}(2017)\citenamefont
  {{Battaglieri}}, \citenamefont {{Belloni}}, \citenamefont {{Chou}},
  \citenamefont {{Cushman}}, \citenamefont {{Echenard}}, \citenamefont
  {{Essig}}, \citenamefont {{Estrada}}, \citenamefont {{Feng}}, \citenamefont
  {{Flaugher}}, \citenamefont {{Fox}}, \citenamefont {{Graham}}, \citenamefont
  {{Hall}}, \citenamefont {{Harnik}}, \citenamefont {{Hewett}}, \citenamefont
  {{Incandela}}, \citenamefont {{Izaguirre}}, \citenamefont {{McKinsey}},
  \citenamefont {{Pyle}}, \citenamefont {{Roe}}, \citenamefont {{Rybka}},
  \citenamefont {{Sikivie}}, \citenamefont {{Tait}}, \citenamefont {{Toro}},
  \citenamefont {{Van De Water}}, \citenamefont {{Weiner}}, \citenamefont
  {{Zurek}}, \citenamefont {{Adelberger}}, \citenamefont {{Afanasev}},
  \citenamefont {{Alexander}}, \citenamefont {{Alexander}}, \citenamefont
  {{Cristian Antochi}}, \citenamefont {{Asner}}, \citenamefont {{Baer}},
  \citenamefont {{Banerjee}}, \citenamefont {{Baracchini}}, \citenamefont
  {{Barbeau}}, \citenamefont {{Barrow}}, \citenamefont {{Bastidon}},
  \citenamefont {{Battat}}, \citenamefont {{Benson}}, \citenamefont {{Berlin}},
  \citenamefont {{Bird}}, \citenamefont {{Blinov}}, \citenamefont {{Boddy}},
  \citenamefont {{Bondi}}, \citenamefont {{Bonivento}}, \citenamefont
  {{Boulay}}, \citenamefont {{Boyce}}, \citenamefont {{Brodeur}}, \citenamefont
  {{Broussard}}, \citenamefont {{Budnik}}, \citenamefont {{Bunting}},
  \citenamefont {{Caffee}}, \citenamefont {{Caiazza}}, \citenamefont
  {{Campbell}}, \citenamefont {{Cao}}, \citenamefont {{Carosi}}, \citenamefont
  {{Carpinelli}}, \citenamefont {{Cavoto}}, \citenamefont {{Celentano}},
  \citenamefont {{Hyeok Chang}}, \citenamefont {{Chattopadhyay}}, \citenamefont
  {{Chavarria}}, \citenamefont {{Chen}}, \citenamefont {{Clark}}, \citenamefont
  {{Clarke}}, \citenamefont {{Colegrove}}, \citenamefont {{Coleman}},
  \citenamefont {{Cooke}}, \citenamefont {{Cooper}}, \citenamefont {{Crisler}},
  \citenamefont {{Crivelli}}, \citenamefont {{D'Eramo}}, \citenamefont
  {{D'Urso}}, \citenamefont {{Dahl}}, \citenamefont {{Dawson}}, \citenamefont
  {{De Napoli}}, \citenamefont {{De Vita}}, \citenamefont {{DeNiverville}},
  \citenamefont {{Derenzo}}, \citenamefont {{Di Crescenzo}}, \citenamefont {{Di
  Marco}}, \citenamefont {{Dienes}}, \citenamefont {{Diwan}}, \citenamefont
  {{Handiipondola Dongwi}}, \citenamefont {{Drlica-Wagner}}, \citenamefont
  {{Ellis}}, \citenamefont {{Chigbo Ezeribe}}, \citenamefont {{Farrar}},
  \citenamefont {{Ferrer}}, \citenamefont {{Figueroa-Feliciano}}, \citenamefont
  {{Filippi}}, \citenamefont {{Fiorillo}}, \citenamefont {{Fornal}},
  \citenamefont {{Freyberger}}, \citenamefont {{Frugiuele}}, \citenamefont
  {{Galbiati}}, \citenamefont {{Galon}}, \citenamefont {{Gardner}},
  \citenamefont {{Geraci}}, \citenamefont {{Gerbier}}, \citenamefont
  {{Graham}}, \citenamefont {{Gschwendtner}}, \citenamefont {{Hearty}},
  \citenamefont {{Heise}}, \citenamefont {{Henning}}, \citenamefont {{Hill}},
  \citenamefont {{Hitlin}}, \citenamefont {{Hochberg}}, \citenamefont
  {{Hogan}}, \citenamefont {{Holtrop}}, \citenamefont {{Hong}}, \citenamefont
  {{Hossbach}}, \citenamefont {{Humensky}}, \citenamefont {{Ilten}},
  \citenamefont {{Irwin}}, \citenamefont {{Jaros}}, \citenamefont {{Johnson}},
  \citenamefont {{Jones}}, \citenamefont {{Kahn}}, \citenamefont
  {{Kalantarians}}, \citenamefont {{Kaplinghat}}, \citenamefont {{Khatiwada}},
  \citenamefont {{Knapen}}, \citenamefont {{Kohl}}, \citenamefont {{Kouvaris}},
  \citenamefont {{Kozaczuk}}, \citenamefont {{Krnjaic}}, \citenamefont
  {{Kubarovsky}}, \citenamefont {{Kuflik}}, \citenamefont {{Kusenko}},
  \citenamefont {{Lang}}, \citenamefont {{Leach}}, \citenamefont {{Lin}},
  \citenamefont {{Lisanti}}, \citenamefont {{Liu}}, \citenamefont {{Liu}},
  \citenamefont {{Liu}}, \citenamefont {{Loomba}}, \citenamefont {{Lykken}},
  \citenamefont {{Mack}}, \citenamefont {{Mans}}, \citenamefont {{Maris}},
  \citenamefont {{Markiewicz}}, \citenamefont {{Marsicano}}, \citenamefont
  {{Martoff}}, \citenamefont {{Mazzitelli}}, \citenamefont {{McCabe}},
  \citenamefont {{McDermott}}, \citenamefont {{McDonald}}, \citenamefont
  {{McKinnon}}, \citenamefont {{Mei}}, \citenamefont {{Melia}}, \citenamefont
  {{Miller}}, \citenamefont {{Miuchi}}, \citenamefont {{Nazeer}}, \citenamefont
  {{Moreno}}, \citenamefont {{Morozov}}, \citenamefont {{Mouton}},
  \citenamefont {{Mueller}}, \citenamefont {{Murphy}}, \citenamefont
  {{Neilson}}, \citenamefont {{Nelson}}, \citenamefont {{Neu}}, \citenamefont
  {{Nosochkov}}, \citenamefont {{O'Hare}}, \citenamefont {{Oblath}},
  \citenamefont {{Orrell}}, \citenamefont {{Ouellet}}, \citenamefont
  {{Pastore}}, \citenamefont {{Paul}}, \citenamefont {{Perelstein}},
  \citenamefont {{Peter}}, \citenamefont {{Phan}}, \citenamefont {{Phinney}},
  \citenamefont {{Pivovaroff}}, \citenamefont {{Pocar}}, \citenamefont
  {{Pospelov}}, \citenamefont {{Pradler}}, \citenamefont {{Privitera}},
  \citenamefont {{Profumo}}, \citenamefont {{Raggi}}, \citenamefont
  {{Rajendran}}, \citenamefont {{Randazzo}}, \citenamefont {{Raubenheimer}},
  \citenamefont {{Regenfus}}, \citenamefont {{Renshaw}}, \citenamefont
  {{Ritz}}, \citenamefont {{Rizzo}}, \citenamefont {{Rosenberg}}, \citenamefont
  {{Rubbia}}, \citenamefont {{Rybolt}}, \citenamefont {{Saab}}, \citenamefont
  {{Safdi}}, \citenamefont {{Santopinto}}, \citenamefont {{Scarff}},
  \citenamefont {{Schneider}}, \citenamefont {{Schuster}}, \citenamefont
  {{Seidel}}, \citenamefont {{Sekiya}}, \citenamefont {{Seong}}, \citenamefont
  {{Simi}}, \citenamefont {{Sipala}}, \citenamefont {{Slatyer}}, \citenamefont
  {{Slone}}, \citenamefont {{Smith}}, \citenamefont {{Smolinsky}},
  \citenamefont {{Snowden-Ifft}}, \citenamefont {{Solt}}, \citenamefont
  {{Sonnenschein}}, \citenamefont {{Sorensen}}, \citenamefont {{Spooner}},
  \citenamefont {{Srivastava}}, \citenamefont {{Stancu}}, \citenamefont
  {{Strigari}}, \citenamefont {{Strube}}, \citenamefont {{Sushkov}},
  \citenamefont {{Szydagis}}, \citenamefont {{Tanedo}}, \citenamefont
  {{Tanner}}, \citenamefont {{Tayloe}}, \citenamefont {{Terrano}},
  \citenamefont {{Thaler}}, \citenamefont {{Thomas}}, \citenamefont {{Thorpe}},
  \citenamefont {{Thorpe}}, \citenamefont {{Tiffenberg}}, \citenamefont
  {{Tran}}, \citenamefont {{Trovato}}, \citenamefont {{Tully}}, \citenamefont
  {{Tyson}}, \citenamefont {{Vachaspati}}, \citenamefont {{Vahsen}},
  \citenamefont {{van Bibber}}, \citenamefont {{Vandenbroucke}}, \citenamefont
  {{Villano}}, \citenamefont {{Volansky}}, \citenamefont {{Wang}},
  \citenamefont {{Ward}}, \citenamefont {{Wester}}, \citenamefont {{Whitbeck}},
  \citenamefont {{Williams}}, \citenamefont {{Wing}}, \citenamefont
  {{Winslow}}, \citenamefont {{Wojtsekhowski}}, \citenamefont {{Yu}},
  \citenamefont {{Yu}}, \citenamefont {{Yu}}, \citenamefont {{Zhang}},
  \citenamefont {{Zhao}},\ and\ \citenamefont {{Zhong}}}]{2017arXiv170704591B}%
  \BibitemOpen
  \bibfield  {author} {\bibinfo {author} {\bibfnamefont {M.}~\bibnamefont
  {{Battaglieri}}}, \bibinfo {author} {\bibfnamefont {A.}~\bibnamefont
  {{Belloni}}}, \bibinfo {author} {\bibfnamefont {A.}~\bibnamefont {{Chou}}},
  \bibinfo {author} {\bibfnamefont {P.}~\bibnamefont {{Cushman}}}, \bibinfo
  {author} {\bibfnamefont {B.}~\bibnamefont {{Echenard}}}, \bibinfo {author}
  {\bibfnamefont {R.}~\bibnamefont {{Essig}}}, \bibinfo {author} {\bibfnamefont
  {J.}~\bibnamefont {{Estrada}}}, \bibinfo {author} {\bibfnamefont {J.~L.}\
  \bibnamefont {{Feng}}}, \bibinfo {author} {\bibfnamefont {B.}~\bibnamefont
  {{Flaugher}}}, \bibinfo {author} {\bibfnamefont {P.~J.}\ \bibnamefont
  {{Fox}}}, \bibinfo {author} {\bibfnamefont {P.}~\bibnamefont {{Graham}}},
  \bibinfo {author} {\bibfnamefont {C.}~\bibnamefont {{Hall}}}, \bibinfo
  {author} {\bibfnamefont {R.}~\bibnamefont {{Harnik}}}, \bibinfo {author}
  {\bibfnamefont {J.}~\bibnamefont {{Hewett}}}, \bibinfo {author}
  {\bibfnamefont {J.}~\bibnamefont {{Incandela}}}, \bibinfo {author}
  {\bibfnamefont {E.}~\bibnamefont {{Izaguirre}}}, \bibinfo {author}
  {\bibfnamefont {D.}~\bibnamefont {{McKinsey}}}, \bibinfo {author}
  {\bibfnamefont {M.}~\bibnamefont {{Pyle}}}, \bibinfo {author} {\bibfnamefont
  {N.}~\bibnamefont {{Roe}}}, \bibinfo {author} {\bibfnamefont
  {G.}~\bibnamefont {{Rybka}}}, \bibinfo {author} {\bibfnamefont
  {P.}~\bibnamefont {{Sikivie}}}, \bibinfo {author} {\bibfnamefont {T.~M.~P.}\
  \bibnamefont {{Tait}}}, \bibinfo {author} {\bibfnamefont {N.}~\bibnamefont
  {{Toro}}}, \bibinfo {author} {\bibfnamefont {R.}~\bibnamefont {{Van De
  Water}}}, \bibinfo {author} {\bibfnamefont {N.}~\bibnamefont {{Weiner}}},
  \bibinfo {author} {\bibfnamefont {K.}~\bibnamefont {{Zurek}}}, \bibinfo
  {author} {\bibfnamefont {E.}~\bibnamefont {{Adelberger}}}, \bibinfo {author}
  {\bibfnamefont {A.}~\bibnamefont {{Afanasev}}}, \bibinfo {author}
  {\bibfnamefont {D.}~\bibnamefont {{Alexander}}}, \bibinfo {author}
  {\bibfnamefont {J.}~\bibnamefont {{Alexander}}}, \bibinfo {author}
  {\bibfnamefont {V.}~\bibnamefont {{Cristian Antochi}}}, \bibinfo {author}
  {\bibfnamefont {D.~M.}\ \bibnamefont {{Asner}}}, \bibinfo {author}
  {\bibfnamefont {H.}~\bibnamefont {{Baer}}}, \bibinfo {author} {\bibfnamefont
  {D.}~\bibnamefont {{Banerjee}}}, \bibinfo {author} {\bibfnamefont
  {E.}~\bibnamefont {{Baracchini}}}, \bibinfo {author} {\bibfnamefont
  {P.}~\bibnamefont {{Barbeau}}}, \bibinfo {author} {\bibfnamefont
  {J.}~\bibnamefont {{Barrow}}}, \bibinfo {author} {\bibfnamefont
  {N.}~\bibnamefont {{Bastidon}}}, \bibinfo {author} {\bibfnamefont
  {J.}~\bibnamefont {{Battat}}}, \bibinfo {author} {\bibfnamefont
  {S.}~\bibnamefont {{Benson}}}, \bibinfo {author} {\bibfnamefont
  {A.}~\bibnamefont {{Berlin}}}, \bibinfo {author} {\bibfnamefont
  {M.}~\bibnamefont {{Bird}}}, \bibinfo {author} {\bibfnamefont
  {N.}~\bibnamefont {{Blinov}}}, \bibinfo {author} {\bibfnamefont {K.~K.}\
  \bibnamefont {{Boddy}}}, \bibinfo {author} {\bibfnamefont {M.}~\bibnamefont
  {{Bondi}}}, \bibinfo {author} {\bibfnamefont {W.~M.}\ \bibnamefont
  {{Bonivento}}}, \bibinfo {author} {\bibfnamefont {M.}~\bibnamefont
  {{Boulay}}}, \bibinfo {author} {\bibfnamefont {J.}~\bibnamefont {{Boyce}}},
  \bibinfo {author} {\bibfnamefont {M.}~\bibnamefont {{Brodeur}}}, \bibinfo
  {author} {\bibfnamefont {L.}~\bibnamefont {{Broussard}}}, \bibinfo {author}
  {\bibfnamefont {R.}~\bibnamefont {{Budnik}}}, \bibinfo {author}
  {\bibfnamefont {P.}~\bibnamefont {{Bunting}}}, \bibinfo {author}
  {\bibfnamefont {M.}~\bibnamefont {{Caffee}}}, \bibinfo {author}
  {\bibfnamefont {S.~S.}\ \bibnamefont {{Caiazza}}}, \bibinfo {author}
  {\bibfnamefont {S.}~\bibnamefont {{Campbell}}}, \bibinfo {author}
  {\bibfnamefont {T.}~\bibnamefont {{Cao}}}, \bibinfo {author} {\bibfnamefont
  {G.}~\bibnamefont {{Carosi}}}, \bibinfo {author} {\bibfnamefont
  {M.}~\bibnamefont {{Carpinelli}}}, \bibinfo {author} {\bibfnamefont
  {G.}~\bibnamefont {{Cavoto}}}, \bibinfo {author} {\bibfnamefont
  {A.}~\bibnamefont {{Celentano}}}, \bibinfo {author} {\bibfnamefont
  {J.}~\bibnamefont {{Hyeok Chang}}}, \bibinfo {author} {\bibfnamefont
  {S.}~\bibnamefont {{Chattopadhyay}}}, \bibinfo {author} {\bibfnamefont
  {A.}~\bibnamefont {{Chavarria}}}, \bibinfo {author} {\bibfnamefont {C.-Y.}\
  \bibnamefont {{Chen}}}, \bibinfo {author} {\bibfnamefont {K.}~\bibnamefont
  {{Clark}}}, \bibinfo {author} {\bibfnamefont {J.}~\bibnamefont {{Clarke}}},
  \bibinfo {author} {\bibfnamefont {O.}~\bibnamefont {{Colegrove}}}, \bibinfo
  {author} {\bibfnamefont {J.}~\bibnamefont {{Coleman}}}, \bibinfo {author}
  {\bibfnamefont {D.}~\bibnamefont {{Cooke}}}, \bibinfo {author} {\bibfnamefont
  {R.}~\bibnamefont {{Cooper}}}, \bibinfo {author} {\bibfnamefont
  {M.}~\bibnamefont {{Crisler}}}, \bibinfo {author} {\bibfnamefont
  {P.}~\bibnamefont {{Crivelli}}}, \bibinfo {author} {\bibfnamefont
  {F.}~\bibnamefont {{D'Eramo}}}, \bibinfo {author} {\bibfnamefont
  {D.}~\bibnamefont {{D'Urso}}}, \bibinfo {author} {\bibfnamefont
  {E.}~\bibnamefont {{Dahl}}}, \bibinfo {author} {\bibfnamefont
  {W.}~\bibnamefont {{Dawson}}}, \bibinfo {author} {\bibfnamefont
  {M.}~\bibnamefont {{De Napoli}}}, \bibinfo {author} {\bibfnamefont
  {R.}~\bibnamefont {{De Vita}}}, \bibinfo {author} {\bibfnamefont
  {P.}~\bibnamefont {{DeNiverville}}}, \bibinfo {author} {\bibfnamefont
  {S.}~\bibnamefont {{Derenzo}}}, \bibinfo {author} {\bibfnamefont
  {A.}~\bibnamefont {{Di Crescenzo}}}, \bibinfo {author} {\bibfnamefont
  {E.}~\bibnamefont {{Di Marco}}}, \bibinfo {author} {\bibfnamefont {K.~R.}\
  \bibnamefont {{Dienes}}}, \bibinfo {author} {\bibfnamefont {M.}~\bibnamefont
  {{Diwan}}}, \bibinfo {author} {\bibfnamefont {D.}~\bibnamefont
  {{Handiipondola Dongwi}}}, \bibinfo {author} {\bibfnamefont {A.}~\bibnamefont
  {{Drlica-Wagner}}}, \bibinfo {author} {\bibfnamefont {S.}~\bibnamefont
  {{Ellis}}}, \bibinfo {author} {\bibfnamefont {A.}~\bibnamefont {{Chigbo
  Ezeribe}}}, \bibinfo {author} {\bibfnamefont {G.}~\bibnamefont {{Farrar}}},
  \bibinfo {author} {\bibfnamefont {F.}~\bibnamefont {{Ferrer}}}, \bibinfo
  {author} {\bibfnamefont {E.}~\bibnamefont {{Figueroa-Feliciano}}}, \bibinfo
  {author} {\bibfnamefont {A.}~\bibnamefont {{Filippi}}}, \bibinfo {author}
  {\bibfnamefont {G.}~\bibnamefont {{Fiorillo}}}, \bibinfo {author}
  {\bibfnamefont {B.}~\bibnamefont {{Fornal}}}, \bibinfo {author}
  {\bibfnamefont {A.}~\bibnamefont {{Freyberger}}}, \bibinfo {author}
  {\bibfnamefont {C.}~\bibnamefont {{Frugiuele}}}, \bibinfo {author}
  {\bibfnamefont {C.}~\bibnamefont {{Galbiati}}}, \bibinfo {author}
  {\bibfnamefont {I.}~\bibnamefont {{Galon}}}, \bibinfo {author} {\bibfnamefont
  {S.}~\bibnamefont {{Gardner}}}, \bibinfo {author} {\bibfnamefont
  {A.}~\bibnamefont {{Geraci}}}, \bibinfo {author} {\bibfnamefont
  {G.}~\bibnamefont {{Gerbier}}}, \bibinfo {author} {\bibfnamefont
  {M.}~\bibnamefont {{Graham}}}, \bibinfo {author} {\bibfnamefont
  {E.}~\bibnamefont {{Gschwendtner}}}, \bibinfo {author} {\bibfnamefont
  {C.}~\bibnamefont {{Hearty}}}, \bibinfo {author} {\bibfnamefont
  {J.}~\bibnamefont {{Heise}}}, \bibinfo {author} {\bibfnamefont
  {R.}~\bibnamefont {{Henning}}}, \bibinfo {author} {\bibfnamefont {R.~J.}\
  \bibnamefont {{Hill}}}, \bibinfo {author} {\bibfnamefont {D.}~\bibnamefont
  {{Hitlin}}}, \bibinfo {author} {\bibfnamefont {Y.}~\bibnamefont
  {{Hochberg}}}, \bibinfo {author} {\bibfnamefont {J.}~\bibnamefont {{Hogan}}},
  \bibinfo {author} {\bibfnamefont {M.}~\bibnamefont {{Holtrop}}}, \bibinfo
  {author} {\bibfnamefont {Z.}~\bibnamefont {{Hong}}}, \bibinfo {author}
  {\bibfnamefont {T.}~\bibnamefont {{Hossbach}}}, \bibinfo {author}
  {\bibfnamefont {T.~B.}\ \bibnamefont {{Humensky}}}, \bibinfo {author}
  {\bibfnamefont {P.}~\bibnamefont {{Ilten}}}, \bibinfo {author} {\bibfnamefont
  {K.}~\bibnamefont {{Irwin}}}, \bibinfo {author} {\bibfnamefont
  {J.}~\bibnamefont {{Jaros}}}, \bibinfo {author} {\bibfnamefont
  {R.}~\bibnamefont {{Johnson}}}, \bibinfo {author} {\bibfnamefont
  {M.}~\bibnamefont {{Jones}}}, \bibinfo {author} {\bibfnamefont
  {Y.}~\bibnamefont {{Kahn}}}, \bibinfo {author} {\bibfnamefont
  {N.}~\bibnamefont {{Kalantarians}}}, \bibinfo {author} {\bibfnamefont
  {M.}~\bibnamefont {{Kaplinghat}}}, \bibinfo {author} {\bibfnamefont
  {R.}~\bibnamefont {{Khatiwada}}}, \bibinfo {author} {\bibfnamefont
  {S.}~\bibnamefont {{Knapen}}}, \bibinfo {author} {\bibfnamefont
  {M.}~\bibnamefont {{Kohl}}}, \bibinfo {author} {\bibfnamefont
  {C.}~\bibnamefont {{Kouvaris}}}, \bibinfo {author} {\bibfnamefont
  {J.}~\bibnamefont {{Kozaczuk}}}, \bibinfo {author} {\bibfnamefont
  {G.}~\bibnamefont {{Krnjaic}}}, \bibinfo {author} {\bibfnamefont
  {V.}~\bibnamefont {{Kubarovsky}}}, \bibinfo {author} {\bibfnamefont
  {E.}~\bibnamefont {{Kuflik}}}, \bibinfo {author} {\bibfnamefont
  {A.}~\bibnamefont {{Kusenko}}}, \bibinfo {author} {\bibfnamefont
  {R.}~\bibnamefont {{Lang}}}, \bibinfo {author} {\bibfnamefont
  {K.}~\bibnamefont {{Leach}}}, \bibinfo {author} {\bibfnamefont
  {T.}~\bibnamefont {{Lin}}}, \bibinfo {author} {\bibfnamefont
  {M.}~\bibnamefont {{Lisanti}}}, \bibinfo {author} {\bibfnamefont
  {J.}~\bibnamefont {{Liu}}}, \bibinfo {author} {\bibfnamefont
  {K.}~\bibnamefont {{Liu}}}, \bibinfo {author} {\bibfnamefont
  {M.}~\bibnamefont {{Liu}}}, \bibinfo {author} {\bibfnamefont
  {D.}~\bibnamefont {{Loomba}}}, \bibinfo {author} {\bibfnamefont
  {J.}~\bibnamefont {{Lykken}}}, \bibinfo {author} {\bibfnamefont
  {K.}~\bibnamefont {{Mack}}}, \bibinfo {author} {\bibfnamefont
  {J.}~\bibnamefont {{Mans}}}, \bibinfo {author} {\bibfnamefont
  {H.}~\bibnamefont {{Maris}}}, \bibinfo {author} {\bibfnamefont
  {T.}~\bibnamefont {{Markiewicz}}}, \bibinfo {author} {\bibfnamefont
  {L.}~\bibnamefont {{Marsicano}}}, \bibinfo {author} {\bibfnamefont {C.~J.}\
  \bibnamefont {{Martoff}}}, \bibinfo {author} {\bibfnamefont {G.}~\bibnamefont
  {{Mazzitelli}}}, \bibinfo {author} {\bibfnamefont {C.}~\bibnamefont
  {{McCabe}}}, \bibinfo {author} {\bibfnamefont {S.~D.}\ \bibnamefont
  {{McDermott}}}, \bibinfo {author} {\bibfnamefont {A.}~\bibnamefont
  {{McDonald}}}, \bibinfo {author} {\bibfnamefont {B.}~\bibnamefont
  {{McKinnon}}}, \bibinfo {author} {\bibfnamefont {D.}~\bibnamefont {{Mei}}},
  \bibinfo {author} {\bibfnamefont {T.}~\bibnamefont {{Melia}}}, \bibinfo
  {author} {\bibfnamefont {G.~A.}\ \bibnamefont {{Miller}}}, \bibinfo {author}
  {\bibfnamefont {K.}~\bibnamefont {{Miuchi}}}, \bibinfo {author}
  {\bibfnamefont {S.~M.~P.}\ \bibnamefont {{Nazeer}}}, \bibinfo {author}
  {\bibfnamefont {O.}~\bibnamefont {{Moreno}}}, \bibinfo {author}
  {\bibfnamefont {V.}~\bibnamefont {{Morozov}}}, \bibinfo {author}
  {\bibfnamefont {F.}~\bibnamefont {{Mouton}}}, \bibinfo {author}
  {\bibfnamefont {H.}~\bibnamefont {{Mueller}}}, \bibinfo {author}
  {\bibfnamefont {A.}~\bibnamefont {{Murphy}}}, \bibinfo {author}
  {\bibfnamefont {R.}~\bibnamefont {{Neilson}}}, \bibinfo {author}
  {\bibfnamefont {T.}~\bibnamefont {{Nelson}}}, \bibinfo {author}
  {\bibfnamefont {C.}~\bibnamefont {{Neu}}}, \bibinfo {author} {\bibfnamefont
  {Y.}~\bibnamefont {{Nosochkov}}}, \bibinfo {author} {\bibfnamefont
  {C.}~\bibnamefont {{O'Hare}}}, \bibinfo {author} {\bibfnamefont
  {N.}~\bibnamefont {{Oblath}}}, \bibinfo {author} {\bibfnamefont
  {J.}~\bibnamefont {{Orrell}}}, \bibinfo {author} {\bibfnamefont
  {J.}~\bibnamefont {{Ouellet}}}, \bibinfo {author} {\bibfnamefont
  {S.}~\bibnamefont {{Pastore}}}, \bibinfo {author} {\bibfnamefont
  {S.}~\bibnamefont {{Paul}}}, \bibinfo {author} {\bibfnamefont
  {M.}~\bibnamefont {{Perelstein}}}, \bibinfo {author} {\bibfnamefont
  {A.}~\bibnamefont {{Peter}}}, \bibinfo {author} {\bibfnamefont
  {N.}~\bibnamefont {{Phan}}}, \bibinfo {author} {\bibfnamefont
  {N.}~\bibnamefont {{Phinney}}}, \bibinfo {author} {\bibfnamefont
  {M.}~\bibnamefont {{Pivovaroff}}}, \bibinfo {author} {\bibfnamefont
  {A.}~\bibnamefont {{Pocar}}}, \bibinfo {author} {\bibfnamefont
  {M.}~\bibnamefont {{Pospelov}}}, \bibinfo {author} {\bibfnamefont
  {J.}~\bibnamefont {{Pradler}}}, \bibinfo {author} {\bibfnamefont
  {P.}~\bibnamefont {{Privitera}}}, \bibinfo {author} {\bibfnamefont
  {S.}~\bibnamefont {{Profumo}}}, \bibinfo {author} {\bibfnamefont
  {M.}~\bibnamefont {{Raggi}}}, \bibinfo {author} {\bibfnamefont
  {S.}~\bibnamefont {{Rajendran}}}, \bibinfo {author} {\bibfnamefont
  {N.}~\bibnamefont {{Randazzo}}}, \bibinfo {author} {\bibfnamefont
  {T.}~\bibnamefont {{Raubenheimer}}}, \bibinfo {author} {\bibfnamefont
  {C.}~\bibnamefont {{Regenfus}}}, \bibinfo {author} {\bibfnamefont
  {A.}~\bibnamefont {{Renshaw}}}, \bibinfo {author} {\bibfnamefont
  {A.}~\bibnamefont {{Ritz}}}, \bibinfo {author} {\bibfnamefont
  {T.}~\bibnamefont {{Rizzo}}}, \bibinfo {author} {\bibfnamefont
  {L.}~\bibnamefont {{Rosenberg}}}, \bibinfo {author} {\bibfnamefont
  {A.}~\bibnamefont {{Rubbia}}}, \bibinfo {author} {\bibfnamefont
  {B.}~\bibnamefont {{Rybolt}}}, \bibinfo {author} {\bibfnamefont
  {T.}~\bibnamefont {{Saab}}}, \bibinfo {author} {\bibfnamefont {B.~R.}\
  \bibnamefont {{Safdi}}}, \bibinfo {author} {\bibfnamefont {E.}~\bibnamefont
  {{Santopinto}}}, \bibinfo {author} {\bibfnamefont {A.}~\bibnamefont
  {{Scarff}}}, \bibinfo {author} {\bibfnamefont {M.}~\bibnamefont
  {{Schneider}}}, \bibinfo {author} {\bibfnamefont {P.}~\bibnamefont
  {{Schuster}}}, \bibinfo {author} {\bibfnamefont {G.}~\bibnamefont
  {{Seidel}}}, \bibinfo {author} {\bibfnamefont {H.}~\bibnamefont {{Sekiya}}},
  \bibinfo {author} {\bibfnamefont {I.}~\bibnamefont {{Seong}}}, \bibinfo
  {author} {\bibfnamefont {G.}~\bibnamefont {{Simi}}}, \bibinfo {author}
  {\bibfnamefont {V.}~\bibnamefont {{Sipala}}}, \bibinfo {author}
  {\bibfnamefont {T.}~\bibnamefont {{Slatyer}}}, \bibinfo {author}
  {\bibfnamefont {O.}~\bibnamefont {{Slone}}}, \bibinfo {author} {\bibfnamefont
  {P.~F.}\ \bibnamefont {{Smith}}}, \bibinfo {author} {\bibfnamefont
  {J.}~\bibnamefont {{Smolinsky}}}, \bibinfo {author} {\bibfnamefont
  {D.}~\bibnamefont {{Snowden-Ifft}}}, \bibinfo {author} {\bibfnamefont
  {M.}~\bibnamefont {{Solt}}}, \bibinfo {author} {\bibfnamefont
  {A.}~\bibnamefont {{Sonnenschein}}}, \bibinfo {author} {\bibfnamefont
  {P.}~\bibnamefont {{Sorensen}}}, \bibinfo {author} {\bibfnamefont
  {N.}~\bibnamefont {{Spooner}}}, \bibinfo {author} {\bibfnamefont
  {B.}~\bibnamefont {{Srivastava}}}, \bibinfo {author} {\bibfnamefont
  {I.}~\bibnamefont {{Stancu}}}, \bibinfo {author} {\bibfnamefont
  {L.}~\bibnamefont {{Strigari}}}, \bibinfo {author} {\bibfnamefont
  {J.}~\bibnamefont {{Strube}}}, \bibinfo {author} {\bibfnamefont {A.~O.}\
  \bibnamefont {{Sushkov}}}, \bibinfo {author} {\bibfnamefont {M.}~\bibnamefont
  {{Szydagis}}}, \bibinfo {author} {\bibfnamefont {P.}~\bibnamefont
  {{Tanedo}}}, \bibinfo {author} {\bibfnamefont {D.}~\bibnamefont {{Tanner}}},
  \bibinfo {author} {\bibfnamefont {R.}~\bibnamefont {{Tayloe}}}, \bibinfo
  {author} {\bibfnamefont {W.}~\bibnamefont {{Terrano}}}, \bibinfo {author}
  {\bibfnamefont {J.}~\bibnamefont {{Thaler}}}, \bibinfo {author}
  {\bibfnamefont {B.}~\bibnamefont {{Thomas}}}, \bibinfo {author}
  {\bibfnamefont {B.}~\bibnamefont {{Thorpe}}}, \bibinfo {author}
  {\bibfnamefont {T.}~\bibnamefont {{Thorpe}}}, \bibinfo {author}
  {\bibfnamefont {J.}~\bibnamefont {{Tiffenberg}}}, \bibinfo {author}
  {\bibfnamefont {N.}~\bibnamefont {{Tran}}}, \bibinfo {author} {\bibfnamefont
  {M.}~\bibnamefont {{Trovato}}}, \bibinfo {author} {\bibfnamefont
  {C.}~\bibnamefont {{Tully}}}, \bibinfo {author} {\bibfnamefont
  {T.}~\bibnamefont {{Tyson}}}, \bibinfo {author} {\bibfnamefont
  {T.}~\bibnamefont {{Vachaspati}}}, \bibinfo {author} {\bibfnamefont
  {S.}~\bibnamefont {{Vahsen}}}, \bibinfo {author} {\bibfnamefont
  {K.}~\bibnamefont {{van Bibber}}}, \bibinfo {author} {\bibfnamefont
  {J.}~\bibnamefont {{Vandenbroucke}}}, \bibinfo {author} {\bibfnamefont
  {A.}~\bibnamefont {{Villano}}}, \bibinfo {author} {\bibfnamefont
  {T.}~\bibnamefont {{Volansky}}}, \bibinfo {author} {\bibfnamefont
  {G.}~\bibnamefont {{Wang}}}, \bibinfo {author} {\bibfnamefont
  {T.}~\bibnamefont {{Ward}}}, \bibinfo {author} {\bibfnamefont
  {W.}~\bibnamefont {{Wester}}}, \bibinfo {author} {\bibfnamefont
  {A.}~\bibnamefont {{Whitbeck}}}, \bibinfo {author} {\bibfnamefont {D.~A.}\
  \bibnamefont {{Williams}}}, \bibinfo {author} {\bibfnamefont
  {M.}~\bibnamefont {{Wing}}}, \bibinfo {author} {\bibfnamefont
  {L.}~\bibnamefont {{Winslow}}}, \bibinfo {author} {\bibfnamefont
  {B.}~\bibnamefont {{Wojtsekhowski}}}, \bibinfo {author} {\bibfnamefont
  {H.-B.}\ \bibnamefont {{Yu}}}, \bibinfo {author} {\bibfnamefont {S.-S.}\
  \bibnamefont {{Yu}}}, \bibinfo {author} {\bibfnamefont {T.-T.}\ \bibnamefont
  {{Yu}}}, \bibinfo {author} {\bibfnamefont {X.}~\bibnamefont {{Zhang}}},
  \bibinfo {author} {\bibfnamefont {Y.}~\bibnamefont {{Zhao}}},\ and\ \bibinfo
  {author} {\bibfnamefont {Y.-M.}\ \bibnamefont {{Zhong}}},\ }\bibfield
  {title} {\bibinfo {title} {{US Cosmic Visions: New Ideas in Dark Matter 2017:
  Community Report}},\ }\href@noop {} {\bibfield  {journal} {\bibinfo
  {journal} {arXiv e-prints}\ ,\ \bibinfo {eid} {arXiv:1707.04591}} (\bibinfo
  {year} {2017})},\ \Eprint {https://arxiv.org/abs/1707.04591}
  {arXiv:1707.04591 [hep-ph]} \BibitemShut {NoStop}%
\bibitem [{\citenamefont {Li}\ \emph {et~al.}(2021)\citenamefont {Li},
  \citenamefont {Shao}, \citenamefont {Wu},\ and\ \citenamefont
  {Yu}}]{Li:2021qer}%
  \BibitemOpen
  \bibfield  {author} {\bibinfo {author} {\bibfnamefont {S.-L.}\ \bibnamefont
  {Li}}, \bibinfo {author} {\bibfnamefont {L.}~\bibnamefont {Shao}}, \bibinfo
  {author} {\bibfnamefont {P.}~\bibnamefont {Wu}},\ and\ \bibinfo {author}
  {\bibfnamefont {H.}~\bibnamefont {Yu}},\ }\bibfield  {title} {\bibinfo
  {title} {{NANOGrav Signal from First-Order Confinement/Deconfinement Phase
  Transition in Different QCD Matters}},\ }\href@noop {} {\  (\bibinfo {year}
  {2021})},\ \Eprint {https://arxiv.org/abs/2101.08012} {arXiv:2101.08012
  [astro-ph.CO]} \BibitemShut {NoStop}%
\bibitem [{\citenamefont {Bian}\ \emph {et~al.}(2020)\citenamefont {Bian},
  \citenamefont {Cai}, \citenamefont {Liu}, \citenamefont {Yang},\ and\
  \citenamefont {Zhou}}]{Bian:2020bps}%
  \BibitemOpen
  \bibfield  {author} {\bibinfo {author} {\bibfnamefont {L.}~\bibnamefont
  {Bian}}, \bibinfo {author} {\bibfnamefont {R.-G.}\ \bibnamefont {Cai}},
  \bibinfo {author} {\bibfnamefont {J.}~\bibnamefont {Liu}}, \bibinfo {author}
  {\bibfnamefont {X.-Y.}\ \bibnamefont {Yang}},\ and\ \bibinfo {author}
  {\bibfnamefont {R.}~\bibnamefont {Zhou}},\ }\bibfield  {title} {\bibinfo
  {title} {{On the gravitational wave sources from the NANOGrav 12.5-yr
  data}},\ }\href@noop {} {\  (\bibinfo {year} {2020})},\ \Eprint
  {https://arxiv.org/abs/2009.13893} {arXiv:2009.13893 [astro-ph.CO]}
  \BibitemShut {NoStop}%
\bibitem [{\citenamefont {{Moore}}\ \emph {et~al.}(2015)\citenamefont
  {{Moore}}, \citenamefont {{Cole}},\ and\ \citenamefont
  {{Berry}}}]{2015CQGra..32a5014M}%
  \BibitemOpen
  \bibfield  {author} {\bibinfo {author} {\bibfnamefont {C.~J.}\ \bibnamefont
  {{Moore}}}, \bibinfo {author} {\bibfnamefont {R.~H.}\ \bibnamefont
  {{Cole}}},\ and\ \bibinfo {author} {\bibfnamefont {C.~P.~L.}\ \bibnamefont
  {{Berry}}},\ }\bibfield  {title} {\bibinfo {title} {{Gravitational-wave
  sensitivity curves}},\ }\href {https://doi.org/10.1088/0264-9381/32/1/015014}
  {\bibfield  {journal} {\bibinfo  {journal} {Classical and Quantum Gravity}\
  }\textbf {\bibinfo {volume} {32}},\ \bibinfo {eid} {015014} (\bibinfo {year}
  {2015})},\ \Eprint {https://arxiv.org/abs/1408.0740} {arXiv:1408.0740
  [gr-qc]} \BibitemShut {NoStop}%
\bibitem [{\citenamefont {{Planck Collaboration}}\ \emph
  {et~al.}(2020)\citenamefont {{Planck Collaboration}}, \citenamefont
  {{Aghanim}}, \citenamefont {{Akrami}}, \citenamefont {{Ashdown}},
  \citenamefont {{Aumont}}, \citenamefont {{Baccigalupi}}, \citenamefont
  {{Ballardini}}, \citenamefont {{Banday}}, \citenamefont {{Barreiro}},
  \citenamefont {{Bartolo}}, \citenamefont {{Basak}}, \citenamefont {{Battye}},
  \citenamefont {{Benabed}}, \citenamefont {{Bernard}}, \citenamefont
  {{Bersanelli}}, \citenamefont {{Bielewicz}}, \citenamefont {{Bock}},
  \citenamefont {{Bond}}, \citenamefont {{Borrill}}, \citenamefont {{Bouchet}},
  \citenamefont {{Boulanger}}, \citenamefont {{Bucher}}, \citenamefont
  {{Burigana}}, \citenamefont {{Butler}}, \citenamefont {{Calabrese}},
  \citenamefont {{Cardoso}}, \citenamefont {{Carron}}, \citenamefont
  {{Challinor}}, \citenamefont {{Chiang}}, \citenamefont {{Chluba}},
  \citenamefont {{Colombo}}, \citenamefont {{Combet}}, \citenamefont
  {{Contreras}}, \citenamefont {{Crill}}, \citenamefont {{Cuttaia}},
  \citenamefont {{de Bernardis}}, \citenamefont {{de Zotti}}, \citenamefont
  {{Delabrouille}}, \citenamefont {{Delouis}}, \citenamefont {{Di Valentino}},
  \citenamefont {{Diego}}, \citenamefont {{Dor{\'e}}}, \citenamefont
  {{Douspis}}, \citenamefont {{Ducout}}, \citenamefont {{Dupac}}, \citenamefont
  {{Dusini}}, \citenamefont {{Efstathiou}}, \citenamefont {{Elsner}},
  \citenamefont {{En{\ss}lin}}, \citenamefont {{Eriksen}}, \citenamefont
  {{Fantaye}}, \citenamefont {{Farhang}}, \citenamefont {{Fergusson}},
  \citenamefont {{Fernandez-Cobos}}, \citenamefont {{Finelli}}, \citenamefont
  {{Forastieri}}, \citenamefont {{Frailis}}, \citenamefont {{Fraisse}},
  \citenamefont {{Franceschi}}, \citenamefont {{Frolov}}, \citenamefont
  {{Galeotta}}, \citenamefont {{Galli}}, \citenamefont {{Ganga}}, \citenamefont
  {{G{\'e}nova-Santos}}, \citenamefont {{Gerbino}}, \citenamefont {{Ghosh}},
  \citenamefont {{Gonz{\'a}lez-Nuevo}}, \citenamefont {{G{\'o}rski}},
  \citenamefont {{Gratton}}, \citenamefont {{Gruppuso}}, \citenamefont
  {{Gudmundsson}}, \citenamefont {{Hamann}}, \citenamefont {{Handley}},
  \citenamefont {{Hansen}}, \citenamefont {{Herranz}}, \citenamefont
  {{Hildebrandt}}, \citenamefont {{Hivon}}, \citenamefont {{Huang}},
  \citenamefont {{Jaffe}}, \citenamefont {{Jones}}, \citenamefont {{Karakci}},
  \citenamefont {{Keih{\"a}nen}}, \citenamefont {{Keskitalo}}, \citenamefont
  {{Kiiveri}}, \citenamefont {{Kim}}, \citenamefont {{Kisner}}, \citenamefont
  {{Knox}}, \citenamefont {{Krachmalnicoff}}, \citenamefont {{Kunz}},
  \citenamefont {{Kurki-Suonio}}, \citenamefont {{Lagache}}, \citenamefont
  {{Lamarre}}, \citenamefont {{Lasenby}}, \citenamefont {{Lattanzi}},
  \citenamefont {{Lawrence}}, \citenamefont {{Le Jeune}}, \citenamefont
  {{Lemos}}, \citenamefont {{Lesgourgues}}, \citenamefont {{Levrier}},
  \citenamefont {{Lewis}}, \citenamefont {{Liguori}}, \citenamefont {{Lilje}},
  \citenamefont {{Lilley}}, \citenamefont {{Lindholm}}, \citenamefont
  {{L{\'o}pez-Caniego}}, \citenamefont {{Lubin}}, \citenamefont {{Ma}},
  \citenamefont {{Mac{\'\i}as-P{\'e}rez}}, \citenamefont {{Maggio}},
  \citenamefont {{Maino}}, \citenamefont {{Mandolesi}}, \citenamefont
  {{Mangilli}}, \citenamefont {{Marcos-Caballero}}, \citenamefont {{Maris}},
  \citenamefont {{Martin}}, \citenamefont {{Martinelli}}, \citenamefont
  {{Mart{\'\i}nez-Gonz{\'a}lez}}, \citenamefont {{Matarrese}}, \citenamefont
  {{Mauri}}, \citenamefont {{McEwen}}, \citenamefont {{Meinhold}},
  \citenamefont {{Melchiorri}}, \citenamefont {{Mennella}}, \citenamefont
  {{Migliaccio}}, \citenamefont {{Millea}}, \citenamefont {{Mitra}},
  \citenamefont {{Miville-Desch{\^e}nes}}, \citenamefont {{Molinari}},
  \citenamefont {{Montier}}, \citenamefont {{Morgante}}, \citenamefont
  {{Moss}}, \citenamefont {{Natoli}}, \citenamefont {{N{\o}rgaard-Nielsen}},
  \citenamefont {{Pagano}}, \citenamefont {{Paoletti}}, \citenamefont
  {{Partridge}}, \citenamefont {{Patanchon}}, \citenamefont {{Peiris}},
  \citenamefont {{Perrotta}}, \citenamefont {{Pettorino}}, \citenamefont
  {{Piacentini}}, \citenamefont {{Polastri}}, \citenamefont {{Polenta}},
  \citenamefont {{Puget}}, \citenamefont {{Rachen}}, \citenamefont
  {{Reinecke}}, \citenamefont {{Remazeilles}}, \citenamefont {{Renzi}},
  \citenamefont {{Rocha}}, \citenamefont {{Rosset}}, \citenamefont {{Roudier}},
  \citenamefont {{Rubi{\~n}o-Mart{\'\i}n}}, \citenamefont {{Ruiz-Granados}},
  \citenamefont {{Salvati}}, \citenamefont {{Sandri}}, \citenamefont
  {{Savelainen}}, \citenamefont {{Scott}}, \citenamefont {{Shellard}},
  \citenamefont {{Sirignano}}, \citenamefont {{Sirri}}, \citenamefont
  {{Spencer}}, \citenamefont {{Sunyaev}}, \citenamefont {{Suur-Uski}},
  \citenamefont {{Tauber}}, \citenamefont {{Tavagnacco}}, \citenamefont
  {{Tenti}}, \citenamefont {{Toffolatti}}, \citenamefont {{Tomasi}},
  \citenamefont {{Trombetti}}, \citenamefont {{Valenziano}}, \citenamefont
  {{Valiviita}}, \citenamefont {{Van Tent}}, \citenamefont {{Vibert}},
  \citenamefont {{Vielva}}, \citenamefont {{Villa}}, \citenamefont
  {{Vittorio}}, \citenamefont {{Wandelt}}, \citenamefont {{Wehus}},
  \citenamefont {{White}}, \citenamefont {{White}}, \citenamefont {{Zacchei}},\
  and\ \citenamefont {{Zonca}}}]{2020A&A...641A...6P}%
  \BibitemOpen
  \bibfield  {author} {\bibinfo {author} {\bibnamefont {{Planck
  Collaboration}}}, \bibinfo {author} {\bibfnamefont {N.}~\bibnamefont
  {{Aghanim}}}, \bibinfo {author} {\bibfnamefont {Y.}~\bibnamefont {{Akrami}}},
  \bibinfo {author} {\bibfnamefont {M.}~\bibnamefont {{Ashdown}}}, \bibinfo
  {author} {\bibfnamefont {J.}~\bibnamefont {{Aumont}}}, \bibinfo {author}
  {\bibfnamefont {C.}~\bibnamefont {{Baccigalupi}}}, \bibinfo {author}
  {\bibfnamefont {M.}~\bibnamefont {{Ballardini}}}, \bibinfo {author}
  {\bibfnamefont {A.~J.}\ \bibnamefont {{Banday}}}, \bibinfo {author}
  {\bibfnamefont {R.~B.}\ \bibnamefont {{Barreiro}}}, \bibinfo {author}
  {\bibfnamefont {N.}~\bibnamefont {{Bartolo}}}, \bibinfo {author}
  {\bibfnamefont {S.}~\bibnamefont {{Basak}}}, \bibinfo {author} {\bibfnamefont
  {R.}~\bibnamefont {{Battye}}}, \bibinfo {author} {\bibfnamefont
  {K.}~\bibnamefont {{Benabed}}}, \bibinfo {author} {\bibfnamefont {J.~P.}\
  \bibnamefont {{Bernard}}}, \bibinfo {author} {\bibfnamefont {M.}~\bibnamefont
  {{Bersanelli}}}, \bibinfo {author} {\bibfnamefont {P.}~\bibnamefont
  {{Bielewicz}}}, \bibinfo {author} {\bibfnamefont {J.~J.}\ \bibnamefont
  {{Bock}}}, \bibinfo {author} {\bibfnamefont {J.~R.}\ \bibnamefont {{Bond}}},
  \bibinfo {author} {\bibfnamefont {J.}~\bibnamefont {{Borrill}}}, \bibinfo
  {author} {\bibfnamefont {F.~R.}\ \bibnamefont {{Bouchet}}}, \bibinfo {author}
  {\bibfnamefont {F.}~\bibnamefont {{Boulanger}}}, \bibinfo {author}
  {\bibfnamefont {M.}~\bibnamefont {{Bucher}}}, \bibinfo {author}
  {\bibfnamefont {C.}~\bibnamefont {{Burigana}}}, \bibinfo {author}
  {\bibfnamefont {R.~C.}\ \bibnamefont {{Butler}}}, \bibinfo {author}
  {\bibfnamefont {E.}~\bibnamefont {{Calabrese}}}, \bibinfo {author}
  {\bibfnamefont {J.~F.}\ \bibnamefont {{Cardoso}}}, \bibinfo {author}
  {\bibfnamefont {J.}~\bibnamefont {{Carron}}}, \bibinfo {author}
  {\bibfnamefont {A.}~\bibnamefont {{Challinor}}}, \bibinfo {author}
  {\bibfnamefont {H.~C.}\ \bibnamefont {{Chiang}}}, \bibinfo {author}
  {\bibfnamefont {J.}~\bibnamefont {{Chluba}}}, \bibinfo {author}
  {\bibfnamefont {L.~P.~L.}\ \bibnamefont {{Colombo}}}, \bibinfo {author}
  {\bibfnamefont {C.}~\bibnamefont {{Combet}}}, \bibinfo {author}
  {\bibfnamefont {D.}~\bibnamefont {{Contreras}}}, \bibinfo {author}
  {\bibfnamefont {B.~P.}\ \bibnamefont {{Crill}}}, \bibinfo {author}
  {\bibfnamefont {F.}~\bibnamefont {{Cuttaia}}}, \bibinfo {author}
  {\bibfnamefont {P.}~\bibnamefont {{de Bernardis}}}, \bibinfo {author}
  {\bibfnamefont {G.}~\bibnamefont {{de Zotti}}}, \bibinfo {author}
  {\bibfnamefont {J.}~\bibnamefont {{Delabrouille}}}, \bibinfo {author}
  {\bibfnamefont {J.~M.}\ \bibnamefont {{Delouis}}}, \bibinfo {author}
  {\bibfnamefont {E.}~\bibnamefont {{Di Valentino}}}, \bibinfo {author}
  {\bibfnamefont {J.~M.}\ \bibnamefont {{Diego}}}, \bibinfo {author}
  {\bibfnamefont {O.}~\bibnamefont {{Dor{\'e}}}}, \bibinfo {author}
  {\bibfnamefont {M.}~\bibnamefont {{Douspis}}}, \bibinfo {author}
  {\bibfnamefont {A.}~\bibnamefont {{Ducout}}}, \bibinfo {author}
  {\bibfnamefont {X.}~\bibnamefont {{Dupac}}}, \bibinfo {author} {\bibfnamefont
  {S.}~\bibnamefont {{Dusini}}}, \bibinfo {author} {\bibfnamefont
  {G.}~\bibnamefont {{Efstathiou}}}, \bibinfo {author} {\bibfnamefont
  {F.}~\bibnamefont {{Elsner}}}, \bibinfo {author} {\bibfnamefont {T.~A.}\
  \bibnamefont {{En{\ss}lin}}}, \bibinfo {author} {\bibfnamefont {H.~K.}\
  \bibnamefont {{Eriksen}}}, \bibinfo {author} {\bibfnamefont {Y.}~\bibnamefont
  {{Fantaye}}}, \bibinfo {author} {\bibfnamefont {M.}~\bibnamefont
  {{Farhang}}}, \bibinfo {author} {\bibfnamefont {J.}~\bibnamefont
  {{Fergusson}}}, \bibinfo {author} {\bibfnamefont {R.}~\bibnamefont
  {{Fernandez-Cobos}}}, \bibinfo {author} {\bibfnamefont {F.}~\bibnamefont
  {{Finelli}}}, \bibinfo {author} {\bibfnamefont {F.}~\bibnamefont
  {{Forastieri}}}, \bibinfo {author} {\bibfnamefont {M.}~\bibnamefont
  {{Frailis}}}, \bibinfo {author} {\bibfnamefont {A.~A.}\ \bibnamefont
  {{Fraisse}}}, \bibinfo {author} {\bibfnamefont {E.}~\bibnamefont
  {{Franceschi}}}, \bibinfo {author} {\bibfnamefont {A.}~\bibnamefont
  {{Frolov}}}, \bibinfo {author} {\bibfnamefont {S.}~\bibnamefont
  {{Galeotta}}}, \bibinfo {author} {\bibfnamefont {S.}~\bibnamefont {{Galli}}},
  \bibinfo {author} {\bibfnamefont {K.}~\bibnamefont {{Ganga}}}, \bibinfo
  {author} {\bibfnamefont {R.~T.}\ \bibnamefont {{G{\'e}nova-Santos}}},
  \bibinfo {author} {\bibfnamefont {M.}~\bibnamefont {{Gerbino}}}, \bibinfo
  {author} {\bibfnamefont {T.}~\bibnamefont {{Ghosh}}}, \bibinfo {author}
  {\bibfnamefont {J.}~\bibnamefont {{Gonz{\'a}lez-Nuevo}}}, \bibinfo {author}
  {\bibfnamefont {K.~M.}\ \bibnamefont {{G{\'o}rski}}}, \bibinfo {author}
  {\bibfnamefont {S.}~\bibnamefont {{Gratton}}}, \bibinfo {author}
  {\bibfnamefont {A.}~\bibnamefont {{Gruppuso}}}, \bibinfo {author}
  {\bibfnamefont {J.~E.}\ \bibnamefont {{Gudmundsson}}}, \bibinfo {author}
  {\bibfnamefont {J.}~\bibnamefont {{Hamann}}}, \bibinfo {author}
  {\bibfnamefont {W.}~\bibnamefont {{Handley}}}, \bibinfo {author}
  {\bibfnamefont {F.~K.}\ \bibnamefont {{Hansen}}}, \bibinfo {author}
  {\bibfnamefont {D.}~\bibnamefont {{Herranz}}}, \bibinfo {author}
  {\bibfnamefont {S.~R.}\ \bibnamefont {{Hildebrandt}}}, \bibinfo {author}
  {\bibfnamefont {E.}~\bibnamefont {{Hivon}}}, \bibinfo {author} {\bibfnamefont
  {Z.}~\bibnamefont {{Huang}}}, \bibinfo {author} {\bibfnamefont {A.~H.}\
  \bibnamefont {{Jaffe}}}, \bibinfo {author} {\bibfnamefont {W.~C.}\
  \bibnamefont {{Jones}}}, \bibinfo {author} {\bibfnamefont {A.}~\bibnamefont
  {{Karakci}}}, \bibinfo {author} {\bibfnamefont {E.}~\bibnamefont
  {{Keih{\"a}nen}}}, \bibinfo {author} {\bibfnamefont {R.}~\bibnamefont
  {{Keskitalo}}}, \bibinfo {author} {\bibfnamefont {K.}~\bibnamefont
  {{Kiiveri}}}, \bibinfo {author} {\bibfnamefont {J.}~\bibnamefont {{Kim}}},
  \bibinfo {author} {\bibfnamefont {T.~S.}\ \bibnamefont {{Kisner}}}, \bibinfo
  {author} {\bibfnamefont {L.}~\bibnamefont {{Knox}}}, \bibinfo {author}
  {\bibfnamefont {N.}~\bibnamefont {{Krachmalnicoff}}}, \bibinfo {author}
  {\bibfnamefont {M.}~\bibnamefont {{Kunz}}}, \bibinfo {author} {\bibfnamefont
  {H.}~\bibnamefont {{Kurki-Suonio}}}, \bibinfo {author} {\bibfnamefont
  {G.}~\bibnamefont {{Lagache}}}, \bibinfo {author} {\bibfnamefont {J.~M.}\
  \bibnamefont {{Lamarre}}}, \bibinfo {author} {\bibfnamefont {A.}~\bibnamefont
  {{Lasenby}}}, \bibinfo {author} {\bibfnamefont {M.}~\bibnamefont
  {{Lattanzi}}}, \bibinfo {author} {\bibfnamefont {C.~R.}\ \bibnamefont
  {{Lawrence}}}, \bibinfo {author} {\bibfnamefont {M.}~\bibnamefont {{Le
  Jeune}}}, \bibinfo {author} {\bibfnamefont {P.}~\bibnamefont {{Lemos}}},
  \bibinfo {author} {\bibfnamefont {J.}~\bibnamefont {{Lesgourgues}}}, \bibinfo
  {author} {\bibfnamefont {F.}~\bibnamefont {{Levrier}}}, \bibinfo {author}
  {\bibfnamefont {A.}~\bibnamefont {{Lewis}}}, \bibinfo {author} {\bibfnamefont
  {M.}~\bibnamefont {{Liguori}}}, \bibinfo {author} {\bibfnamefont {P.~B.}\
  \bibnamefont {{Lilje}}}, \bibinfo {author} {\bibfnamefont {M.}~\bibnamefont
  {{Lilley}}}, \bibinfo {author} {\bibfnamefont {V.}~\bibnamefont
  {{Lindholm}}}, \bibinfo {author} {\bibfnamefont {M.}~\bibnamefont
  {{L{\'o}pez-Caniego}}}, \bibinfo {author} {\bibfnamefont {P.~M.}\
  \bibnamefont {{Lubin}}}, \bibinfo {author} {\bibfnamefont {Y.~Z.}\
  \bibnamefont {{Ma}}}, \bibinfo {author} {\bibfnamefont {J.~F.}\ \bibnamefont
  {{Mac{\'\i}as-P{\'e}rez}}}, \bibinfo {author} {\bibfnamefont
  {G.}~\bibnamefont {{Maggio}}}, \bibinfo {author} {\bibfnamefont
  {D.}~\bibnamefont {{Maino}}}, \bibinfo {author} {\bibfnamefont
  {N.}~\bibnamefont {{Mandolesi}}}, \bibinfo {author} {\bibfnamefont
  {A.}~\bibnamefont {{Mangilli}}}, \bibinfo {author} {\bibfnamefont
  {A.}~\bibnamefont {{Marcos-Caballero}}}, \bibinfo {author} {\bibfnamefont
  {M.}~\bibnamefont {{Maris}}}, \bibinfo {author} {\bibfnamefont {P.~G.}\
  \bibnamefont {{Martin}}}, \bibinfo {author} {\bibfnamefont {M.}~\bibnamefont
  {{Martinelli}}}, \bibinfo {author} {\bibfnamefont {E.}~\bibnamefont
  {{Mart{\'\i}nez-Gonz{\'a}lez}}}, \bibinfo {author} {\bibfnamefont
  {S.}~\bibnamefont {{Matarrese}}}, \bibinfo {author} {\bibfnamefont
  {N.}~\bibnamefont {{Mauri}}}, \bibinfo {author} {\bibfnamefont {J.~D.}\
  \bibnamefont {{McEwen}}}, \bibinfo {author} {\bibfnamefont {P.~R.}\
  \bibnamefont {{Meinhold}}}, \bibinfo {author} {\bibfnamefont
  {A.}~\bibnamefont {{Melchiorri}}}, \bibinfo {author} {\bibfnamefont
  {A.}~\bibnamefont {{Mennella}}}, \bibinfo {author} {\bibfnamefont
  {M.}~\bibnamefont {{Migliaccio}}}, \bibinfo {author} {\bibfnamefont
  {M.}~\bibnamefont {{Millea}}}, \bibinfo {author} {\bibfnamefont
  {S.}~\bibnamefont {{Mitra}}}, \bibinfo {author} {\bibfnamefont {M.~A.}\
  \bibnamefont {{Miville-Desch{\^e}nes}}}, \bibinfo {author} {\bibfnamefont
  {D.}~\bibnamefont {{Molinari}}}, \bibinfo {author} {\bibfnamefont
  {L.}~\bibnamefont {{Montier}}}, \bibinfo {author} {\bibfnamefont
  {G.}~\bibnamefont {{Morgante}}}, \bibinfo {author} {\bibfnamefont
  {A.}~\bibnamefont {{Moss}}}, \bibinfo {author} {\bibfnamefont
  {P.}~\bibnamefont {{Natoli}}}, \bibinfo {author} {\bibfnamefont {H.~U.}\
  \bibnamefont {{N{\o}rgaard-Nielsen}}}, \bibinfo {author} {\bibfnamefont
  {L.}~\bibnamefont {{Pagano}}}, \bibinfo {author} {\bibfnamefont
  {D.}~\bibnamefont {{Paoletti}}}, \bibinfo {author} {\bibfnamefont
  {B.}~\bibnamefont {{Partridge}}}, \bibinfo {author} {\bibfnamefont
  {G.}~\bibnamefont {{Patanchon}}}, \bibinfo {author} {\bibfnamefont {H.~V.}\
  \bibnamefont {{Peiris}}}, \bibinfo {author} {\bibfnamefont {F.}~\bibnamefont
  {{Perrotta}}}, \bibinfo {author} {\bibfnamefont {V.}~\bibnamefont
  {{Pettorino}}}, \bibinfo {author} {\bibfnamefont {F.}~\bibnamefont
  {{Piacentini}}}, \bibinfo {author} {\bibfnamefont {L.}~\bibnamefont
  {{Polastri}}}, \bibinfo {author} {\bibfnamefont {G.}~\bibnamefont
  {{Polenta}}}, \bibinfo {author} {\bibfnamefont {J.~L.}\ \bibnamefont
  {{Puget}}}, \bibinfo {author} {\bibfnamefont {J.~P.}\ \bibnamefont
  {{Rachen}}}, \bibinfo {author} {\bibfnamefont {M.}~\bibnamefont
  {{Reinecke}}}, \bibinfo {author} {\bibfnamefont {M.}~\bibnamefont
  {{Remazeilles}}}, \bibinfo {author} {\bibfnamefont {A.}~\bibnamefont
  {{Renzi}}}, \bibinfo {author} {\bibfnamefont {G.}~\bibnamefont {{Rocha}}},
  \bibinfo {author} {\bibfnamefont {C.}~\bibnamefont {{Rosset}}}, \bibinfo
  {author} {\bibfnamefont {G.}~\bibnamefont {{Roudier}}}, \bibinfo {author}
  {\bibfnamefont {J.~A.}\ \bibnamefont {{Rubi{\~n}o-Mart{\'\i}n}}}, \bibinfo
  {author} {\bibfnamefont {B.}~\bibnamefont {{Ruiz-Granados}}}, \bibinfo
  {author} {\bibfnamefont {L.}~\bibnamefont {{Salvati}}}, \bibinfo {author}
  {\bibfnamefont {M.}~\bibnamefont {{Sandri}}}, \bibinfo {author}
  {\bibfnamefont {M.}~\bibnamefont {{Savelainen}}}, \bibinfo {author}
  {\bibfnamefont {D.}~\bibnamefont {{Scott}}}, \bibinfo {author} {\bibfnamefont
  {E.~P.~S.}\ \bibnamefont {{Shellard}}}, \bibinfo {author} {\bibfnamefont
  {C.}~\bibnamefont {{Sirignano}}}, \bibinfo {author} {\bibfnamefont
  {G.}~\bibnamefont {{Sirri}}}, \bibinfo {author} {\bibfnamefont {L.~D.}\
  \bibnamefont {{Spencer}}}, \bibinfo {author} {\bibfnamefont {R.}~\bibnamefont
  {{Sunyaev}}}, \bibinfo {author} {\bibfnamefont {A.~S.}\ \bibnamefont
  {{Suur-Uski}}}, \bibinfo {author} {\bibfnamefont {J.~A.}\ \bibnamefont
  {{Tauber}}}, \bibinfo {author} {\bibfnamefont {D.}~\bibnamefont
  {{Tavagnacco}}}, \bibinfo {author} {\bibfnamefont {M.}~\bibnamefont
  {{Tenti}}}, \bibinfo {author} {\bibfnamefont {L.}~\bibnamefont
  {{Toffolatti}}}, \bibinfo {author} {\bibfnamefont {M.}~\bibnamefont
  {{Tomasi}}}, \bibinfo {author} {\bibfnamefont {T.}~\bibnamefont
  {{Trombetti}}}, \bibinfo {author} {\bibfnamefont {L.}~\bibnamefont
  {{Valenziano}}}, \bibinfo {author} {\bibfnamefont {J.}~\bibnamefont
  {{Valiviita}}}, \bibinfo {author} {\bibfnamefont {B.}~\bibnamefont {{Van
  Tent}}}, \bibinfo {author} {\bibfnamefont {L.}~\bibnamefont {{Vibert}}},
  \bibinfo {author} {\bibfnamefont {P.}~\bibnamefont {{Vielva}}}, \bibinfo
  {author} {\bibfnamefont {F.}~\bibnamefont {{Villa}}}, \bibinfo {author}
  {\bibfnamefont {N.}~\bibnamefont {{Vittorio}}}, \bibinfo {author}
  {\bibfnamefont {B.~D.}\ \bibnamefont {{Wandelt}}}, \bibinfo {author}
  {\bibfnamefont {I.~K.}\ \bibnamefont {{Wehus}}}, \bibinfo {author}
  {\bibfnamefont {M.}~\bibnamefont {{White}}}, \bibinfo {author} {\bibfnamefont
  {S.~D.~M.}\ \bibnamefont {{White}}}, \bibinfo {author} {\bibfnamefont
  {A.}~\bibnamefont {{Zacchei}}},\ and\ \bibinfo {author} {\bibfnamefont
  {A.}~\bibnamefont {{Zonca}}},\ }\bibfield  {title} {\bibinfo {title} {{Planck
  2018 results. VI. Cosmological parameters}},\ }\href
  {https://doi.org/10.1051/0004-6361/201833910} {\bibfield  {journal} {\bibinfo
   {journal} {\aap}\ }\textbf {\bibinfo {volume} {641}},\ \bibinfo {eid} {A6}
  (\bibinfo {year} {2020})},\ \Eprint {https://arxiv.org/abs/1807.06209}
  {arXiv:1807.06209 [astro-ph.CO]} \BibitemShut {NoStop}%
\bibitem [{\citenamefont {{Phinney}}(2001)}]{p01}%
  \BibitemOpen
  \bibfield  {author} {\bibinfo {author} {\bibfnamefont {E.~S.}\ \bibnamefont
  {{Phinney}}},\ }\bibfield  {title} {\bibinfo {title} {{A Practical Theorem on
  Gravitational Wave Backgrounds}},\ }\href@noop {} {\bibfield  {journal}
  {\bibinfo  {journal} {ArXiv Astrophysics e-prints}\ } (\bibinfo {year}
  {2001})},\ \Eprint {https://arxiv.org/abs/astro-ph/0108028}
  {astro-ph/0108028} \BibitemShut {NoStop}%
\bibitem [{\citenamefont {{Hellings}}\ and\ \citenamefont
  {{Downs}}(1983)}]{hd83}%
  \BibitemOpen
  \bibfield  {author} {\bibinfo {author} {\bibfnamefont {R.~W.}\ \bibnamefont
  {{Hellings}}}\ and\ \bibinfo {author} {\bibfnamefont {G.~S.}\ \bibnamefont
  {{Downs}}},\ }\bibfield  {title} {\bibinfo {title} {{Upper limits on the
  isotropic gravitational radiation background from pulsar timing analysis}},\
  }\href {https://doi.org/10.1086/183954} {\bibfield  {journal} {\bibinfo
  {journal} {\apjl}\ }\textbf {\bibinfo {volume} {265}},\ \bibinfo {pages}
  {L39} (\bibinfo {year} {1983})}\BibitemShut {NoStop}%
\bibitem [{\citenamefont {Jinno}\ and\ \citenamefont
  {Takimoto}(2017)}]{1605.01403}%
  \BibitemOpen
  \bibfield  {author} {\bibinfo {author} {\bibfnamefont {R.}~\bibnamefont
  {Jinno}}\ and\ \bibinfo {author} {\bibfnamefont {M.}~\bibnamefont
  {Takimoto}},\ }\bibfield  {title} {\bibinfo {title} {{Gravitational waves
  from bubble collisions: An analytic derivation}},\ }\href
  {https://doi.org/10.1103/PhysRevD.95.024009} {\bibfield  {journal} {\bibinfo
  {journal} {Phys. Rev. D}\ }\textbf {\bibinfo {volume} {95}},\ \bibinfo
  {pages} {024009} (\bibinfo {year} {2017})},\ \Eprint
  {https://arxiv.org/abs/1605.01403} {arXiv:1605.01403 [astro-ph.CO]}
  \BibitemShut {NoStop}%
\bibitem [{\citenamefont {Hindmarsh}\ \emph {et~al.}(2017)\citenamefont
  {Hindmarsh}, \citenamefont {Huber}, \citenamefont {Rummukainen},\ and\
  \citenamefont {Weir}}]{1704.05871}%
  \BibitemOpen
  \bibfield  {author} {\bibinfo {author} {\bibfnamefont {M.}~\bibnamefont
  {Hindmarsh}}, \bibinfo {author} {\bibfnamefont {S.~J.}\ \bibnamefont
  {Huber}}, \bibinfo {author} {\bibfnamefont {K.}~\bibnamefont {Rummukainen}},\
  and\ \bibinfo {author} {\bibfnamefont {D.~J.}\ \bibnamefont {Weir}},\
  }\bibfield  {title} {\bibinfo {title} {{Shape of the acoustic gravitational
  wave power spectrum from a first order phase transition}},\ }\href
  {https://doi.org/10.1103/PhysRevD.96.103520} {\bibfield  {journal} {\bibinfo
  {journal} {Phys. Rev. D}\ }\textbf {\bibinfo {volume} {96}},\ \bibinfo
  {pages} {103520} (\bibinfo {year} {2017})},\ \bibinfo {note} {[Erratum:
  Phys.Rev.D 101, 089902 (2020)]},\ \Eprint {https://arxiv.org/abs/1704.05871}
  {arXiv:1704.05871 [astro-ph.CO]} \BibitemShut {NoStop}%
\bibitem [{\citenamefont {Lewicki}\ and\ \citenamefont
  {Vaskonen}(2020{\natexlab{a}})}]{2012.07826}%
  \BibitemOpen
  \bibfield  {author} {\bibinfo {author} {\bibfnamefont {M.}~\bibnamefont
  {Lewicki}}\ and\ \bibinfo {author} {\bibfnamefont {V.}~\bibnamefont
  {Vaskonen}},\ }\bibfield  {title} {\bibinfo {title} {{Gravitational waves
  from colliding vacuum bubbles in gauge theories}},\ }\href@noop {} {\
  (\bibinfo {year} {2020}{\natexlab{a}})},\ \Eprint
  {https://arxiv.org/abs/2012.07826} {arXiv:2012.07826 [astro-ph.CO]}
  \BibitemShut {NoStop}%
\bibitem [{\citenamefont {Cutting}\ \emph {et~al.}(2021)\citenamefont
  {Cutting}, \citenamefont {Escartin}, \citenamefont {Hindmarsh},\ and\
  \citenamefont {Weir}}]{2005.13537}%
  \BibitemOpen
  \bibfield  {author} {\bibinfo {author} {\bibfnamefont {D.}~\bibnamefont
  {Cutting}}, \bibinfo {author} {\bibfnamefont {E.~G.}\ \bibnamefont
  {Escartin}}, \bibinfo {author} {\bibfnamefont {M.}~\bibnamefont
  {Hindmarsh}},\ and\ \bibinfo {author} {\bibfnamefont {D.~J.}\ \bibnamefont
  {Weir}},\ }\bibfield  {title} {\bibinfo {title} {{Gravitational waves from
  vacuum first order phase transitions II: from thin to thick walls}},\ }\href
  {https://doi.org/10.1103/PhysRevD.103.023531} {\bibfield  {journal} {\bibinfo
   {journal} {Phys. Rev. D}\ }\textbf {\bibinfo {volume} {103}},\ \bibinfo
  {pages} {023531} (\bibinfo {year} {2021})},\ \Eprint
  {https://arxiv.org/abs/2005.13537} {arXiv:2005.13537 [astro-ph.CO]}
  \BibitemShut {NoStop}%
\bibitem [{\citenamefont {Caprini}\ \emph {et~al.}(2016)\citenamefont {Caprini}
  \emph {et~al.}}]{1512.06239}%
  \BibitemOpen
  \bibfield  {author} {\bibinfo {author} {\bibfnamefont {C.}~\bibnamefont
  {Caprini}} \emph {et~al.},\ }\bibfield  {title} {\bibinfo {title} {{Science
  with the space-based interferometer eLISA. II: Gravitational waves from
  cosmological phase transitions}},\ }\href
  {https://doi.org/10.1088/1475-7516/2016/04/001} {\bibfield  {journal}
  {\bibinfo  {journal} {JCAP}\ }\textbf {\bibinfo {volume} {04}},\ \bibinfo
  {pages} {001}},\ \Eprint {https://arxiv.org/abs/1512.06239} {arXiv:1512.06239
  [astro-ph.CO]} \BibitemShut {NoStop}%
\bibitem [{\citenamefont {Roper~Pol}\ \emph {et~al.}(2020)\citenamefont
  {Roper~Pol}, \citenamefont {Mandal}, \citenamefont {Brandenburg},
  \citenamefont {Kahniashvili},\ and\ \citenamefont {Kosowsky}}]{1903.08585}%
  \BibitemOpen
  \bibfield  {author} {\bibinfo {author} {\bibfnamefont {A.}~\bibnamefont
  {Roper~Pol}}, \bibinfo {author} {\bibfnamefont {S.}~\bibnamefont {Mandal}},
  \bibinfo {author} {\bibfnamefont {A.}~\bibnamefont {Brandenburg}}, \bibinfo
  {author} {\bibfnamefont {T.}~\bibnamefont {Kahniashvili}},\ and\ \bibinfo
  {author} {\bibfnamefont {A.}~\bibnamefont {Kosowsky}},\ }\bibfield  {title}
  {\bibinfo {title} {{Numerical simulations of gravitational waves from
  early-universe turbulence}},\ }\href
  {https://doi.org/10.1103/PhysRevD.102.083512} {\bibfield  {journal} {\bibinfo
   {journal} {Phys. Rev. D}\ }\textbf {\bibinfo {volume} {102}},\ \bibinfo
  {pages} {083512} (\bibinfo {year} {2020})},\ \Eprint
  {https://arxiv.org/abs/1903.08585} {arXiv:1903.08585 [astro-ph.CO]}
  \BibitemShut {NoStop}%
\bibitem [{\citenamefont {Brandenburg}\ \emph {et~al.}(2021)\citenamefont
  {Brandenburg}, \citenamefont {Clarke}, \citenamefont {He},\ and\
  \citenamefont {Kahniashvili}}]{2102.12428}%
  \BibitemOpen
  \bibfield  {author} {\bibinfo {author} {\bibfnamefont {A.}~\bibnamefont
  {Brandenburg}}, \bibinfo {author} {\bibfnamefont {E.}~\bibnamefont {Clarke}},
  \bibinfo {author} {\bibfnamefont {Y.}~\bibnamefont {He}},\ and\ \bibinfo
  {author} {\bibfnamefont {T.}~\bibnamefont {Kahniashvili}},\ }\bibfield
  {title} {\bibinfo {title} {{Can we observe the QCD phase transition-generated
  gravitational waves through pulsar timing arrays?}},\ }\href
  {https://doi.org/10.1103/PhysRevD.104.043513} {\bibfield  {journal} {\bibinfo
   {journal} {Phys. Rev. D}\ }\textbf {\bibinfo {volume} {104}},\ \bibinfo
  {pages} {043513} (\bibinfo {year} {2021})},\ \Eprint
  {https://arxiv.org/abs/2102.12428} {arXiv:2102.12428 [astro-ph.CO]}
  \BibitemShut {NoStop}%
\bibitem [{\citenamefont {Binetruy}\ \emph {et~al.}(2012)\citenamefont
  {Binetruy}, \citenamefont {Bohe}, \citenamefont {Caprini},\ and\
  \citenamefont {Dufaux}}]{1201.0983}%
  \BibitemOpen
  \bibfield  {author} {\bibinfo {author} {\bibfnamefont {P.}~\bibnamefont
  {Binetruy}}, \bibinfo {author} {\bibfnamefont {A.}~\bibnamefont {Bohe}},
  \bibinfo {author} {\bibfnamefont {C.}~\bibnamefont {Caprini}},\ and\ \bibinfo
  {author} {\bibfnamefont {J.-F.}\ \bibnamefont {Dufaux}},\ }\bibfield  {title}
  {\bibinfo {title} {{Cosmological Backgrounds of Gravitational Waves and
  eLISA/NGO: Phase Transitions, Cosmic Strings and Other Sources}},\ }\href
  {https://doi.org/10.1088/1475-7516/2012/06/027} {\bibfield  {journal}
  {\bibinfo  {journal} {JCAP}\ }\textbf {\bibinfo {volume} {06}},\ \bibinfo
  {pages} {027}},\ \Eprint {https://arxiv.org/abs/1201.0983} {arXiv:1201.0983
  [gr-qc]} \BibitemShut {NoStop}%
\bibitem [{\citenamefont {Guo}\ \emph {et~al.}(2021)\citenamefont {Guo},
  \citenamefont {Sinha}, \citenamefont {Vagie},\ and\ \citenamefont
  {White}}]{2007.08537}%
  \BibitemOpen
  \bibfield  {author} {\bibinfo {author} {\bibfnamefont {H.-K.}\ \bibnamefont
  {Guo}}, \bibinfo {author} {\bibfnamefont {K.}~\bibnamefont {Sinha}}, \bibinfo
  {author} {\bibfnamefont {D.}~\bibnamefont {Vagie}},\ and\ \bibinfo {author}
  {\bibfnamefont {G.}~\bibnamefont {White}},\ }\bibfield  {title} {\bibinfo
  {title} {{Phase Transitions in an Expanding Universe: Stochastic
  Gravitational Waves in Standard and Non-Standard Histories}},\ }\href
  {https://doi.org/10.1088/1475-7516/2021/01/001} {\bibfield  {journal}
  {\bibinfo  {journal} {JCAP}\ }\textbf {\bibinfo {volume} {01}},\ \bibinfo
  {pages} {001}},\ \Eprint {https://arxiv.org/abs/2007.08537} {arXiv:2007.08537
  [hep-ph]} \BibitemShut {NoStop}%
\bibitem [{\citenamefont {Ellis}\ \emph {et~al.}(2020)\citenamefont {Ellis},
  \citenamefont {Lewicki},\ and\ \citenamefont {No}}]{2003.07360}%
  \BibitemOpen
  \bibfield  {author} {\bibinfo {author} {\bibfnamefont {J.}~\bibnamefont
  {Ellis}}, \bibinfo {author} {\bibfnamefont {M.}~\bibnamefont {Lewicki}},\
  and\ \bibinfo {author} {\bibfnamefont {J.~M.}\ \bibnamefont {No}},\
  }\bibfield  {title} {\bibinfo {title} {{Gravitational waves from first-order
  cosmological phase transitions: lifetime of the sound wave source}},\ }\href
  {https://doi.org/10.1088/1475-7516/2020/07/050} {\bibfield  {journal}
  {\bibinfo  {journal} {JCAP}\ }\textbf {\bibinfo {volume} {07}},\ \bibinfo
  {pages} {050}},\ \Eprint {https://arxiv.org/abs/2003.07360} {arXiv:2003.07360
  [hep-ph]} \BibitemShut {NoStop}%
\bibitem [{\citenamefont {Weir}(2018)}]{1705.01783}%
  \BibitemOpen
  \bibfield  {author} {\bibinfo {author} {\bibfnamefont {D.~J.}\ \bibnamefont
  {Weir}},\ }\bibfield  {title} {\bibinfo {title} {{Gravitational waves from a
  first order electroweak phase transition: a brief review}},\ }\href
  {https://doi.org/10.1098/rsta.2017.0126} {\bibfield  {journal} {\bibinfo
  {journal} {Phil. Trans. Roy. Soc. Lond. A}\ }\textbf {\bibinfo {volume}
  {376}},\ \bibinfo {pages} {20170126} (\bibinfo {year} {2018})},\ \Eprint
  {https://arxiv.org/abs/1705.01783} {arXiv:1705.01783 [hep-ph]} \BibitemShut
  {NoStop}%
\bibitem [{\citenamefont {Hindmarsh}\ and\ \citenamefont
  {Hijazi}(2019)}]{1909.10040}%
  \BibitemOpen
  \bibfield  {author} {\bibinfo {author} {\bibfnamefont {M.}~\bibnamefont
  {Hindmarsh}}\ and\ \bibinfo {author} {\bibfnamefont {M.}~\bibnamefont
  {Hijazi}},\ }\bibfield  {title} {\bibinfo {title} {{Gravitational waves from
  first order cosmological phase transitions in the Sound Shell Model}},\
  }\href {https://doi.org/10.1088/1475-7516/2019/12/062} {\bibfield  {journal}
  {\bibinfo  {journal} {JCAP}\ }\textbf {\bibinfo {volume} {12}},\ \bibinfo
  {pages} {062}},\ \Eprint {https://arxiv.org/abs/1909.10040} {arXiv:1909.10040
  [astro-ph.CO]} \BibitemShut {NoStop}%
\bibitem [{\citenamefont {Espinosa}\ \emph {et~al.}(2010)\citenamefont
  {Espinosa}, \citenamefont {Konstandin}, \citenamefont {No},\ and\
  \citenamefont {Servant}}]{1004.4187}%
  \BibitemOpen
  \bibfield  {author} {\bibinfo {author} {\bibfnamefont {J.~R.}\ \bibnamefont
  {Espinosa}}, \bibinfo {author} {\bibfnamefont {T.}~\bibnamefont
  {Konstandin}}, \bibinfo {author} {\bibfnamefont {J.~M.}\ \bibnamefont {No}},\
  and\ \bibinfo {author} {\bibfnamefont {G.}~\bibnamefont {Servant}},\
  }\bibfield  {title} {\bibinfo {title} {{Energy Budget of Cosmological
  First-order Phase Transitions}},\ }\href
  {https://doi.org/10.1088/1475-7516/2010/06/028} {\bibfield  {journal}
  {\bibinfo  {journal} {JCAP}\ }\textbf {\bibinfo {volume} {06}},\ \bibinfo
  {pages} {028}},\ \Eprint {https://arxiv.org/abs/1004.4187} {arXiv:1004.4187
  [hep-ph]} \BibitemShut {NoStop}%
\bibitem [{\citenamefont {Kosowsky}\ \emph {et~al.}(1992)\citenamefont
  {Kosowsky}, \citenamefont {Turner},\ and\ \citenamefont
  {Watkins}}]{Kosowsky:1991ua}%
  \BibitemOpen
  \bibfield  {author} {\bibinfo {author} {\bibfnamefont {A.}~\bibnamefont
  {Kosowsky}}, \bibinfo {author} {\bibfnamefont {M.~S.}\ \bibnamefont
  {Turner}},\ and\ \bibinfo {author} {\bibfnamefont {R.}~\bibnamefont
  {Watkins}},\ }\bibfield  {title} {\bibinfo {title} {{Gravitational radiation
  from colliding vacuum bubbles}},\ }\href
  {https://doi.org/10.1103/PhysRevD.45.4514} {\bibfield  {journal} {\bibinfo
  {journal} {Phys. Rev. D}\ }\textbf {\bibinfo {volume} {45}},\ \bibinfo
  {pages} {4514} (\bibinfo {year} {1992})}\BibitemShut {NoStop}%
\bibitem [{\citenamefont {Cutting}\ \emph {et~al.}(2018)\citenamefont
  {Cutting}, \citenamefont {Hindmarsh},\ and\ \citenamefont
  {Weir}}]{1802.05712}%
  \BibitemOpen
  \bibfield  {author} {\bibinfo {author} {\bibfnamefont {D.}~\bibnamefont
  {Cutting}}, \bibinfo {author} {\bibfnamefont {M.}~\bibnamefont {Hindmarsh}},\
  and\ \bibinfo {author} {\bibfnamefont {D.~J.}\ \bibnamefont {Weir}},\
  }\bibfield  {title} {\bibinfo {title} {{Gravitational waves from vacuum
  first-order phase transitions: from the envelope to the lattice}},\ }\href
  {https://doi.org/10.1103/PhysRevD.97.123513} {\bibfield  {journal} {\bibinfo
  {journal} {Phys. Rev. D}\ }\textbf {\bibinfo {volume} {97}},\ \bibinfo
  {pages} {123513} (\bibinfo {year} {2018})},\ \Eprint
  {https://arxiv.org/abs/1802.05712} {arXiv:1802.05712 [astro-ph.CO]}
  \BibitemShut {NoStop}%
\bibitem [{\citenamefont {Lewicki}\ and\ \citenamefont
  {Vaskonen}(2020{\natexlab{b}})}]{2007.04967}%
  \BibitemOpen
  \bibfield  {author} {\bibinfo {author} {\bibfnamefont {M.}~\bibnamefont
  {Lewicki}}\ and\ \bibinfo {author} {\bibfnamefont {V.}~\bibnamefont
  {Vaskonen}},\ }\bibfield  {title} {\bibinfo {title} {{Gravitational wave
  spectra from strongly supercooled phase transitions}},\ }\href
  {https://doi.org/10.1140/epjc/s10052-020-08589-1} {\bibfield  {journal}
  {\bibinfo  {journal} {Eur. Phys. J. C}\ }\textbf {\bibinfo {volume} {80}},\
  \bibinfo {pages} {1003} (\bibinfo {year} {2020}{\natexlab{b}})},\ \Eprint
  {https://arxiv.org/abs/2007.04967} {arXiv:2007.04967 [astro-ph.CO]}
  \BibitemShut {NoStop}%
\bibitem [{\citenamefont {{Schwarz}}(1978)}]{1978AnSta...6..461S}%
  \BibitemOpen
  \bibfield  {author} {\bibinfo {author} {\bibfnamefont {G.}~\bibnamefont
  {{Schwarz}}},\ }\bibfield  {title} {\bibinfo {title} {{Estimating the
  Dimension of a Model}},\ }\href@noop {} {\bibfield  {journal} {\bibinfo
  {journal} {Annals of Statistics}\ }\textbf {\bibinfo {volume} {6}},\ \bibinfo
  {pages} {461} (\bibinfo {year} {1978})}\BibitemShut {NoStop}%
\bibitem [{\citenamefont {Kass}\ and\ \citenamefont
  {Raftery}(1995)}]{doi:10.1080/01621459.1995.10476572}%
  \BibitemOpen
  \bibfield  {author} {\bibinfo {author} {\bibfnamefont {R.~E.}\ \bibnamefont
  {Kass}}\ and\ \bibinfo {author} {\bibfnamefont {A.~E.}\ \bibnamefont
  {Raftery}},\ }\bibfield  {title} {\bibinfo {title} {Bayes factors},\ }\href
  {https://doi.org/10.1080/01621459.1995.10476572} {\bibfield  {journal}
  {\bibinfo  {journal} {Journal of the American Statistical Association}\
  }\textbf {\bibinfo {volume} {90}},\ \bibinfo {pages} {773} (\bibinfo {year}
  {1995})}\BibitemShut {NoStop}%
\bibitem [{\citenamefont {Cutting}\ \emph {et~al.}(2020)\citenamefont
  {Cutting}, \citenamefont {Hindmarsh},\ and\ \citenamefont
  {Weir}}]{1906.00480}%
  \BibitemOpen
  \bibfield  {author} {\bibinfo {author} {\bibfnamefont {D.}~\bibnamefont
  {Cutting}}, \bibinfo {author} {\bibfnamefont {M.}~\bibnamefont {Hindmarsh}},\
  and\ \bibinfo {author} {\bibfnamefont {D.~J.}\ \bibnamefont {Weir}},\
  }\bibfield  {title} {\bibinfo {title} {{Vorticity, kinetic energy, and
  suppressed gravitational wave production in strong first order phase
  transitions}},\ }\href {https://doi.org/10.1103/PhysRevLett.125.021302}
  {\bibfield  {journal} {\bibinfo  {journal} {Phys. Rev. Lett.}\ }\textbf
  {\bibinfo {volume} {125}},\ \bibinfo {pages} {021302} (\bibinfo {year}
  {2020})},\ \Eprint {https://arxiv.org/abs/1906.00480} {arXiv:1906.00480
  [hep-ph]} \BibitemShut {NoStop}%
\bibitem [{\citenamefont {{Particle Data Group}}\ \emph
  {et~al.}(2020)\citenamefont {{Particle Data Group}}, \citenamefont {{Zyla}},
  \citenamefont {{Barnett}}, \citenamefont {{Beringer}}, \citenamefont
  {{Dahl}}, \citenamefont {{Dwyer}}, \citenamefont {{Groom}}, \citenamefont
  {{Lin}}, \citenamefont {{Lugovsky}}, \citenamefont {{Pianori}}, \citenamefont
  {{Robinson}}, \citenamefont {{Wohl}}, \citenamefont {{Yao}}, \citenamefont
  {{Agashe}}, \citenamefont {{Aielli}}, \citenamefont {{Allanach}},
  \citenamefont {{Amsler}}, \citenamefont {{Antonelli}}, \citenamefont
  {{Aschenauer}}, \citenamefont {{Asner}}, \citenamefont {{Baer}},
  \citenamefont {{Banerjee}}, \citenamefont {{Baudis}}, \citenamefont
  {{Bauer}}, \citenamefont {{Beatty}}, \citenamefont {{Belousov}},
  \citenamefont {{Bethke}}, \citenamefont {{Bettini}}, \citenamefont
  {{Biebel}}, \citenamefont {{Black}}, \citenamefont {{Blucher}}, \citenamefont
  {{Buchmuller}}, \citenamefont {{Burkert}}, \citenamefont {{Bychkov}},
  \citenamefont {{Cahn}}, \citenamefont {{Carena}}, \citenamefont {{Ceccucci}},
  \citenamefont {{Cerri}}, \citenamefont {{Chakraborty}}, \citenamefont
  {{Chivukula}}, \citenamefont {{Cowan}}, \citenamefont {{D'Ambrosio}},
  \citenamefont {{Damour}}, \citenamefont {{de Florian}}, \citenamefont {{de
  Gouv{\^e}a}}, \citenamefont {{DeGrand}}, \citenamefont {{de Jong}},
  \citenamefont {{Dissertori}}, \citenamefont {{Dobrescu}}, \citenamefont
  {{D'Onofrio}}, \citenamefont {{Doser}}, \citenamefont {{Drees}},
  \citenamefont {{Dreiner}}, \citenamefont {{Eerola}}, \citenamefont {{Egede}},
  \citenamefont {{Eidelman}}, \citenamefont {{Ellis}}, \citenamefont {{Erler}},
  \citenamefont {{Ezhela}}, \citenamefont {{Fetscher}}, \citenamefont
  {{Fields}}, \citenamefont {{Foster}}, \citenamefont {{Freitas}},
  \citenamefont {{Gallagher}}, \citenamefont {{Garren}}, \citenamefont
  {{Gerber}}, \citenamefont {{Gerbier}}, \citenamefont {{Gershon}},
  \citenamefont {{Gershtein}}, \citenamefont {{Gherghetta}}, \citenamefont
  {{Godizov}}, \citenamefont {{Gonzalez-Garcia}}, \citenamefont {{Goodman}},
  \citenamefont {{Grab}}, \citenamefont {{Gritsan}}, \citenamefont {{Grojean}},
  \citenamefont {{Gr{\"u}newald}}, \citenamefont {{Gurtu}}, \citenamefont
  {{Gutsche}}, \citenamefont {{Haber}}, \citenamefont {{Hanhart}},
  \citenamefont {{Hashimoto}}, \citenamefont {{Hayato}}, \citenamefont
  {{Hebecker}}, \citenamefont {{Heinemeyer}}, \citenamefont {{Heltsley}},
  \citenamefont {{Hern{\'a}ndez-Rey}}, \citenamefont {{Hikasa}}, \citenamefont
  {{Hisano}}, \citenamefont {{H{\"o}cker}}, \citenamefont {{Holder}},
  \citenamefont {{Holtkamp}}, \citenamefont {{Huston}}, \citenamefont
  {{Hyodo}}, \citenamefont {{Johnson}}, \citenamefont {{Kado}}, \citenamefont
  {{Karliner}}, \citenamefont {{Katz}}, \citenamefont {{Kenzie}}, \citenamefont
  {{Khoze}}, \citenamefont {{Klein}}, \citenamefont {{Klempt}}, \citenamefont
  {{Kowalewski}}, \citenamefont {{Krauss}}, \citenamefont {{Kreps}},
  \citenamefont {{Krusche}}, \citenamefont {{Kwon}}, \citenamefont {{Lahav}},
  \citenamefont {{Laiho}}, \citenamefont {{Lellouch}}, \citenamefont
  {{Lesgourgues}}, \citenamefont {{Liddle}}, \citenamefont {{Ligeti}},
  \citenamefont {{Lippmann}}, \citenamefont {{Liss}}, \citenamefont
  {{Littenberg}}, \citenamefont {{Lourengo}}, \citenamefont {{Lugovsky}},
  \citenamefont {{Lusiani}}, \citenamefont {{Makida}}, \citenamefont
  {{Maltoni}}, \citenamefont {{Mannel}}, \citenamefont {{Manohar}},
  \citenamefont {{Marciano}}, \citenamefont {{Masoni}}, \citenamefont
  {{Matthews}}, \citenamefont {{Mei{\ss}ner}}, \citenamefont {{Mikhasenko}},
  \citenamefont {{Miller}}, \citenamefont {{Milstead}}, \citenamefont
  {{Mitchell}}, \citenamefont {{M{\"o}nig}}, \citenamefont {{Molaro}},
  \citenamefont {{Moortgat}}, \citenamefont {{Moskovic}}, \citenamefont
  {{Nakamura}}, \citenamefont {{Narain}}, \citenamefont {{Nason}},
  \citenamefont {{Navas}}, \citenamefont {{Neubert}}, \citenamefont {{Nevski}},
  \citenamefont {{Nir}}, \citenamefont {{Olive}}, \citenamefont {{Patrignani}},
  \citenamefont {{Peacock}}, \citenamefont {{Petcov}}, \citenamefont
  {{Petrov}}, \citenamefont {{Pich}}, \citenamefont {{Piepke}}, \citenamefont
  {{Pomarol}}, \citenamefont {{Profumo}}, \citenamefont {{Quadt}},
  \citenamefont {{Rabbertz}}, \citenamefont {{Rademacker}}, \citenamefont
  {{Raffelt}}, \citenamefont {{Ramani}}, \citenamefont {{Ramsey-Musolf}},
  \citenamefont {{Ratcliff}}, \citenamefont {{Richardson}}, \citenamefont
  {{Ringwald}}, \citenamefont {{Roesler}}, \citenamefont {{Rolli}},
  \citenamefont {{Romaniouk}}, \citenamefont {{Rosenberg}}, \citenamefont
  {{Rosner}}, \citenamefont {{Rybka}}, \citenamefont {{Ryskin}}, \citenamefont
  {{Ryutin}}, \citenamefont {{Sakai}}, \citenamefont {{Salam}}, \citenamefont
  {{Sarkar}}, \citenamefont {{Sauli}}, \citenamefont {{Schneider}},
  \citenamefont {{Scholberg}}, \citenamefont {{Schwartz}}, \citenamefont
  {{Schwiening}}, \citenamefont {{Scott}}, \citenamefont {{Sharma}},
  \citenamefont {{Sharpe}}, \citenamefont {{Shutt}}, \citenamefont {{Silari}},
  \citenamefont {{Sj{\"o}strand}}, \citenamefont {{Skands}}, \citenamefont
  {{Skwarnicki}}, \citenamefont {{Smoot}}, \citenamefont {{Soffer}},
  \citenamefont {{Sozzi}}, \citenamefont {{Spanier}}, \citenamefont
  {{Spiering}}, \citenamefont {{Stahl}}, \citenamefont {{Stone}}, \citenamefont
  {{Sumino}}, \citenamefont {{Sumiyoshi}}, \citenamefont {{Syphers}},
  \citenamefont {{Takahashi}}, \citenamefont {{Tanabashi}}, \citenamefont
  {{Tanaka}}, \citenamefont {{Ta{\v{s}}evsk{\'y}}}, \citenamefont {{Terashi}},
  \citenamefont {{Terning}}, \citenamefont {{Thoma}}, \citenamefont {{Thorne}},
  \citenamefont {{Tiator}}, \citenamefont {{Titov}}, \citenamefont
  {{Tkachenko}}, \citenamefont {{Tovey}}, \citenamefont {{Trabelsi}},
  \citenamefont {{Urquijo}}, \citenamefont {{Valencia}}, \citenamefont {{Van de
  Water}}, \citenamefont {{Varelas}}, \citenamefont {{Venanzoni}},
  \citenamefont {{Verde}}, \citenamefont {{Vincter}}, \citenamefont {{Vogel}},
  \citenamefont {{Vogelsang}}, \citenamefont {{Vogt}}, \citenamefont
  {{Vorobyev}}, \citenamefont {{Wakely}}, \citenamefont {{Walkowiak}},
  \citenamefont {{Walter}}, \citenamefont {{Wands}}, \citenamefont {{Wascko}},
  \citenamefont {{Weinberg}}, \citenamefont {{Weinberg}}, \citenamefont
  {{White}}, \citenamefont {{Wiencke}}, \citenamefont {{Willocq}},
  \citenamefont {{Woody}}, \citenamefont {{Workman}}, \citenamefont
  {{Yokoyama}}, \citenamefont {{Yoshida}}, \citenamefont {{Zanderighi}},
  \citenamefont {{Zeller}}, \citenamefont {{Zenin}}, \citenamefont {{Zhu}},
  \citenamefont {{Zhu}}, \citenamefont {{Zimmermann}}, \citenamefont
  {{Anderson}}, \citenamefont {{Basaglia}}, \citenamefont {{Lugovsky}},
  \citenamefont {{Schaffner}},\ and\ \citenamefont
  {{Zheng}}}]{2020PTEP.2020h3C01P}%
  \BibitemOpen
  \bibfield  {author} {\bibinfo {author} {\bibnamefont {{Particle Data
  Group}}}, \bibinfo {author} {\bibfnamefont {P.~A.}\ \bibnamefont {{Zyla}}},
  \bibinfo {author} {\bibfnamefont {R.~M.}\ \bibnamefont {{Barnett}}}, \bibinfo
  {author} {\bibfnamefont {J.}~\bibnamefont {{Beringer}}}, \bibinfo {author}
  {\bibfnamefont {O.}~\bibnamefont {{Dahl}}}, \bibinfo {author} {\bibfnamefont
  {D.~A.}\ \bibnamefont {{Dwyer}}}, \bibinfo {author} {\bibfnamefont {D.~E.}\
  \bibnamefont {{Groom}}}, \bibinfo {author} {\bibfnamefont {C.~J.}\
  \bibnamefont {{Lin}}}, \bibinfo {author} {\bibfnamefont {K.~S.}\ \bibnamefont
  {{Lugovsky}}}, \bibinfo {author} {\bibfnamefont {E.}~\bibnamefont
  {{Pianori}}}, \bibinfo {author} {\bibfnamefont {D.~J.}\ \bibnamefont
  {{Robinson}}}, \bibinfo {author} {\bibfnamefont {C.~G.}\ \bibnamefont
  {{Wohl}}}, \bibinfo {author} {\bibfnamefont {W.~M.}\ \bibnamefont {{Yao}}},
  \bibinfo {author} {\bibfnamefont {K.}~\bibnamefont {{Agashe}}}, \bibinfo
  {author} {\bibfnamefont {G.}~\bibnamefont {{Aielli}}}, \bibinfo {author}
  {\bibfnamefont {B.~C.}\ \bibnamefont {{Allanach}}}, \bibinfo {author}
  {\bibfnamefont {C.}~\bibnamefont {{Amsler}}}, \bibinfo {author}
  {\bibfnamefont {M.}~\bibnamefont {{Antonelli}}}, \bibinfo {author}
  {\bibfnamefont {E.~C.}\ \bibnamefont {{Aschenauer}}}, \bibinfo {author}
  {\bibfnamefont {D.~M.}\ \bibnamefont {{Asner}}}, \bibinfo {author}
  {\bibfnamefont {H.}~\bibnamefont {{Baer}}}, \bibinfo {author} {\bibfnamefont
  {S.}~\bibnamefont {{Banerjee}}}, \bibinfo {author} {\bibfnamefont
  {L.}~\bibnamefont {{Baudis}}}, \bibinfo {author} {\bibfnamefont {C.~W.}\
  \bibnamefont {{Bauer}}}, \bibinfo {author} {\bibfnamefont {J.~J.}\
  \bibnamefont {{Beatty}}}, \bibinfo {author} {\bibfnamefont {V.~I.}\
  \bibnamefont {{Belousov}}}, \bibinfo {author} {\bibfnamefont
  {S.}~\bibnamefont {{Bethke}}}, \bibinfo {author} {\bibfnamefont
  {A.}~\bibnamefont {{Bettini}}}, \bibinfo {author} {\bibfnamefont
  {O.}~\bibnamefont {{Biebel}}}, \bibinfo {author} {\bibfnamefont {K.~M.}\
  \bibnamefont {{Black}}}, \bibinfo {author} {\bibfnamefont {E.}~\bibnamefont
  {{Blucher}}}, \bibinfo {author} {\bibfnamefont {O.}~\bibnamefont
  {{Buchmuller}}}, \bibinfo {author} {\bibfnamefont {V.}~\bibnamefont
  {{Burkert}}}, \bibinfo {author} {\bibfnamefont {M.~A.}\ \bibnamefont
  {{Bychkov}}}, \bibinfo {author} {\bibfnamefont {R.~N.}\ \bibnamefont
  {{Cahn}}}, \bibinfo {author} {\bibfnamefont {M.}~\bibnamefont {{Carena}}},
  \bibinfo {author} {\bibfnamefont {A.}~\bibnamefont {{Ceccucci}}}, \bibinfo
  {author} {\bibfnamefont {A.}~\bibnamefont {{Cerri}}}, \bibinfo {author}
  {\bibfnamefont {D.}~\bibnamefont {{Chakraborty}}}, \bibinfo {author}
  {\bibfnamefont {R.~S.}\ \bibnamefont {{Chivukula}}}, \bibinfo {author}
  {\bibfnamefont {G.}~\bibnamefont {{Cowan}}}, \bibinfo {author} {\bibfnamefont
  {G.}~\bibnamefont {{D'Ambrosio}}}, \bibinfo {author} {\bibfnamefont
  {T.}~\bibnamefont {{Damour}}}, \bibinfo {author} {\bibfnamefont
  {D.}~\bibnamefont {{de Florian}}}, \bibinfo {author} {\bibfnamefont
  {A.}~\bibnamefont {{de Gouv{\^e}a}}}, \bibinfo {author} {\bibfnamefont
  {T.}~\bibnamefont {{DeGrand}}}, \bibinfo {author} {\bibfnamefont
  {P.}~\bibnamefont {{de Jong}}}, \bibinfo {author} {\bibfnamefont
  {G.}~\bibnamefont {{Dissertori}}}, \bibinfo {author} {\bibfnamefont {B.~A.}\
  \bibnamefont {{Dobrescu}}}, \bibinfo {author} {\bibfnamefont
  {M.}~\bibnamefont {{D'Onofrio}}}, \bibinfo {author} {\bibfnamefont
  {M.}~\bibnamefont {{Doser}}}, \bibinfo {author} {\bibfnamefont
  {M.}~\bibnamefont {{Drees}}}, \bibinfo {author} {\bibfnamefont {H.~K.}\
  \bibnamefont {{Dreiner}}}, \bibinfo {author} {\bibfnamefont {P.}~\bibnamefont
  {{Eerola}}}, \bibinfo {author} {\bibfnamefont {U.}~\bibnamefont {{Egede}}},
  \bibinfo {author} {\bibfnamefont {S.}~\bibnamefont {{Eidelman}}}, \bibinfo
  {author} {\bibfnamefont {J.}~\bibnamefont {{Ellis}}}, \bibinfo {author}
  {\bibfnamefont {J.}~\bibnamefont {{Erler}}}, \bibinfo {author} {\bibfnamefont
  {V.~V.}\ \bibnamefont {{Ezhela}}}, \bibinfo {author} {\bibfnamefont
  {W.}~\bibnamefont {{Fetscher}}}, \bibinfo {author} {\bibfnamefont {B.~D.}\
  \bibnamefont {{Fields}}}, \bibinfo {author} {\bibfnamefont {B.}~\bibnamefont
  {{Foster}}}, \bibinfo {author} {\bibfnamefont {A.}~\bibnamefont {{Freitas}}},
  \bibinfo {author} {\bibfnamefont {H.}~\bibnamefont {{Gallagher}}}, \bibinfo
  {author} {\bibfnamefont {L.}~\bibnamefont {{Garren}}}, \bibinfo {author}
  {\bibfnamefont {H.~J.}\ \bibnamefont {{Gerber}}}, \bibinfo {author}
  {\bibfnamefont {G.}~\bibnamefont {{Gerbier}}}, \bibinfo {author}
  {\bibfnamefont {T.}~\bibnamefont {{Gershon}}}, \bibinfo {author}
  {\bibfnamefont {Y.}~\bibnamefont {{Gershtein}}}, \bibinfo {author}
  {\bibfnamefont {T.}~\bibnamefont {{Gherghetta}}}, \bibinfo {author}
  {\bibfnamefont {A.~A.}\ \bibnamefont {{Godizov}}}, \bibinfo {author}
  {\bibfnamefont {M.~C.}\ \bibnamefont {{Gonzalez-Garcia}}}, \bibinfo {author}
  {\bibfnamefont {M.}~\bibnamefont {{Goodman}}}, \bibinfo {author}
  {\bibfnamefont {C.}~\bibnamefont {{Grab}}}, \bibinfo {author} {\bibfnamefont
  {A.~V.}\ \bibnamefont {{Gritsan}}}, \bibinfo {author} {\bibfnamefont
  {C.}~\bibnamefont {{Grojean}}}, \bibinfo {author} {\bibfnamefont
  {M.}~\bibnamefont {{Gr{\"u}newald}}}, \bibinfo {author} {\bibfnamefont
  {A.}~\bibnamefont {{Gurtu}}}, \bibinfo {author} {\bibfnamefont
  {T.}~\bibnamefont {{Gutsche}}}, \bibinfo {author} {\bibfnamefont {H.~E.}\
  \bibnamefont {{Haber}}}, \bibinfo {author} {\bibfnamefont {C.}~\bibnamefont
  {{Hanhart}}}, \bibinfo {author} {\bibfnamefont {S.}~\bibnamefont
  {{Hashimoto}}}, \bibinfo {author} {\bibfnamefont {Y.}~\bibnamefont
  {{Hayato}}}, \bibinfo {author} {\bibfnamefont {A.}~\bibnamefont
  {{Hebecker}}}, \bibinfo {author} {\bibfnamefont {S.}~\bibnamefont
  {{Heinemeyer}}}, \bibinfo {author} {\bibfnamefont {B.}~\bibnamefont
  {{Heltsley}}}, \bibinfo {author} {\bibfnamefont {J.~J.}\ \bibnamefont
  {{Hern{\'a}ndez-Rey}}}, \bibinfo {author} {\bibfnamefont {K.}~\bibnamefont
  {{Hikasa}}}, \bibinfo {author} {\bibfnamefont {J.}~\bibnamefont {{Hisano}}},
  \bibinfo {author} {\bibfnamefont {A.}~\bibnamefont {{H{\"o}cker}}}, \bibinfo
  {author} {\bibfnamefont {J.}~\bibnamefont {{Holder}}}, \bibinfo {author}
  {\bibfnamefont {A.}~\bibnamefont {{Holtkamp}}}, \bibinfo {author}
  {\bibfnamefont {J.}~\bibnamefont {{Huston}}}, \bibinfo {author}
  {\bibfnamefont {T.}~\bibnamefont {{Hyodo}}}, \bibinfo {author} {\bibfnamefont
  {K.~F.}\ \bibnamefont {{Johnson}}}, \bibinfo {author} {\bibfnamefont
  {M.}~\bibnamefont {{Kado}}}, \bibinfo {author} {\bibfnamefont
  {M.}~\bibnamefont {{Karliner}}}, \bibinfo {author} {\bibfnamefont {U.~F.}\
  \bibnamefont {{Katz}}}, \bibinfo {author} {\bibfnamefont {M.}~\bibnamefont
  {{Kenzie}}}, \bibinfo {author} {\bibfnamefont {V.~A.}\ \bibnamefont
  {{Khoze}}}, \bibinfo {author} {\bibfnamefont {S.~R.}\ \bibnamefont
  {{Klein}}}, \bibinfo {author} {\bibfnamefont {E.}~\bibnamefont {{Klempt}}},
  \bibinfo {author} {\bibfnamefont {R.~V.}\ \bibnamefont {{Kowalewski}}},
  \bibinfo {author} {\bibfnamefont {F.}~\bibnamefont {{Krauss}}}, \bibinfo
  {author} {\bibfnamefont {M.}~\bibnamefont {{Kreps}}}, \bibinfo {author}
  {\bibfnamefont {B.}~\bibnamefont {{Krusche}}}, \bibinfo {author}
  {\bibfnamefont {Y.}~\bibnamefont {{Kwon}}}, \bibinfo {author} {\bibfnamefont
  {O.}~\bibnamefont {{Lahav}}}, \bibinfo {author} {\bibfnamefont
  {J.}~\bibnamefont {{Laiho}}}, \bibinfo {author} {\bibfnamefont {L.~P.}\
  \bibnamefont {{Lellouch}}}, \bibinfo {author} {\bibfnamefont
  {J.}~\bibnamefont {{Lesgourgues}}}, \bibinfo {author} {\bibfnamefont {A.~R.}\
  \bibnamefont {{Liddle}}}, \bibinfo {author} {\bibfnamefont {Z.}~\bibnamefont
  {{Ligeti}}}, \bibinfo {author} {\bibfnamefont {C.}~\bibnamefont
  {{Lippmann}}}, \bibinfo {author} {\bibfnamefont {T.~M.}\ \bibnamefont
  {{Liss}}}, \bibinfo {author} {\bibfnamefont {L.}~\bibnamefont
  {{Littenberg}}}, \bibinfo {author} {\bibfnamefont {C.}~\bibnamefont
  {{Lourengo}}}, \bibinfo {author} {\bibfnamefont {S.~B.}\ \bibnamefont
  {{Lugovsky}}}, \bibinfo {author} {\bibfnamefont {A.}~\bibnamefont
  {{Lusiani}}}, \bibinfo {author} {\bibfnamefont {Y.}~\bibnamefont {{Makida}}},
  \bibinfo {author} {\bibfnamefont {F.}~\bibnamefont {{Maltoni}}}, \bibinfo
  {author} {\bibfnamefont {T.}~\bibnamefont {{Mannel}}}, \bibinfo {author}
  {\bibfnamefont {A.~V.}\ \bibnamefont {{Manohar}}}, \bibinfo {author}
  {\bibfnamefont {W.~J.}\ \bibnamefont {{Marciano}}}, \bibinfo {author}
  {\bibfnamefont {A.}~\bibnamefont {{Masoni}}}, \bibinfo {author}
  {\bibfnamefont {J.}~\bibnamefont {{Matthews}}}, \bibinfo {author}
  {\bibfnamefont {U.~G.}\ \bibnamefont {{Mei{\ss}ner}}}, \bibinfo {author}
  {\bibfnamefont {M.}~\bibnamefont {{Mikhasenko}}}, \bibinfo {author}
  {\bibfnamefont {D.~J.}\ \bibnamefont {{Miller}}}, \bibinfo {author}
  {\bibfnamefont {D.}~\bibnamefont {{Milstead}}}, \bibinfo {author}
  {\bibfnamefont {R.~E.}\ \bibnamefont {{Mitchell}}}, \bibinfo {author}
  {\bibfnamefont {K.}~\bibnamefont {{M{\"o}nig}}}, \bibinfo {author}
  {\bibfnamefont {P.}~\bibnamefont {{Molaro}}}, \bibinfo {author}
  {\bibfnamefont {F.}~\bibnamefont {{Moortgat}}}, \bibinfo {author}
  {\bibfnamefont {M.}~\bibnamefont {{Moskovic}}}, \bibinfo {author}
  {\bibfnamefont {K.}~\bibnamefont {{Nakamura}}}, \bibinfo {author}
  {\bibfnamefont {M.}~\bibnamefont {{Narain}}}, \bibinfo {author}
  {\bibfnamefont {P.}~\bibnamefont {{Nason}}}, \bibinfo {author} {\bibfnamefont
  {S.}~\bibnamefont {{Navas}}}, \bibinfo {author} {\bibfnamefont
  {M.}~\bibnamefont {{Neubert}}}, \bibinfo {author} {\bibfnamefont
  {P.}~\bibnamefont {{Nevski}}}, \bibinfo {author} {\bibfnamefont
  {Y.}~\bibnamefont {{Nir}}}, \bibinfo {author} {\bibfnamefont {K.~A.}\
  \bibnamefont {{Olive}}}, \bibinfo {author} {\bibfnamefont {C.}~\bibnamefont
  {{Patrignani}}}, \bibinfo {author} {\bibfnamefont {J.~A.}\ \bibnamefont
  {{Peacock}}}, \bibinfo {author} {\bibfnamefont {S.~T.}\ \bibnamefont
  {{Petcov}}}, \bibinfo {author} {\bibfnamefont {V.~A.}\ \bibnamefont
  {{Petrov}}}, \bibinfo {author} {\bibfnamefont {A.}~\bibnamefont {{Pich}}},
  \bibinfo {author} {\bibfnamefont {A.}~\bibnamefont {{Piepke}}}, \bibinfo
  {author} {\bibfnamefont {A.}~\bibnamefont {{Pomarol}}}, \bibinfo {author}
  {\bibfnamefont {S.}~\bibnamefont {{Profumo}}}, \bibinfo {author}
  {\bibfnamefont {A.}~\bibnamefont {{Quadt}}}, \bibinfo {author} {\bibfnamefont
  {K.}~\bibnamefont {{Rabbertz}}}, \bibinfo {author} {\bibfnamefont
  {J.}~\bibnamefont {{Rademacker}}}, \bibinfo {author} {\bibfnamefont
  {G.}~\bibnamefont {{Raffelt}}}, \bibinfo {author} {\bibfnamefont
  {H.}~\bibnamefont {{Ramani}}}, \bibinfo {author} {\bibfnamefont
  {M.}~\bibnamefont {{Ramsey-Musolf}}}, \bibinfo {author} {\bibfnamefont
  {B.~N.}\ \bibnamefont {{Ratcliff}}}, \bibinfo {author} {\bibfnamefont
  {P.}~\bibnamefont {{Richardson}}}, \bibinfo {author} {\bibfnamefont
  {A.}~\bibnamefont {{Ringwald}}}, \bibinfo {author} {\bibfnamefont
  {S.}~\bibnamefont {{Roesler}}}, \bibinfo {author} {\bibfnamefont
  {S.}~\bibnamefont {{Rolli}}}, \bibinfo {author} {\bibfnamefont
  {A.}~\bibnamefont {{Romaniouk}}}, \bibinfo {author} {\bibfnamefont {L.~J.}\
  \bibnamefont {{Rosenberg}}}, \bibinfo {author} {\bibfnamefont {J.~L.}\
  \bibnamefont {{Rosner}}}, \bibinfo {author} {\bibfnamefont {G.}~\bibnamefont
  {{Rybka}}}, \bibinfo {author} {\bibfnamefont {M.}~\bibnamefont {{Ryskin}}},
  \bibinfo {author} {\bibfnamefont {R.~A.}\ \bibnamefont {{Ryutin}}}, \bibinfo
  {author} {\bibfnamefont {Y.}~\bibnamefont {{Sakai}}}, \bibinfo {author}
  {\bibfnamefont {G.~P.}\ \bibnamefont {{Salam}}}, \bibinfo {author}
  {\bibfnamefont {S.}~\bibnamefont {{Sarkar}}}, \bibinfo {author}
  {\bibfnamefont {F.}~\bibnamefont {{Sauli}}}, \bibinfo {author} {\bibfnamefont
  {O.}~\bibnamefont {{Schneider}}}, \bibinfo {author} {\bibfnamefont
  {K.}~\bibnamefont {{Scholberg}}}, \bibinfo {author} {\bibfnamefont {A.~J.}\
  \bibnamefont {{Schwartz}}}, \bibinfo {author} {\bibfnamefont
  {J.}~\bibnamefont {{Schwiening}}}, \bibinfo {author} {\bibfnamefont
  {D.}~\bibnamefont {{Scott}}}, \bibinfo {author} {\bibfnamefont
  {V.}~\bibnamefont {{Sharma}}}, \bibinfo {author} {\bibfnamefont {S.~R.}\
  \bibnamefont {{Sharpe}}}, \bibinfo {author} {\bibfnamefont {T.}~\bibnamefont
  {{Shutt}}}, \bibinfo {author} {\bibfnamefont {M.}~\bibnamefont {{Silari}}},
  \bibinfo {author} {\bibfnamefont {T.}~\bibnamefont {{Sj{\"o}strand}}},
  \bibinfo {author} {\bibfnamefont {P.}~\bibnamefont {{Skands}}}, \bibinfo
  {author} {\bibfnamefont {T.}~\bibnamefont {{Skwarnicki}}}, \bibinfo {author}
  {\bibfnamefont {G.~F.}\ \bibnamefont {{Smoot}}}, \bibinfo {author}
  {\bibfnamefont {A.}~\bibnamefont {{Soffer}}}, \bibinfo {author}
  {\bibfnamefont {M.~S.}\ \bibnamefont {{Sozzi}}}, \bibinfo {author}
  {\bibfnamefont {S.}~\bibnamefont {{Spanier}}}, \bibinfo {author}
  {\bibfnamefont {C.}~\bibnamefont {{Spiering}}}, \bibinfo {author}
  {\bibfnamefont {A.}~\bibnamefont {{Stahl}}}, \bibinfo {author} {\bibfnamefont
  {S.~L.}\ \bibnamefont {{Stone}}}, \bibinfo {author} {\bibfnamefont
  {Y.}~\bibnamefont {{Sumino}}}, \bibinfo {author} {\bibfnamefont
  {T.}~\bibnamefont {{Sumiyoshi}}}, \bibinfo {author} {\bibfnamefont {M.~J.}\
  \bibnamefont {{Syphers}}}, \bibinfo {author} {\bibfnamefont {F.}~\bibnamefont
  {{Takahashi}}}, \bibinfo {author} {\bibfnamefont {M.}~\bibnamefont
  {{Tanabashi}}}, \bibinfo {author} {\bibfnamefont {J.}~\bibnamefont
  {{Tanaka}}}, \bibinfo {author} {\bibfnamefont {M.}~\bibnamefont
  {{Ta{\v{s}}evsk{\'y}}}}, \bibinfo {author} {\bibfnamefont {K.}~\bibnamefont
  {{Terashi}}}, \bibinfo {author} {\bibfnamefont {J.}~\bibnamefont
  {{Terning}}}, \bibinfo {author} {\bibfnamefont {U.}~\bibnamefont {{Thoma}}},
  \bibinfo {author} {\bibfnamefont {R.~S.}\ \bibnamefont {{Thorne}}}, \bibinfo
  {author} {\bibfnamefont {L.}~\bibnamefont {{Tiator}}}, \bibinfo {author}
  {\bibfnamefont {M.}~\bibnamefont {{Titov}}}, \bibinfo {author} {\bibfnamefont
  {N.~P.}\ \bibnamefont {{Tkachenko}}}, \bibinfo {author} {\bibfnamefont
  {D.~R.}\ \bibnamefont {{Tovey}}}, \bibinfo {author} {\bibfnamefont
  {K.}~\bibnamefont {{Trabelsi}}}, \bibinfo {author} {\bibfnamefont
  {P.}~\bibnamefont {{Urquijo}}}, \bibinfo {author} {\bibfnamefont
  {G.}~\bibnamefont {{Valencia}}}, \bibinfo {author} {\bibfnamefont
  {R.}~\bibnamefont {{Van de Water}}}, \bibinfo {author} {\bibfnamefont
  {N.}~\bibnamefont {{Varelas}}}, \bibinfo {author} {\bibfnamefont
  {G.}~\bibnamefont {{Venanzoni}}}, \bibinfo {author} {\bibfnamefont
  {L.}~\bibnamefont {{Verde}}}, \bibinfo {author} {\bibfnamefont {M.~G.}\
  \bibnamefont {{Vincter}}}, \bibinfo {author} {\bibfnamefont {P.}~\bibnamefont
  {{Vogel}}}, \bibinfo {author} {\bibfnamefont {W.}~\bibnamefont
  {{Vogelsang}}}, \bibinfo {author} {\bibfnamefont {A.}~\bibnamefont {{Vogt}}},
  \bibinfo {author} {\bibfnamefont {V.}~\bibnamefont {{Vorobyev}}}, \bibinfo
  {author} {\bibfnamefont {S.~P.}\ \bibnamefont {{Wakely}}}, \bibinfo {author}
  {\bibfnamefont {W.}~\bibnamefont {{Walkowiak}}}, \bibinfo {author}
  {\bibfnamefont {C.~W.}\ \bibnamefont {{Walter}}}, \bibinfo {author}
  {\bibfnamefont {D.}~\bibnamefont {{Wands}}}, \bibinfo {author} {\bibfnamefont
  {M.~O.}\ \bibnamefont {{Wascko}}}, \bibinfo {author} {\bibfnamefont {D.~H.}\
  \bibnamefont {{Weinberg}}}, \bibinfo {author} {\bibfnamefont {E.~J.}\
  \bibnamefont {{Weinberg}}}, \bibinfo {author} {\bibfnamefont
  {M.}~\bibnamefont {{White}}}, \bibinfo {author} {\bibfnamefont {L.~R.}\
  \bibnamefont {{Wiencke}}}, \bibinfo {author} {\bibfnamefont {S.}~\bibnamefont
  {{Willocq}}}, \bibinfo {author} {\bibfnamefont {C.~L.}\ \bibnamefont
  {{Woody}}}, \bibinfo {author} {\bibfnamefont {R.~L.}\ \bibnamefont
  {{Workman}}}, \bibinfo {author} {\bibfnamefont {M.}~\bibnamefont
  {{Yokoyama}}}, \bibinfo {author} {\bibfnamefont {R.}~\bibnamefont
  {{Yoshida}}}, \bibinfo {author} {\bibfnamefont {G.}~\bibnamefont
  {{Zanderighi}}}, \bibinfo {author} {\bibfnamefont {G.~P.}\ \bibnamefont
  {{Zeller}}}, \bibinfo {author} {\bibfnamefont {O.~V.}\ \bibnamefont
  {{Zenin}}}, \bibinfo {author} {\bibfnamefont {R.~Y.}\ \bibnamefont {{Zhu}}},
  \bibinfo {author} {\bibfnamefont {S.~L.}\ \bibnamefont {{Zhu}}}, \bibinfo
  {author} {\bibfnamefont {F.}~\bibnamefont {{Zimmermann}}}, \bibinfo {author}
  {\bibfnamefont {J.}~\bibnamefont {{Anderson}}}, \bibinfo {author}
  {\bibfnamefont {T.}~\bibnamefont {{Basaglia}}}, \bibinfo {author}
  {\bibfnamefont {V.~S.}\ \bibnamefont {{Lugovsky}}}, \bibinfo {author}
  {\bibfnamefont {P.}~\bibnamefont {{Schaffner}}},\ and\ \bibinfo {author}
  {\bibfnamefont {W.}~\bibnamefont {{Zheng}}},\ }\bibfield  {title} {\bibinfo
  {title} {{Review of Particle Physics}},\ }\href
  {https://doi.org/10.1093/ptep/ptaa104} {\bibfield  {journal} {\bibinfo
  {journal} {Progress of Theoretical and Experimental Physics}\ }\textbf
  {\bibinfo {volume} {2020}},\ \bibinfo {eid} {083C01} (\bibinfo {year}
  {2020})}\BibitemShut {NoStop}%
\bibitem [{\citenamefont {{Ellis}}\ \emph {et~al.}(2019)\citenamefont
  {{Ellis}}, \citenamefont {{Vallisneri}}, \citenamefont {{Taylor}},\ and\
  \citenamefont {{Baker}}}]{enterprise}%
  \BibitemOpen
  \bibfield  {author} {\bibinfo {author} {\bibfnamefont {J.~A.}\ \bibnamefont
  {{Ellis}}}, \bibinfo {author} {\bibfnamefont {M.}~\bibnamefont
  {{Vallisneri}}}, \bibinfo {author} {\bibfnamefont {S.~R.}\ \bibnamefont
  {{Taylor}}},\ and\ \bibinfo {author} {\bibfnamefont {P.~T.}\ \bibnamefont
  {{Baker}}},\ }\href@noop {} {\bibinfo {title} {{ENTERPRISE: Enhanced
  Numerical Toolbox Enabling a Robust PulsaR Inference SuitE}}} (\bibinfo
  {year} {2019}),\ \Eprint {https://arxiv.org/abs/1912.015} {ascl:1912.015}
  \BibitemShut {NoStop}%
\bibitem [{\citenamefont {{Taylor}}\ \emph {et~al.}(2018)\citenamefont
  {{Taylor}}, \citenamefont {{Baker}}, \citenamefont {{Hazboun}}, \citenamefont
  {{Simon}},\ and\ \citenamefont {{Vigeland}}}]{enterprise_ext}%
  \BibitemOpen
  \bibfield  {author} {\bibinfo {author} {\bibfnamefont {S.~R.}\ \bibnamefont
  {{Taylor}}}, \bibinfo {author} {\bibfnamefont {P.~T.}\ \bibnamefont
  {{Baker}}}, \bibinfo {author} {\bibfnamefont {J.~S.}\ \bibnamefont
  {{Hazboun}}}, \bibinfo {author} {\bibfnamefont {J.~J.}\ \bibnamefont
  {{Simon}}},\ and\ \bibinfo {author} {\bibfnamefont {S.~J.}\ \bibnamefont
  {{Vigeland}}},\ }\href {https://github.com/nanograv/enterprise_extensions}
  {\bibinfo {title} {enterprise extensions}} (\bibinfo {year}
  {2018})\BibitemShut {NoStop}%
\bibitem [{\citenamefont {{Hazboun}}\ \emph {et~al.}(2019)\citenamefont
  {{Hazboun}}, \citenamefont {{Romano}},\ and\ \citenamefont
  {{Smith}}}]{hasasia}%
  \BibitemOpen
  \bibfield  {author} {\bibinfo {author} {\bibfnamefont {J.}~\bibnamefont
  {{Hazboun}}}, \bibinfo {author} {\bibfnamefont {J.}~\bibnamefont
  {{Romano}}},\ and\ \bibinfo {author} {\bibfnamefont {T.}~\bibnamefont
  {{Smith}}},\ }\bibfield  {title} {\bibinfo {title} {{Hasasia: A Python
  package for Pulsar Timing Array Sensitivity Curves}},\ }\href
  {https://doi.org/10.21105/joss.01775} {\bibfield  {journal} {\bibinfo
  {journal} {The Journal of Open Source Software}\ }\textbf {\bibinfo {volume}
  {4}},\ \bibinfo {eid} {1775} (\bibinfo {year} {2019})}\BibitemShut {NoStop}%
\bibitem [{\citenamefont {{Vallisneri}}(2020)}]{libstempo}%
  \BibitemOpen
  \bibfield  {author} {\bibinfo {author} {\bibfnamefont {M.}~\bibnamefont
  {{Vallisneri}}},\ }\href@noop {} {\bibinfo {title} {{libstempo: Python
  wrapper for Tempo2}}} (\bibinfo {year} {2020}),\ \Eprint
  {https://arxiv.org/abs/2002.017} {ascl:2002.017} \BibitemShut {NoStop}%
\bibitem [{\citenamefont {{Hunter}}(2007)}]{matplotlib}%
  \BibitemOpen
  \bibfield  {author} {\bibinfo {author} {\bibfnamefont {J.~D.}\ \bibnamefont
  {{Hunter}}},\ }\bibfield  {title} {\bibinfo {title} {{Matplotlib: A 2D
  Graphics Environment}},\ }\href {https://doi.org/10.1109/MCSE.2007.55}
  {\bibfield  {journal} {\bibinfo  {journal} {Computing in Science and
  Engineering}\ }\textbf {\bibinfo {volume} {9}},\ \bibinfo {pages} {90}
  (\bibinfo {year} {2007})}\BibitemShut {NoStop}%
\bibitem [{\citenamefont {Ellis}\ and\ \citenamefont {van
  Haasteren}(2017)}]{ptmcmc}%
  \BibitemOpen
  \bibfield  {author} {\bibinfo {author} {\bibfnamefont {J.}~\bibnamefont
  {Ellis}}\ and\ \bibinfo {author} {\bibfnamefont {R.}~\bibnamefont {van
  Haasteren}},\ }\href {https://doi.org/10.5281/zenodo.1037579} {\bibinfo
  {title} {jellis18/ptmcmcsampler: Official release}} (\bibinfo {year}
  {2017})\BibitemShut {NoStop}%
\bibitem [{\citenamefont {{Nice}}\ \emph {et~al.}(2015)\citenamefont {{Nice}},
  \citenamefont {{Demorest}}, \citenamefont {{Stairs}}, \citenamefont
  {{Manchester}}, \citenamefont {{Taylor}}, \citenamefont {{Peters}},
  \citenamefont {{Weisberg}}, \citenamefont {{Irwin}}, \citenamefont {{Wex}},\
  and\ \citenamefont {{Huang}}}]{tempo}%
  \BibitemOpen
  \bibfield  {author} {\bibinfo {author} {\bibfnamefont {D.}~\bibnamefont
  {{Nice}}}, \bibinfo {author} {\bibfnamefont {P.}~\bibnamefont {{Demorest}}},
  \bibinfo {author} {\bibfnamefont {I.}~\bibnamefont {{Stairs}}}, \bibinfo
  {author} {\bibfnamefont {R.}~\bibnamefont {{Manchester}}}, \bibinfo {author}
  {\bibfnamefont {J.}~\bibnamefont {{Taylor}}}, \bibinfo {author}
  {\bibfnamefont {W.}~\bibnamefont {{Peters}}}, \bibinfo {author}
  {\bibfnamefont {J.}~\bibnamefont {{Weisberg}}}, \bibinfo {author}
  {\bibfnamefont {A.}~\bibnamefont {{Irwin}}}, \bibinfo {author} {\bibfnamefont
  {N.}~\bibnamefont {{Wex}}},\ and\ \bibinfo {author} {\bibfnamefont
  {Y.}~\bibnamefont {{Huang}}},\ }\href@noop {} {\bibinfo {title} {{Tempo:
  Pulsar timing data analysis}}} (\bibinfo {year} {2015}),\ \Eprint
  {https://arxiv.org/abs/1509.002} {ascl:1509.002} \BibitemShut {NoStop}%
\bibitem [{\citenamefont {{Hobbs}}\ and\ \citenamefont
  {{Edwards}}(2012)}]{tempo2}%
  \BibitemOpen
  \bibfield  {author} {\bibinfo {author} {\bibfnamefont {G.}~\bibnamefont
  {{Hobbs}}}\ and\ \bibinfo {author} {\bibfnamefont {R.}~\bibnamefont
  {{Edwards}}},\ }\href@noop {} {\bibinfo {title} {{Tempo2: Pulsar Timing
  Package}}} (\bibinfo {year} {2012}),\ \Eprint
  {https://arxiv.org/abs/1210.015} {ascl:1210.015} \BibitemShut {NoStop}%
\bibitem [{\citenamefont {{Luo}}\ \emph {et~al.}(2019)\citenamefont {{Luo}},
  \citenamefont {{Ransom}}, \citenamefont {{Demorest}}, \citenamefont {{van
  Haasteren}}, \citenamefont {{Ray}}, \citenamefont {{Stovall}}, \citenamefont
  {{Bachetti}}, \citenamefont {{Archibald}}, \citenamefont {{Kerr}},
  \citenamefont {{Colen}},\ and\ \citenamefont {{Jenet}}}]{pint}%
  \BibitemOpen
  \bibfield  {author} {\bibinfo {author} {\bibfnamefont {J.}~\bibnamefont
  {{Luo}}}, \bibinfo {author} {\bibfnamefont {S.}~\bibnamefont {{Ransom}}},
  \bibinfo {author} {\bibfnamefont {P.}~\bibnamefont {{Demorest}}}, \bibinfo
  {author} {\bibfnamefont {R.}~\bibnamefont {{van Haasteren}}}, \bibinfo
  {author} {\bibfnamefont {P.}~\bibnamefont {{Ray}}}, \bibinfo {author}
  {\bibfnamefont {K.}~\bibnamefont {{Stovall}}}, \bibinfo {author}
  {\bibfnamefont {M.}~\bibnamefont {{Bachetti}}}, \bibinfo {author}
  {\bibfnamefont {A.}~\bibnamefont {{Archibald}}}, \bibinfo {author}
  {\bibfnamefont {M.}~\bibnamefont {{Kerr}}}, \bibinfo {author} {\bibfnamefont
  {J.}~\bibnamefont {{Colen}}},\ and\ \bibinfo {author} {\bibfnamefont
  {F.}~\bibnamefont {{Jenet}}},\ }\href@noop {} {\bibinfo {title} {{PINT:
  High-precision pulsar timing analysis package}}} (\bibinfo {year} {2019}),\
  \Eprint {https://arxiv.org/abs/1902.007} {ascl:1902.007} \BibitemShut
  {NoStop}%
\end{thebibliography}%
